\documentclass[preprint,prd,showpacs,showkeys]{revtex4}
\usepackage{graphicx}
\usepackage{color}
\usepackage{amsmath}
\usepackage{amssymb, graphics}
\usepackage{multirow}
\usepackage{latexsym}
\usepackage{bm}
\usepackage{wrapfig}
\usepackage{fancybox}
\def\be{\begin{equation}}
\def\ee{\end{equation}}
\def\bea{\begin{eqnarray}}
\def\eea{\end{eqnarray}}
\begin{document}

\title{Geometrothermodynamics of five dimensional black holes in Einstein-Gauss-Bonnet
theory }

\author{Safia Taj$^{1,3}$, Hernando Quevedo$^{2,3}$ and Alberto S\'anchez$^4$}
\email{safiataaj@gmail.com, quevedo@nucleares.unam.mx}
\affiliation{
$^1$Center for Advanced Mathematics and Physics, National University of Sciences and Technology, H-12, Islamabad, Pakistan\\
$^2$Instituto de Ciencias Nucleares, Universidad Nacional Aut\'onoma de M\'exico, AP 70543, M\'exico, DF 04510, Mexico\\
$^3$Dipartimento di Fisica and ICRA, Universit\`a di Roma "La Sapienza", I-00185 Roma, Italy\\
$^4$Departamento de Posgrado, CIIDET, AP 752, Quer\'etaro, QRO 76000, Mexico
}

\date{\today}

\pacs{04.25.Nx; 04.80.Cc; 04.50.Kd}

\begin{abstract}
We investigate the thermodynamic properties of 5D static and
spherically symmetric black holes in (i)
Einstein-Maxwell-Gauss-Bonnet theory, (ii)
Einstein-Maxwell-Gauss-Bonnet theory with negative cosmological
constant, and in (iii) Einstein-Yang-Mills-Gauss-Bonnet theory. To
formulate the thermodynamics of these black holes we use the
Bekenstein-Hawking entropy relation and, alternatively, a modified
entropy formula which follows from the first law of thermodynamics
of black holes. The results of both approaches are not equivalent.
Using the formalism of geometrothermodynamics, we introduce in the
manifold of equilibrium states a Legendre invariant metric for each
black hole and for each thermodynamic approach, and show that the
thermodynamic curvature diverges at those points where the
temperature vanishes and the heat capacity diverges.

\keywords{Geometrothermodynamics, Einstein-Gauss-Bonnet theory,
Phase transition}

\end{abstract}

\maketitle

\section{Introduction}
\label{intro}

Black holes can be regarded as thermodynamical systems \cite{A1,A2}
which radiate Hawking thermal radiation of temperature proportional
to the surface gravity on the horizon of the black hole, and
Bekenstein-Hawking entropy proportional to its horizon area
\cite{A3,A4}. Indeed, these quantities satisfy the first law of
black hole thermodynamics \cite{A5}. Finding the microscopic
description of this entropy is one of the most challenging questions
in theoretical physics. The solution to this problem  still remains obscure.
According to
the no-hair theorems of Einstein-Maxwell theory, electro-vacuum
black holes are completely described by three parameters only: mass
$M$, angular momentum $J$, and electric charge $Q$. All these parameters
define the heat capacity whose sign, in turn, determines the
thermodynamical stability of the black hole.

During the last few decades several attempts have been made in order
to introduce Riemannian geometric concepts in ordinary
thermodynamics. First, Weinhold \cite{A6} introduced on the space of
equilibrium states a metric whose components are given as the
Hessian of the internal thermodynamic energy. Later, Ruppeiner
\cite{A7} introduced a metric which is defined as minus the Hessian
of the entropy, and is conformally equivalent to Weinhold's metric,
with the inverse of the temperature as the conformal factor. 
The physical meaning of Ruppeiner's metric lays in the fluctuation
theory of equilibrium thermodynamics. It turns out that the second
moments of fluctuation are related to the components of
Ruppeiner's metric.

One of
the aims of the application of geometry in thermodynamics is to
describe phase transitions in terms of curvature singularities so
that the curvature can be interpreted as a measure of thermodynamic
interaction. The study of the relation between the phase space and
the metric structures of the space of equilibrium states led to the
result that Weinhold's and Ruppeiner's thermodynamic metrics are not
invariant under Legendre transformations \cite{A8}.
On the other hand, ordinary thermodynamics is invariant with respect 
to Legendre transformations, i. e., the physical properties of a 
thermodynamic system
do not change  when different thermodynamic potentials are used 
\cite{callen}. One might then wonder whether the use of 
non Legendre invariant metric structures in ordinary thermodynamics 
would always lead to results that do not depend on the thermodynamic 
potential. Indeed, several examples are known in the literature 
in which a change of thermodynamic potential leads to a modification 
of the thermodynamic geometry \cite{contra}. Some puzzling results 
arise also in connection with the use of different metrics in the 
equilibrium space \cite{puzzles}, in the sense that for the same 
thermodynamic system the resulting geometry
can be either flat or curved, depending on the chosen metric. 
Recently, in the analysis of Einstein-Gauss-Bonnet (EGB) black holes 
inconsistencies were also found \cite{egbinc}. In this work, 
we will focus on the study of EGB black holes and will clarify 
these inconsistencies.

Recently, the formalism of geometrothermodynamics (GTD) was
developed in order to unify in a consistent manner the geometric
properties of the phase space and the space of equilibrium states
\cite{A9}. Legendre invariance plays an important role in this
formalism. In particular, it allows us to derive Legendre invariant
metrics for the space of equilibrium states. The thermodynamic phase
space ${\cal T}$ is assumed to be coordinatized by the set of independent
coordinates ${\Phi,E^{a}, I^{a}}, a = 1, ..., n$, where $\Phi$
represents the thermodynamic potential, and $E^{a}$ and $I^{a}$ are
the extensive and intensive thermodynamic variables, respectively 
The positive integer $n$ indicates the number of thermodynamic degrees
of freedom of the black hole configuration. Moreover, the phase
space is endowed with
the Gibbs one-form $\Theta = d\Phi - \delta_{ab}I^{a}dE^{b}$, with  $%
\delta_{ab}={\rm diag}(1,...,1)$, satisfying the condition
$\Theta\wedge(d\Theta)^{n}\neq0$. Consider also on ${\cal T}$ 
a Riemannian metric $G$ which must be invariant with respect to 
Legendre transformations of the form $(%
\Phi,E^{a},I^{a})\rightarrow(\tilde{\Phi},\tilde{E^{a}},\tilde{I^{a}})$,
with $\Phi=\tilde{\Phi}-\delta_{ab}\tilde{E^{a}}\tilde{I^{b}}$, $E^{a}=-%
\tilde{I^{a}}$, $I^{a}=\tilde{E^{a}}$ \cite{A10}. We impose the Legendre 
invariance condition in order to incorporate in GTD the invariance 
of ordinary thermodynamics with respect to changes of the thermodynamic potential 
\cite{callen}. It turns out that
the condition of Legendre invariance is not sufficient to fix uniquely the metric $G$.
In fact, Legendre invariance generates the metric
\begin{equation}
G = \left(d\Phi - I_a dE^a\right)^2  +\Lambda
\left(\xi_{ab}E^{a}I^{b}\right)\left(\chi_{cd}dE^{c}dI^{d}\right) \  ,
\label{gup1}
\end{equation}
where $\Lambda$ is an arbitrary real constant, and $\xi_{ab}$ and $\chi_{ab}$ are diagonal constant
tensors. Clearly, the diagonal components of these tensors can be normalized by rescaling the coordinates
$E^a$ and $I^a$, and by choosing the multiplicative constant $\Lambda$ appropriately. 
Then, one can express  $\xi_{ab}$ and $\chi_{ab}$ in terms of the 
usual Euclidean metric, $\delta_{ab}={\rm diag}(1,...,1)$, and the pseudo-Euclidean metric,  
$\eta_{ab} = {\rm diag}(-1, 1, ..., 1)$. Physical conditions must be invoked to fix the final form of these
tensors \cite{qstv10a}. Indeed, it turns out that the choice 
$ \xi_{ab}=\delta_{ab}$ and $\chi_{ab}=\delta_{ab}$ 
$(\xi_{ab}=\delta_{ab}$ and $\chi_{ab}=\eta_{ab})$ 
leads to a metric which 
describes systems characterized by first (second) order phase transitions.
Moreover, the choice  
$\xi_{ab}= \left(\delta_{ab}-\eta_{ab}\right)/2$ 
allows us also to correctly handle the zero-temperature limit in a geometric way. We see that
Legendre invariance leaves free only the signature of $\chi_{ab}$. The signature, in turn, is fixed 
by the order of the phase transition under consideration 
\footnote{This simple observation turns out to be the starting
point to formulate an invariant classification of phase transitions 
(H. Quevedo, {\it An invariant classification of phase transitions} (2012), in preparation) 
which can be used,
in particular, to clarify certain puzzling results found recently in 
black hole  phase transitions by N. Pidokrajt, and J. Ward, arXiv:gr-qc/1106.2831.}.
Since in this work we will analyze 
second order phase transitions of EGB black holes \cite{egbinc}, we choose the metric  as
\begin{equation}
G=(d\Phi-\delta_{ab}I^{a}dE^{b})^{2}+\frac{1}{2}(\delta_{ab}-\eta_{ab}) E^{a}I^{b}
(\eta_{cd}dE^{c}dI^{d})\ .
\end{equation}

The set $({\cal T},\Theta,G)$ defines a Legendre invariant manifold with a contact
Riemannian structure. The equilibrium space ${\cal E} \subset T$
is defined by the map $\varphi: {\cal E}\rightarrow {\cal T}$ or, in
local coordinates, $\varphi: (E^{a})
\mapsto ({\Phi,E^{a},I^{a}})$, satisfying the condition $%
\varphi^{\ast}(\Theta)=0$, i. e., on ${\cal E}$ it holds the first
law of thermodynamics, $d\Phi = \delta_{ab}I^{a}dE^{b}$, and the
conditions of thermodynamic equilibrium
$I^{a}=\delta^{ab}\partial{\Phi}/\partial{E^{b}}$,
which relate the extensive variables $E^{a}$ with the intensive ones $I^{a}$.
Then, the pullback $\varphi^{\ast}$ induces on ${\cal E}$, by means of $g
= \varphi^{\ast}(G)$, the thermodynamic metric
\begin{equation}
g=\frac{1}{2}\left[  E^{a}\left( \frac{\partial{\Phi}}{\partial{E^{a}}} -
\eta_{ab}\delta^{bc}\frac{\partial{\Phi}}{\partial{E^{c}}}\right)\right]
\left(\eta
_{ab}\delta^{bc}\frac{\partial^{2}\Phi}{\partial {E^{c}}\partial{E^{d}}}
dE^a dE ^d
\right).
\label{gdown}
\end{equation}
For the construction of this thermodynamic metric it is only
necessary to know explicitly the thermodynamic potential in terms of
the extensive variables $\Phi=\Phi(E^{a})$. It is worth mentioning
that GTD allows us to  implement easily different thermodynamic
representations \cite{q1,q2,q3,q4,q5,q6}. 

Since a thermodynamic system
is uniquely determined by the fundamental equation $\Phi=\Phi(E^{a})$ \cite{callen},
the equilibrium space contains the information about the thermodynamic geometry 
of the system. In this connection, two geometric concepts are very important: the distance
and the curvature of ${\cal E}$. The metric $g$ determines the thermodynamic distance 
$ds^2= g_{ab} d E^a d E^b$ along the curve which connects two equilibrium states $E^a$ and $E^a + dE^a$, i. e.,
two points of the equilibrium manifold ${\cal E}$ \cite{A9}.  
As has been shown in \cite{vqs10}, the curves satisfying the variational principle $\delta \int ds =0$
and the entropy condition determine thermodynamic geodesics, i. e., extremal curves which describe
quasi-static thermodynamic processes. 

Notice that in general the signature of $g_{ab}$ is not fixed and so the equilibrium manifold can be either Riemannian or pseudo-Riemannian. This implies that the thermodynamic distance $ds^2$ can be either positive, negative or zero. As a consequence, one can show \cite{ig,zhao}
that there exists a casual structure in ${\cal E}$ that permits to connect thermodynamically one state with another state, but forbids the connection with some other states. The boundary of causality is determined by the states satisfying the condition $ds^2=0$, i. e., states that can be connected by a reversible process.
   
Recently,  it was pointed out in  \cite{brasil1} that the thermodynamic metric $g_{ab}$ can be used to define in the equilibrium manifold ${\cal E}$ the probability $P(E^a)$ to find a system within the interval $E^a+dE^a$ as
\be
P(E^a) = \frac{\sqrt{|{\rm det}(g_{ab})|}}{(2\pi)^{n/2}} \exp\left(\frac{1}{2}g_{ab}dE^a dE^b\right)\ .
\ee
Then, $g^{ab}$ turns out to be related to the second
moments of fluctuation, in a way that resembles the physical interpretation of Ruppeiner's metric in fluctuation theory. This opens the possibility to relate the thermodynamic curvature of ${\cal E}$ with the correlation length of the thermodynamic system described by ${\cal E}$. In this case, the determinant and the principal minors of $g_{ab}$ must be positive definite, a property that could be related to the causal structure of ${\cal E}$ mentioned above. A more detailed analysis will be necessary to clarify this point.

Moreover, the curvature of ${\cal E}$ is interpreted as a measure
of the thermodynamic interaction of the system with curvature singularities at those points where phase transitions 
take place \cite{korean}. This interpretation resembles the role of curvature in 
gravity and gauge theories, i. e., the curvature is a measure of the field interaction 
with curvature singularities indicating the break down of the theory.

The above interpretation of the thermodynamic curvature in GTD has been proved in several ordinary thermodynamic 
systems and also in the context of black holes. In the case of the ideal gas, the curvature 
vanishes and the thermodynamic geodesics are straight lines. This is in accordance with our physical expectations
since the ideal gas has no internal thermodynamic interaction. In the case of the van der Waals gas, which 
has a non-trivial  thermodynamic interaction \cite{callen},  
the equilibrium space is curved and the curvature singularities correspond to the points where phase 
transitions take place. The corresponding thermodynamic geodesics are curved and represent quasi-static processes 
which, under certain initial conditions, end at those points where phase transitions occur \cite{quevquev11},
in agreement with the well-known relation between geodesic incompleteness and curvature singularities.
For other ordinary thermodynamic systems, like several generalizations of the ideal gas and the Ising model, 
we obtained similar results. In the context of black holes \cite{qstv10a}, the results are
compatible with those obtained in ordinary black hole thermodynamics, i. e., black holes with thermodynamic 
interaction possess a curved equilibrium space with singularities at those places where phase transitions occur.
In the present work, the correctness of this interpretation will be shown for black holes 
in a particular set of higher dimensional gravity theories.

Since a system with thermodynamic interaction can be either stable or unstable, we expect that the curvature of ${\cal E}$ reproduces this behavior and so it can be different from zero, regardless of the stability properties of the system. In fact, we have found so far in GTD only two systems with a flat curvature, namely, the ideal gas \cite{A9,ig} and a particular topological black hole with flat horizon in Ho\v rava-Lifshitz gravity \cite{tqsv12}. Both cases correspond to stable systems. In some sense, this can be expected intuitively since in a stable system with vanishing curvature there is no thermodynamic interaction that could drive the evolution into another state. On the contrary, an unstable system must naturally evolve into a different state, a process that demands a non-zero thermodynamic interaction and, consequently, a non-flat curvature. In all the thermodynamic systems we have analyzed in GTD so far, an unstable system is always characterized by a non-zero curvature. We will show in this work that this is also true in the case of Einstein-Gauss-Bonnet black holes.

In five dimensions, the most general theory leading to second order
field equations for the metric is the so-called
Einstein-Gauss-Bonnet (EGB) theory, which contains quadratic powers
of the curvature. The most general action of the EGB theory
is obtained by adding the Gauss-Bonnet (GB)
invariant and the matter Lagrangian $L_{matter}$
to the Einstein-Hilbert action
\begin{equation}
I= \kappa\int d^{5}x \sqrt{g}[R +\alpha(R^{2}-4R^{\mu\nu}R_{\mu%
\nu}+R^{\alpha\beta\gamma\delta}R_{\alpha\beta\gamma\delta}) + L_{matter}],
\label{egbaction}
\end{equation}
where $\kappa$ is related to the Newton constant, and $\alpha$ is
the Gauss-Bonnet coupling constant. GB extensions of General
Relativity  have been motivated from a string theoretical point of
view as a version of higher-dimensional gravity, since this sort of
modification also appears in low energy effective actions of string
theory. The pioneering work in this regard belongs to Boulware and
Deser \cite{A12}. They obtained the most general static black hole
solutions in EGB theory. The GB term has some remarkable features.
For instance, in higher dimensions, it is the most general quadratic
correction which preserves the property that the equations of motion
involve only second order derivatives of the metric \cite{A13}.
However, in 4D, the GB term is topological in nature and it does not
enter the dynamics \cite{A14}. The GB term is important from both
physical and geometrical points of view; it naturally arises as the
next leading order of the $\alpha$-expansion of the heterotic
superstring theory ($\alpha^{-1}$ is the string tension) \cite{A15},
and plays a fundamental role in Chern-Simons gravitational theories
\cite{A16}. Several aspects  of such extended gravity theories have
been extensively studied in \cite{A17}.

In this work, we study the thermodynamics of static spherically
symmetric black holes in five dimensional
Einstein-Maxwell-Gauss-Bonnet (EMGB) with and without cosmological
constant, and in the Einstein-Yang-Mills-Gauss-Bonnet (EYMGB)
theory. We use two different approaches to formulate  black holes
thermodynamics. The first one is based upon the use of the
Bekenstein-Hawking entropy relation, and the second one uses a
different formula for the entropy which follows from the first law
of black hole thermodynamics. We will see that the two approaches
lead to completely different thermodynamics that have effects on the
stability properties and the phase transition structure of black
holes. The approach of GTD is used to show that there exists a
Legendre invariant thermodynamic metric for the equilibrium space
which in all the cases considered here describes correctly the
thermodynamic interaction in terms of the curvature of the
equilibrium manifold, and the phase transitions and the point with
zero temperature  in terms of curvature singularities.

This paper is organized as follows. Section \ref{sec:ht}  deals with
the thermodynamics and GTD of a particular asymptotically de Sitter
black hole solution \cite{A18} of EMGB theory. Section \ref{sec:qm}
is dedicated to the study of an asymptotically anti de Sitter black
hole solution of EMGB theory with cosmological constant. A black
hole spacetime of EYMGB theory is investigated in Section
\ref{sec:fock}. Finally,  Section \ref{sec:con} is devoted to
conclusions and remarks. Throughout this work we use Planck units in which $c=G=\hbar=k_{_B}=1$.

\section{Spherically symmetric black hole in EMGB gravity}
\label{sec:ht}

In the case of the EGB gravity minimally coupled to the electromagnetic field,
the matter component of the action (\ref{egbaction}) is given by
\be
 L_{matter}= F_{\alpha\beta}F^{\alpha\beta}\ , \quad F_{\alpha\beta}=
 A_{\beta,\alpha} -A_ {\alpha,\beta}\ . \ee

Spherically symmetric black holes of the EMGB theory have been
investigated very intensively as possible scenarios for the
realization of the low energy limit of certain string theories. A
particular solution which contains as a special case a black hole
spacetime was obtained in \cite{A18} (see also \cite{odin1,A21,A22}) by using the following 5D
static spherically symmetric line element
\begin{equation}
ds^{2}= -f(r)dt^{2}+\frac{dr^{2}}{f(r)}+r^{2}d\Omega _{3}^{2}\ ,
\label{lelemgb}
\end{equation}%
where $d\Omega _{3}^{2}$ is the metric of a 3D hypersurface with
constant curvature $6k$ which has the explicit form
\begin{equation}
d\Omega _{3}^{2}=
\begin{cases}
d\theta_{1}^{2}+\sin^{2}\theta_{1}(d\theta_{2}^{2}+\sin^{2}\theta_{2}
d\theta_{3}^{2}) \ ,&(k=1) \\
d\theta_{1}^{2}+\sinh^{2}\theta_{1}(d\theta_{2}^{2}+\sin^{2}\theta_{2}
d\theta_{3}^{2})\ ,&(k=-1)  \\
\alpha^{-1}dx^{2}+ d \phi_1^2 + d\phi_2^2 \ , &(k=0)\ .
\end{cases}
\end{equation}
Here, the coordinate $x$ has the dimension of length while the
angular coordinates $(\theta_{1},\theta_{2})\in[0,\pi]$ and
$(\theta_{3},\phi_{1},\phi_{2})\in[0,2\pi]$.

A particular solution is given by the metric function
\begin{equation}
f(r) = k+\frac{r^{2}}{4\alpha}\left[
1\pm\sqrt{1+\frac{8\alpha(M+2\alpha |k|)}{r^{4}}-\frac{8\alpha
Q^{2}}{3r^{6}}}\right],
\label{fung}
\end{equation}%
where the geometric mass $M+2\alpha|k|$ is different from that of
Einstein gravity for $k=\pm 1$. This solution is well defined if the
expression within the square root is positive definite. For the
solution (\ref{fung}) of the EMGB theory to describe a black hole it
is necessary that the condition $f(r)=0$ be satisfied. For the
special case $k=+1$, the roots of this equation are \be r_\pm =
\frac{1}{2}\left[\sqrt{M+\frac{2Q}{\sqrt{3}}} \pm
\sqrt{M-\frac{2Q}{\sqrt{3}}}\ \right] \ , \ee independently of the
value of the coupling constant $\alpha$. It turns out that in some
cases these radii determine naked singularities \cite{deh04}.
However, the specific case with $\alpha<0$ and $k=+1$ corresponds to
a solution which is asymptotically de Sitter, and represents a black
hole  with an event horizon situated at
\begin{equation}
r_{H}=\frac{1}{2}\left[\sqrt{M+\frac{2Q}{\sqrt{3}}}+
\sqrt{M-\frac{2Q}{\sqrt{3}}}\, \right]\ ,
\end{equation}
provided $(\frac{Q}{M})^{2}\leq\frac{3}{4}$. In this section, we will limit
ourselves to the study of this black hole spacetime.

It is interesting to mention that this specific black hole solution
is asymptotically de Sitter although the cosmological constant does
not appear explicitly in the action (\ref{egbaction}). This is a
particular characteristic of the EGB theory in five dimensions
\cite{deh04}. Moreover, the fact that the radius of the event
horizon does not depend on the value of the coupling constant
$\alpha$ leads to interesting thermodynamic consequences. In fact,
we will see that if we use the Bekenstein-Hawking entropy
relationship the thermodynamics of the black hole  differs
completely from the one obtained by using a modified entropy in
which the coupling constant appears explicitly.

\subsection{Analysis with the Bekenstein-Hawking entropy relation}
\label{sec:bh1}

Since the surface area of the event horizon is given by
\begin{equation}
A=r_{H
}^{3}\int_{\theta=0}^{\pi}\int_{\phi=0}^{\pi}\int_{\psi=0}^{2\pi}
{\sin^{2}\theta} \sin\phi d\theta d\phi d\psi=2\pi^{2}r_{H }^{3} \ ,
\end{equation}
the Bekenstein-Hawking entropy of the black hole is
$\frac{k_{B}A}{4G\hbar}=\frac {k_{B}\pi ^{2}}{2G\hbar}r_{H }^{3}$
\cite{A19}. Choosing the constants appropriately, the entropy takes
the form $S=r_{H }^{3}$, representing the fundamental equation that
contains all the thermodynamic information. In the mass
representation, $M=M(S,Q)$, for the black hole solution presented
above this fundamental equation can be rewritten as
\begin{equation}
M = S^{\frac{2}{3}}+\frac{1}{3}\frac{Q^{2}}{S^{\frac{2}{3}}}\ .
\label{feegb}
\end{equation}

\subsubsection{Thermodynamics}

\begin{figure}
  \includegraphics[width=5cm]{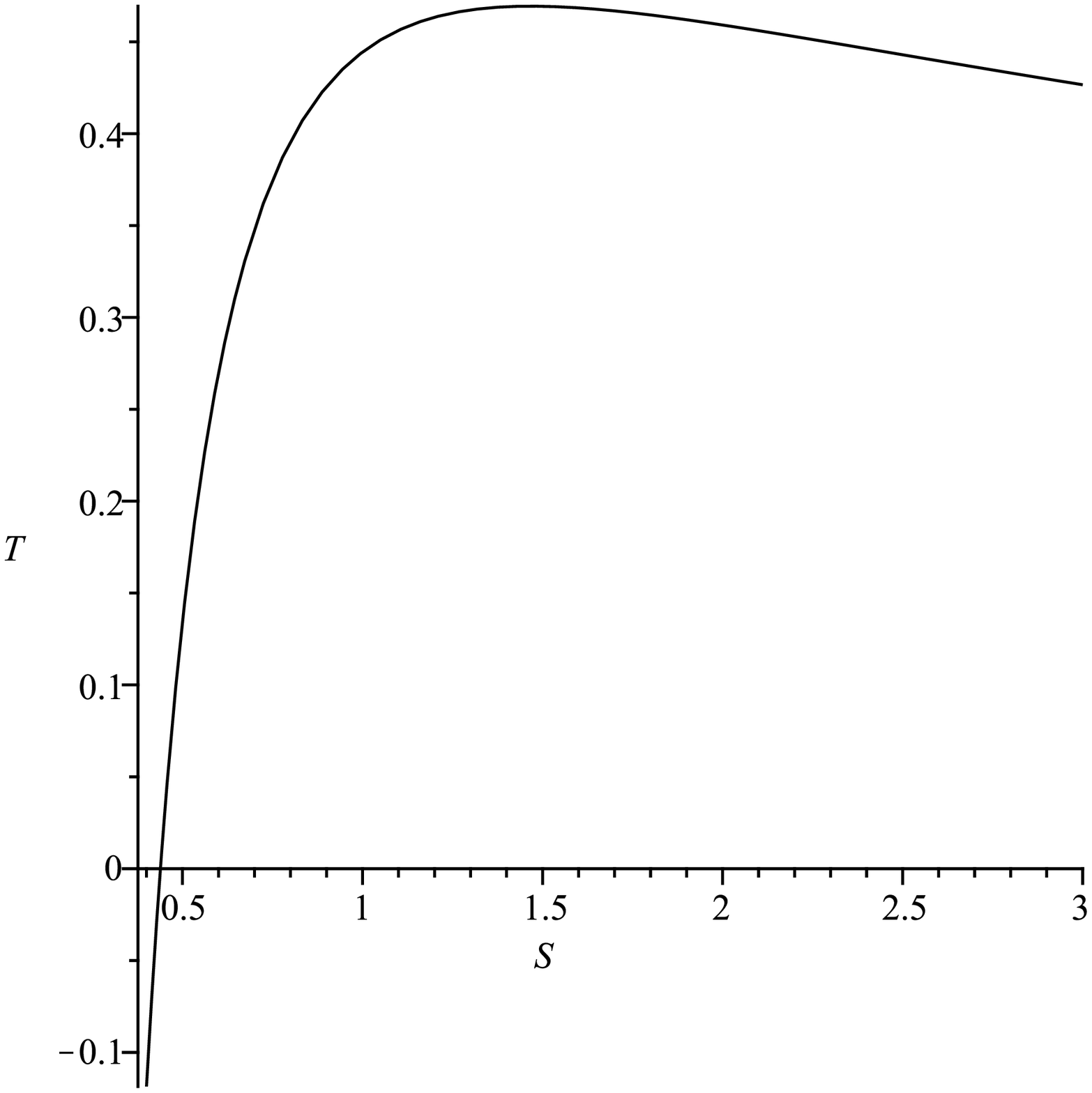}
  \includegraphics[width=5cm]{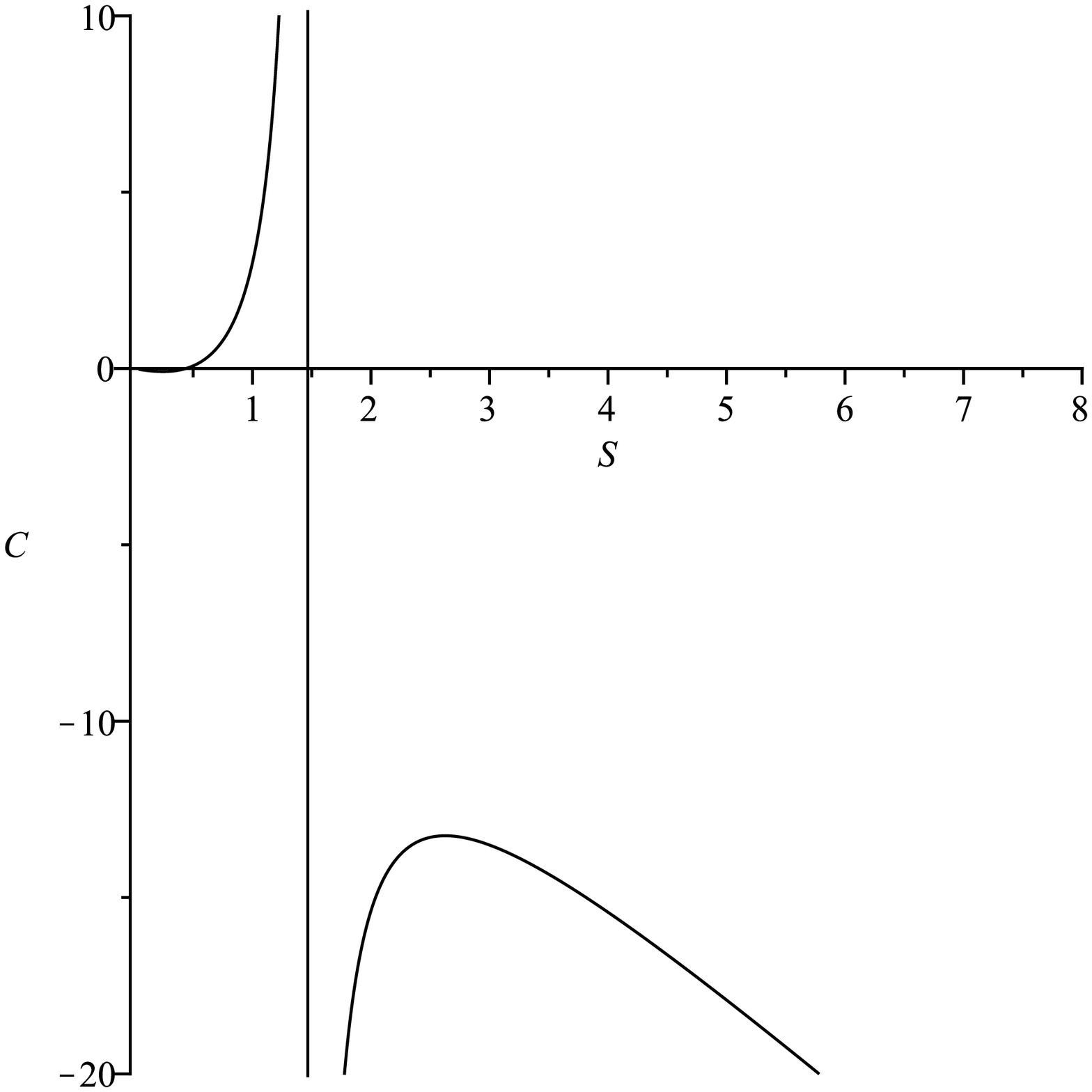}
    \includegraphics[width=5cm]{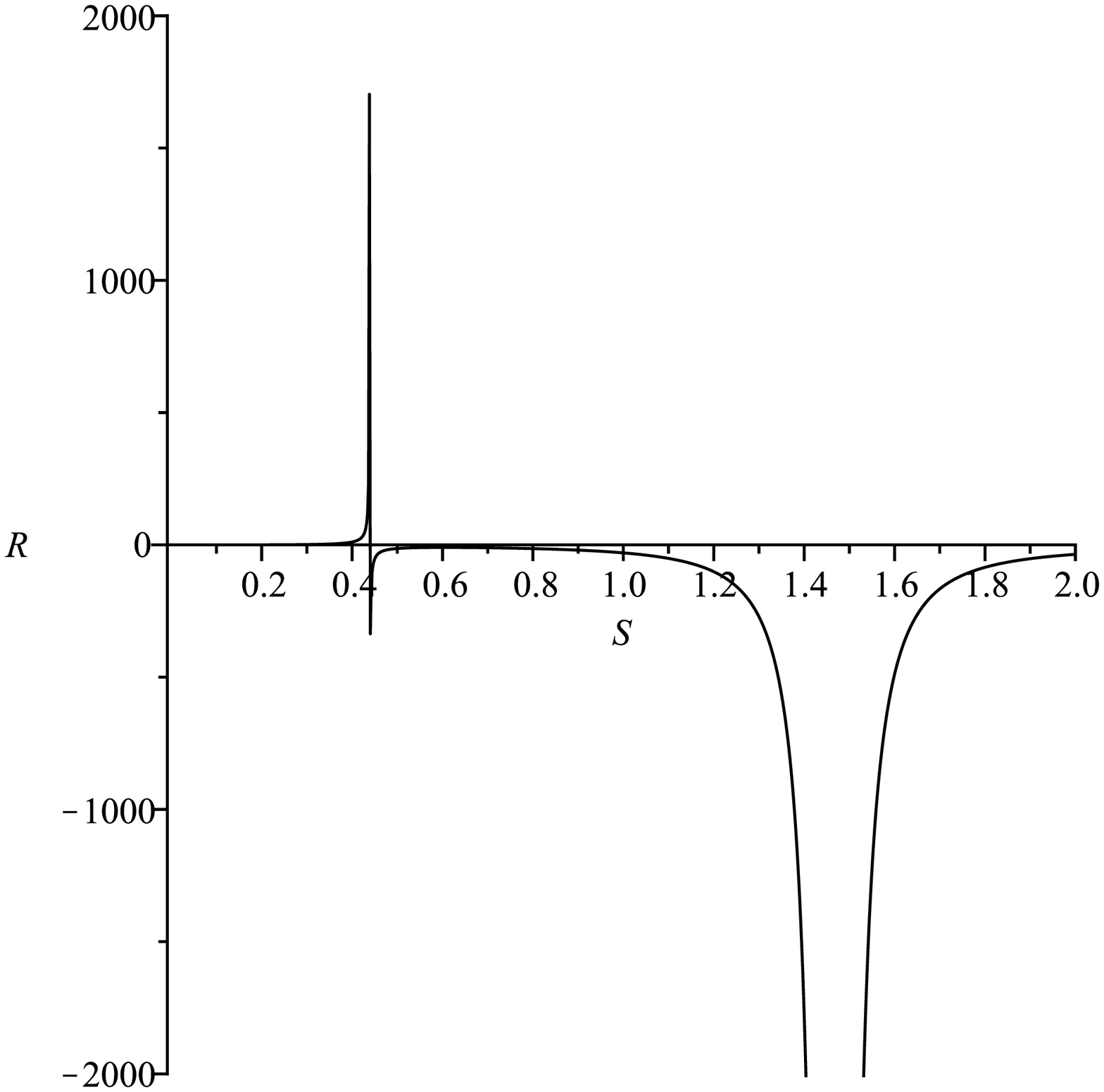}
  \caption{Temperature $T$, heat capacity $C_Q$ and thermodynamic curvature $R$ of  a
  charged black hole in EGB theory as a function of the Bekenstein-Hawking
  entropy $S$ for $Q=1$.
  The curvature singularities are located at the points $T=0$ and $C_Q\rightarrow
\infty$.}
  \label{fig1}
\end{figure}

Using the energy conservation law of the black hole (i. e., $dM = TdS
+\phi dQ$), one obtains the temperature and electric potential of the
black hole on the event horizon as
\begin{equation}
T = \frac{2}{9}\frac{3S^{\frac{4}{3}}-Q^{2}}{S^{\frac{5}{3}}},
\label{temp}
\end{equation}
and
\begin{equation}
\phi = \frac{2}{3}\frac{Q}{S^{\frac{2}{3}}}.
\label{phi}
\end{equation}
In the positive domain $(S^{4/3}> Q^2/3)$, the temperature increases
rapidly as a function of the entropy $S$ until it reaches its
maximum value at $S^{4/3} = 5Q^2/3$. Then, as the entropy increases,
the temperature becomes a monotonically decreasing function. This
behavior is shown in Fig. \ref{fig1}.

According to Davies \cite{A20}, the phase
transition structure of the black hole can be derived from the behavior of the heat
capacity. Strictly speaking, this implies that first we must introduce the concept of ``heat", say $Q_{heat}$, for a black 
hole. This is a complicated question that has not been answered so far, in particular, due to the lack of a physically reasonable statistical model for black holes \footnote{It is widely believed that to solve this problem it is necessary to start from a consistent quantum theory of gravity that is still unknown.}. For this reason, we use here the analogy with ordinary thermodynamics as follows. Using the 
the thermodynamic potential $M(S,Q)$ in which the first law of thermodynamics is expressed as $dM = TdS + \phi dQ$, we define the ``heat" through the relationship $dQ_{heat}\equiv TdS$ so that $dM=dQ_{heat} + \phi dQ$. 
Then, following the standard approach of ordinary thermodynamics,  we introduce the heat capacity  $C_Q \equiv \left(\frac{\partial Q_{heat}}{\partial T}\right)_Q = \left(\frac{\partial M}{\partial T}\right)_Q$ that  
in this case is given by
\begin{equation}
C_{Q}=-3S\left(\frac{3S^{\frac{4}{3}}-Q^{2}}{3S^{\frac{4}{3}}-5Q^{2}}\right).
\label{hcegb}
\end{equation}
In the physical region with $3S^{4/3}-Q^{2}>0$, i. e., the region
with positive temperature, the heat capacity is positive in the
interval $ Q^2 <3 S^{4/3} < 5 Q^2 $, indicating that the black hole
is stable in this region. At the maximum value of the temperature
which occurs at $3S^{4/3}-5Q^{2}=0$, the heat capacity diverges and
changes spontaneously its sign from positive to negative. This
indicates the presence of a second order phase transition which is
accompanied by a transition into a region of instability (see
Fig.\ref{fig1}). 

Usually, when applied to ordinary thermodynamic systems, the above analysis is associated with a particular statistical ensemble, 
once the internal energy of the system is well defined. In the case of black holes, however, there are several possibilities to define the internal energy. For this reason, sometimes the potential $M(S,Q)$, or equivalently $S(M,Q)$, is associated with the microcanonical ensemble \cite{wei09}, the canonical ensemble  with fixed potential \cite{myung08} and, in principle, one could also associate it with the grand canonical ensemble because the system is allowed to exchange charge. Here, we will use this last option.

Using this notation, we can say that the phase transition structure we have found above for the EMGB black hole is based upon the analysis of the grand canonical ensemble. On the other hand,   
 it is known that the critical points of the heat capacities may depend on the ensemble. In the case of the EMGB black hole we are considering here, from the thermodynamic potential $M(S,Q)$ we can define the potentials 
\be
H\equiv M-\phi Q\ , \quad F\equiv M - TS \ ,\quad G\equiv M - TS - \phi Q \ ,
\ee
that are usually denoted as the enthalpy, the Helmholtz free energy, and the Gibbs free energy, respectively. The enthalpy satisfies the first law of thermodynamics, $dH = TdS - Q d\phi$, and can be considered as determining the canonical ensemble. Then, if we assume again that the ``heat" of the black hole is defined by $dQ_{heat}=TdS$, the heat capacity for fixed $\phi$ is given by
\be
C_\phi \equiv \left(\frac{\partial Q_{heat}}{\partial T}\right)_\phi =
\left(\frac{\partial H}{\partial T}\right)_\phi =- 3S \ . 
\ee
We see that in this ensemble the heat capacity is always negative for $S>0$, indicating that the black hole is unstable. In the limit $S\rightarrow 0$, the thermodynamic description breaks down, because it also implies that $H\rightarrow 0$ and $M\rightarrow \infty$ 
\footnote{
Using the ``heat" as starting point, one can only introduce $C_Q$ and $C_\phi$. However, since these capacities can also be expressed as $C_X = \left(T\frac{\partial S}{\partial T}\right)_X$, some authors use the additional capacities  $\tilde C_ X =\left(\phi \frac{\partial Q}{\partial \phi}\right)_X$ to investigate phase transitions of black holes.}. 

\begin{figure}
  \includegraphics[width=7cm]{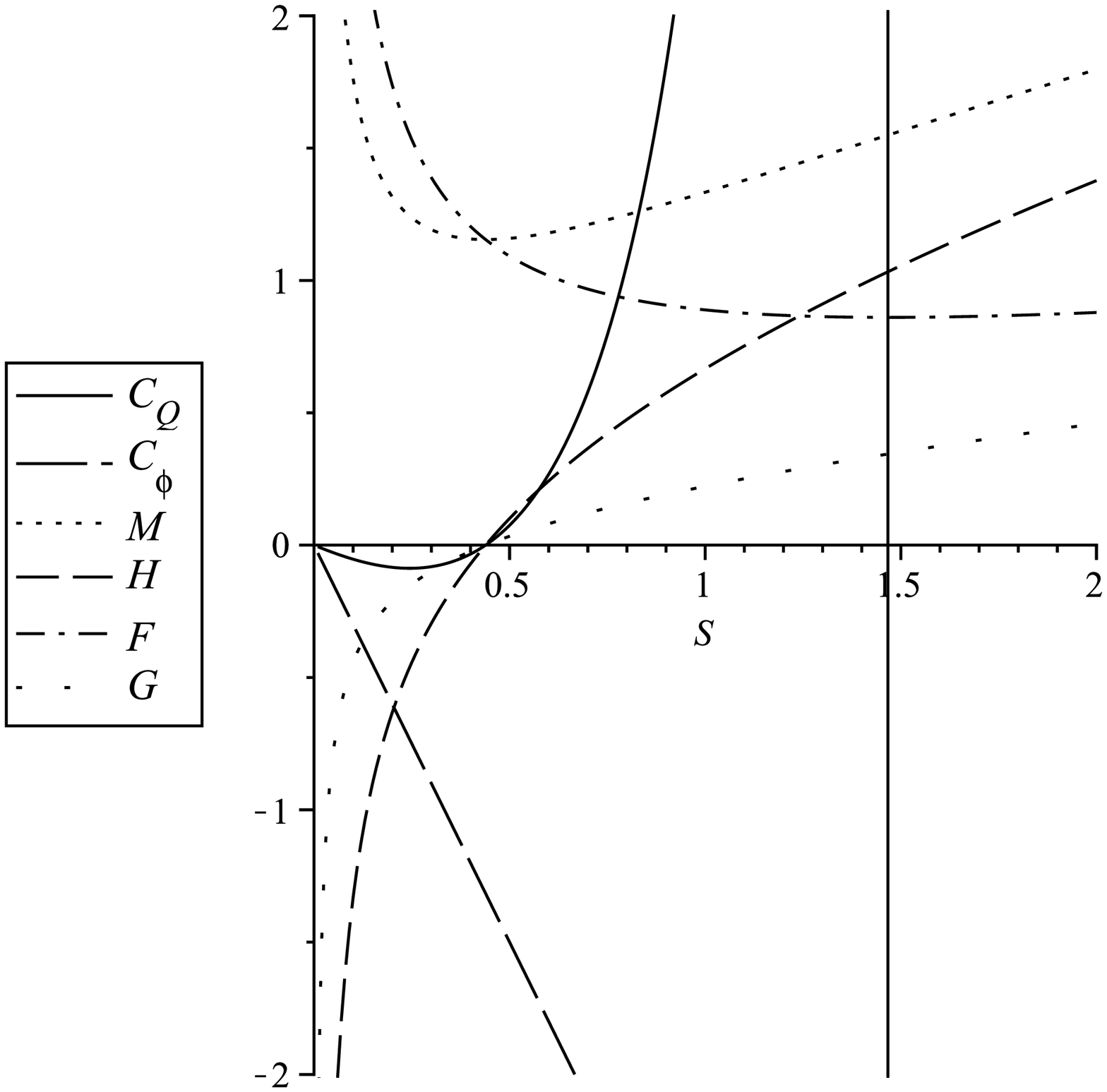}
  \qquad
  \includegraphics[width=7cm]{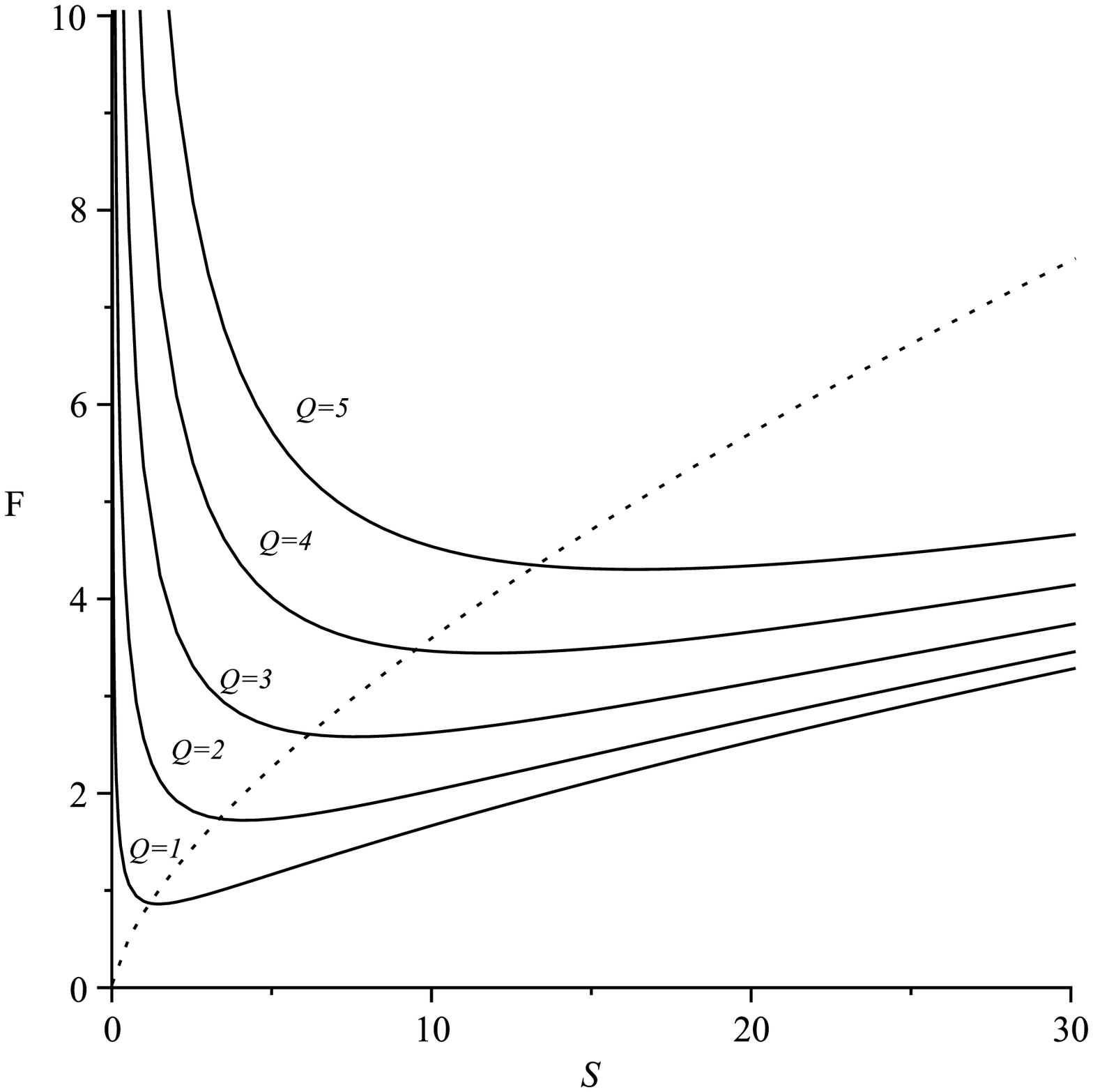}
    \caption{Left plot: Heat capacities and thermodynamic potentials of the EMGB black hole for a fixed value of the charge $Q=1$.
    Right plot: The Helmholtz free energy for different values of the charge $Q$. The dotted curve denotes the location of the minima where the phase transitions occur.
    }
 \label{fig1d}
\end{figure}

Consequently, the phase structure predicted by $C_\phi$ is completely different from that of $C_Q$. In this context, it is interesting to investigate the behavior of the thermodynamic potentials at the points of phase transitions. 
From Eq.(\ref{feegb}), we can calculate explicitly all the potentials and investigate their properties. In Fig. \ref{fig1d}, we illustrate the behavior of the potentials near the point where the heat capacity $C_Q$ diverges. At this point, all the potentials are well behaved  and only the Helmholtz energy possesses a minimum. This behavior is illustrated in the right plot of Fig. \ref{fig1d} that shows the stable minima of the Helmholtz energy for different values of the charge. 
We conclude that the phase transitions corresponding to divergences of $C_Q$ occur in a region of stability of $F$. Nevertheless, after the phase transition the black hole becomes unstable. 

These are the main features of the thermodynamic
behavior of the charged spherically symmetric black hole
(\ref{fung}).

\subsubsection{Geometrothermodynamics}

For our geometrothermodynamic approach to black hole thermodynamics
all what is needed is the fundamental equation, $M=M(S,Q)$
as given in Eq.(\ref{feegb}).
 Then, from the
general metric (\ref{gdown}) with $\Phi=M$ and $E^a=(S,Q)$ we obtain
the thermodynamic metric of the equilibrium manifold:
\begin{equation}
g =\frac{4}{27}\frac{3S^{4/3} - Q^2}{S^{4/3}}\left(\frac{3S^{4/3} - 5Q^2}{9S^2}
dS^2 + dQ^2\right)\ .
\label{metegbc}
\end{equation}
The corresponding scalar curvature is given by
\begin{equation}
R=-\frac{243\ S^{8/3}}{(3S^{4/3}-Q^2)(3S^{4/3}-5Q^2)^2}\ .
\label{Rmass}
\end{equation}
A first singularity is situated at the roots of the equation
$3S^{4/3}-Q^2=0$, i. e., at the points where the temperature
vanishes. The second singularity corresponds to the roots of $3S^{4/3}-5Q^2=0$.
 According to the
expression for the heat capacity $C_Q$ given in Eq.(\ref{hcegb}), these are exactly the points
where phase transitions take place and the temperature reaches its maximum value.
It follows that the geometrothermodynamic curvature of the metric (\ref{metegbc})
reproduces correctly the thermodynamic behavior near the points of zero temperature
as well as near the points of phase transitions (see Fig.\ref{fig1}).
The above results show that GTD correctly describes the thermodynamic properties of the black hole under consideration in the grand canonical ensemble in which $C_Q$ was considered.  

Notice that the curvature of the metric (\ref{metegbc}) cannot reproduce the phase transition structure predicted by $C_\phi$, because it represents a different definition of phase transitions based on the use of a different ensemble, namely, the canonical ensemble. Nevertheless, in GTD we can also reproduce the results obtained from the analysis of $C_\phi$ by using the fact that the general thermodynamic metric (\ref{gdown}) can be applied to any potential in any representation. Below we will show this explicitly.  

Consider  the canonical ensemble with the thermodynamic potential
\be
H\equiv  M- \phi Q = S^{2/3} \left( 1 -\frac{3}{4}\phi^2\right) \ , 
\label{M1}
\ee
in which the first law reads $dH = T dS - Q d\phi$. Then, the corresponding temperature can be expressed as 
\be
T= \frac{2}{3} S^{-1/3}\left( 1 -\frac{3}{4}\phi^2\right) \ ,  
\ee
and the heat capacity $C_\phi$, as mentioned above, is always negative for $S>0$, i. e., the black hole is unstable. 
Since the fundamental equation in this ensemble is $H=H(S,\phi)$, we choose the thermodynamic potential $\Phi = H$ and the independent thermodynamic variables $E^a=(S,\phi)$ to construct the thermodynamic metric of the equilibrium manifold.  Then, introducing Eq.(\ref{M1}) into the metric (\ref{gdown}), we get
\be
g= \left( 1 -\frac{3}{4}\phi^2\right) \left[ \frac{4}{27} S^{-2/3} \left( 1 -\frac{3}{4}\phi^2\right)dS^2 - S^{4/3} d\phi^2\right]\ ,
\ee
from which we obtain the curvature scalar
\be
R = \frac{3}{S^{4/3} \left( 1 -\frac{3}{4}\phi^2\right)^3}\ .
\ee
In GTD, we interpret the non-vanishing of the curvature as due to the presence of thermodynamic interaction in the system, regardless of its stability properties. In this case, we see that a non-zero curvature describes the thermodynamics of a completely unstable black hole. 
Indeed, from the above expression for the curvature scalar, it follows that there is a first singularity for $S\rightarrow 0$ that corresponds to the limit $C_\phi\rightarrow 0$ at which, as mentioned above, the thermodynamic description breaks down. The second singularity is located at $\phi^2 = 4/3$ and corresponds to the limit of vanishing temperature. We conclude that the curvature singularities of the equilibrium manifold correspond to the locations where the thermodynamic description of the black hole in this ensemble breaks down. 

The above results corroborate the well-known fact that the phase transition structure depends on the chosen ensemble. In this context, it is interesting to investigate the thermodynamic properties that follow from the study of the Gibbs function. From Eqs.(\ref{temp}) and (\ref{phi}), we obtain
\be
S= \frac{8}{27 T^3}\left( 1-\frac{3}{4}\phi^2\right)^3\ ,\quad Q=\frac{2\phi}{3T^2} \left( 1-\frac{3}{4}\phi^2\right)^2\ ,
\label{SQ}
\ee
so that 
\be
G = M-TS-\phi Q= \frac{4}{27 T^2} \left( 1-\frac{3}{4}\phi^2\right)^2\ .
\label{gibbs}
\ee
Thus, the Gibbs energy is a well-behaved function in the interval $\phi^2<4/3$ of positive temperature. 
To study this ensemble in GTD, we consider the thermodynamic metric (\ref{gdown}) with $\Phi=G$ and $E^a=(T,\phi)$, and introduce in the resulting metric the fundamental equation (\ref{gibbs}). Then, we obtain the expression
\be
g= \frac{4}{81 T^4}\left( 1-\frac{3}{4}\phi^2\right)^4\left[\frac{16}{3 T^2} \left( 1-\frac{3}{4}\phi^2\right)^2 dT^2 + (4-15\phi^2) d\phi^2\right] \ ,
\ee
for which the following curvature scalar can be derived
\be
R = - \frac{729\,T^4}{ \left( 1-\frac{3}{4}\phi^2\right)^5 (4 - 15 \phi^2)^2}\ .
\ee
The first curvature singularity appears at $\phi^2= 4/3$, and corresponds to the limit $T\rightarrow 0$ at which the thermodynamic approach breaks down. 
The second singularity at $\phi^2=4/15$, which coincides with the point where the second derivative of $G$ vanishes, indicates the presence of a second-order phase transition. This result is in accordance with the one obtained for the thermodynamic potential $M=M(S,Q)$ in Eq.(\ref{Rmass}). Indeed, using the Eqs.(\ref{SQ}), one can show that $3S^{4/3}-5Q^2\propto 4-15\phi^2$, and this is exactly the term in the denominator of (\ref{Rmass}) that vanishes as $C_Q\rightarrow \infty$.

The thermodynamic geometry of this black hole was also studied in
\cite{A18} using the Ruppeiner geometry. It turns out that
the Ruppeiner metric is flat in this case and, consequently, cannot
reproduce the behavior at the places where phase transitions occur or the 
temperature vanishes.

\subsection{Geometrothermodynamics with a modified entropy relation}
\label{sec:mod}

Usually the entropy of black holes satisfies the so-called area
formula, i.e, the black hole entropy equals one-quarter of the
horizon area. In gravity theories in higher dimensions and with
higher  powered curvature terms, however, the entropy of black holes
does not necessarily satisfy the area formula and other
possibilities can be considered to define entropy. For instance, in
\cite{A23} a simple method was suggested to obtain the black hole
entropy, by assuming that black holes, considered as genuine
thermodynamic systems,  must obey the first law of thermodynamics.
That is, we suppose that a black hole solution,  parameterized by
the mass $M$ or, alternatively, by the outer horizon radius $r_{H}$,
and the temperature $T$, satisfies the first law of thermodynamics
$dM = T dS + \mu_i dQ^i$, where $Q^i$ are the additional charges of
the black hole and $\mu_i$ are the corresponding chemical
potentials. If the mass and the temperature can be calculated by
using standard methods, the integration of the first law yields the
modified entropy formula
\begin{equation}
S=\int_0^{r_H} {T^{-1}\left(\frac{\partial M}{\partial
r_{H}}\right)_{Q_i} dr_{H}} +S_0\ , \label{smodf}
\end{equation}
where the additive integration constant $S_0$ can be fixed by
imposing the condition that the entropy goes to zero in the case of
an extreme black hole or when the area of the horizon vanishes.
Notice that in the integration (\ref{smodf}) the charges $Q_i$ must
be considered as constants. In \cite{odin2,crs04}, the modified entropy
was computed for an $n-$dimensional generalization of the black hole
solution (\ref{fung}) with the result
\begin{equation}
S=\frac{\Omega_{K}r^{n-2}_{H}}{4G}\left[ 1 + \frac{2\tilde{\alpha}k(n-2 )}
{(n-4)r^{2}_{H} }\right]\ ,
\label{smodn}
\end{equation}
where $\Omega_k$ denotes the spatial volume element, and $\tilde \alpha =
(n-2)(n-3)\alpha $.
In the case $n=5$, we are considering here, the modified entropy reduces to
\begin{equation}
S= r_{H}^{3}+6k\tilde{\alpha}r_{H}\ ,
\label{smod5}
\end{equation}
where  suitable units were chosen and we set $S_0=0$ for simplicity.
Notice that the contribution of the correction term vanishes for
$k=0$ so that the GB term has no effect on the expression for the
entropy which reduces in this case to the standard area formula.
Moreover, the modified entropy formula does not contain the
additional charges $Q_i$ explicitly, but only implicitly through the
horizon radius $r_H$. So we can assume the validity of the modified
entropy (\ref{smod5}) regardless of the nature of the additional
charges.

For the black hole solution (\ref{fung}) the modified entropy formula (with $k=1$)
gives
\begin{equation}
S=\frac{1}{8}\left[\sqrt{M+\frac{2Q}{\sqrt{3}}}+\sqrt{M-\frac{2Q}{\sqrt{3}}}
\right]^3+3\tilde{\alpha}\left[\sqrt{M+\frac{2Q}{\sqrt{3}}}+\sqrt{M-\frac{2Q}
{\sqrt{3}}}\right]\ . \label{feqmod1}
\end{equation}

In this case the fundamental equation is of the form
$S=S(M,Q)$ and cannot be  rewritten explicitly as $M=M(S,Q)$. This means
that for the further analysis we must use the entropy representation.
This is not a problem for the formalism of GTD which can be applied to
any arbitrary representation. In fact, for the entropy representation we only
need to consider the fundamental one-form as
\begin{equation}
\Theta_{S} = dS - \frac{1}{ T} dM + \frac{\phi}{T}dQ ,
\end{equation}
so that the thermodynamic potential is $\Phi=S$,
the coordinates of the equilibrium manifold are $E^a=(M,Q)$,
and the equilibrium conditions become

\begin{figure}
  \includegraphics[width=5cm]{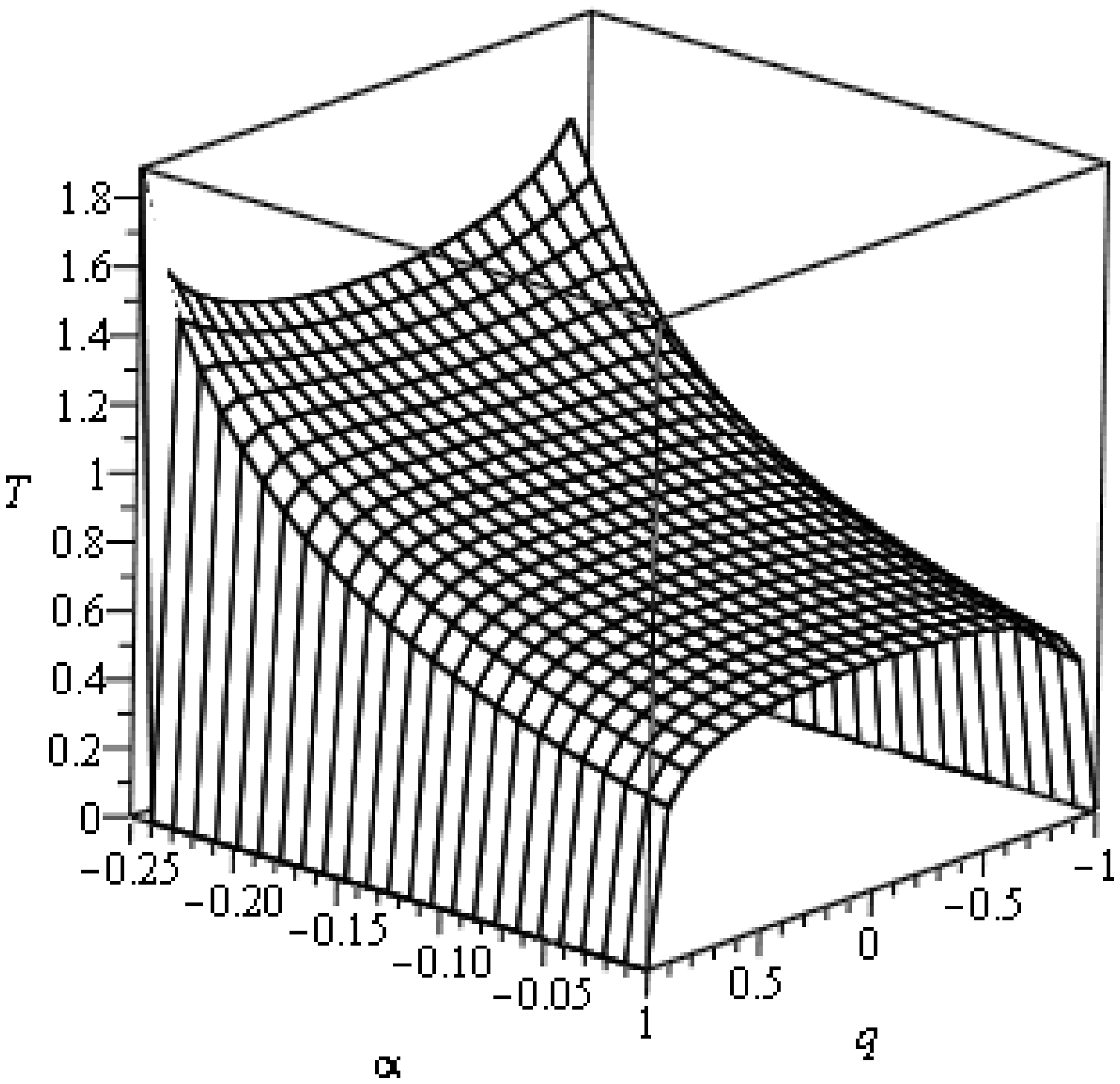}
  \includegraphics[width=5cm]{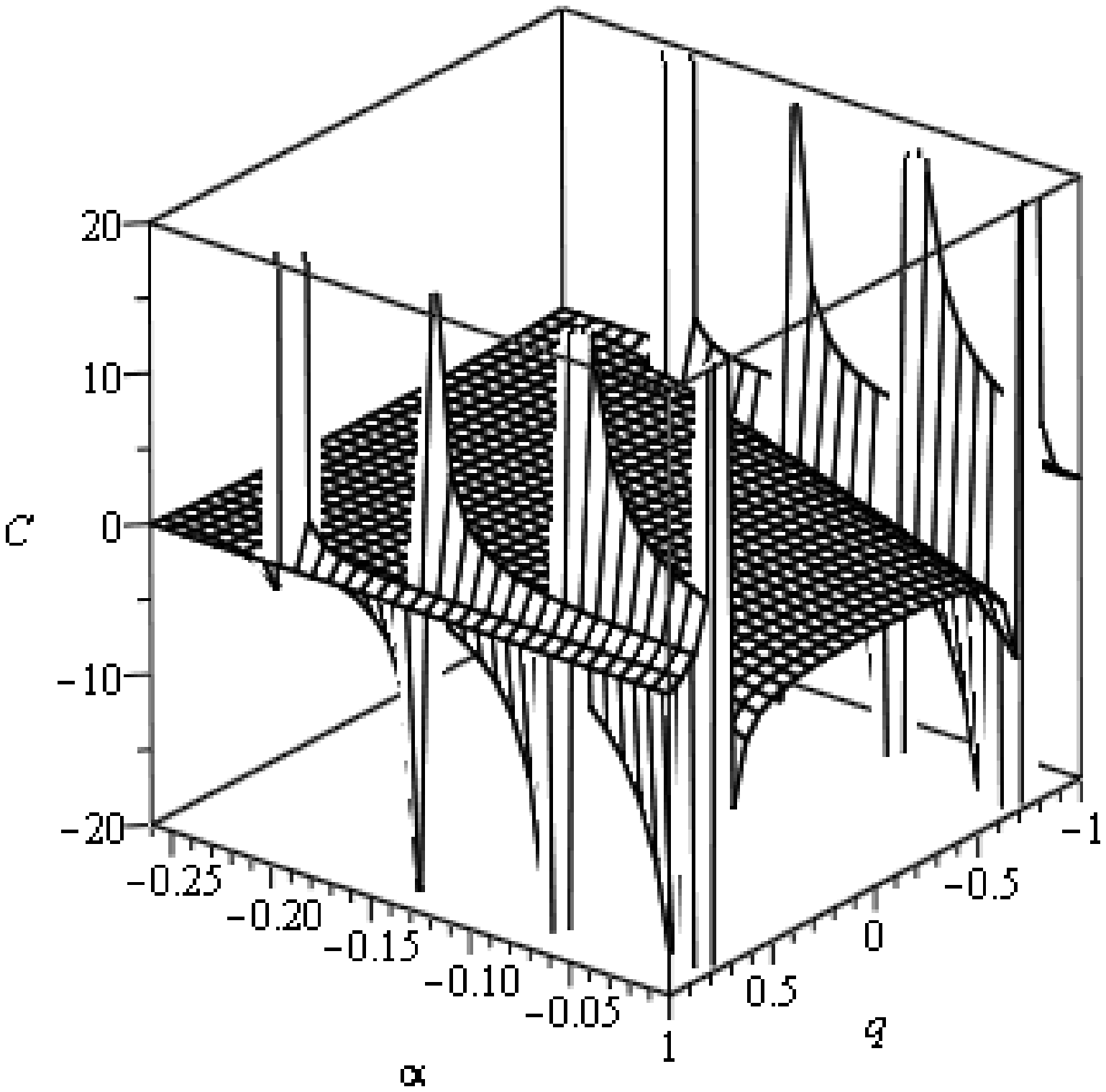}
  \includegraphics[width=5cm]{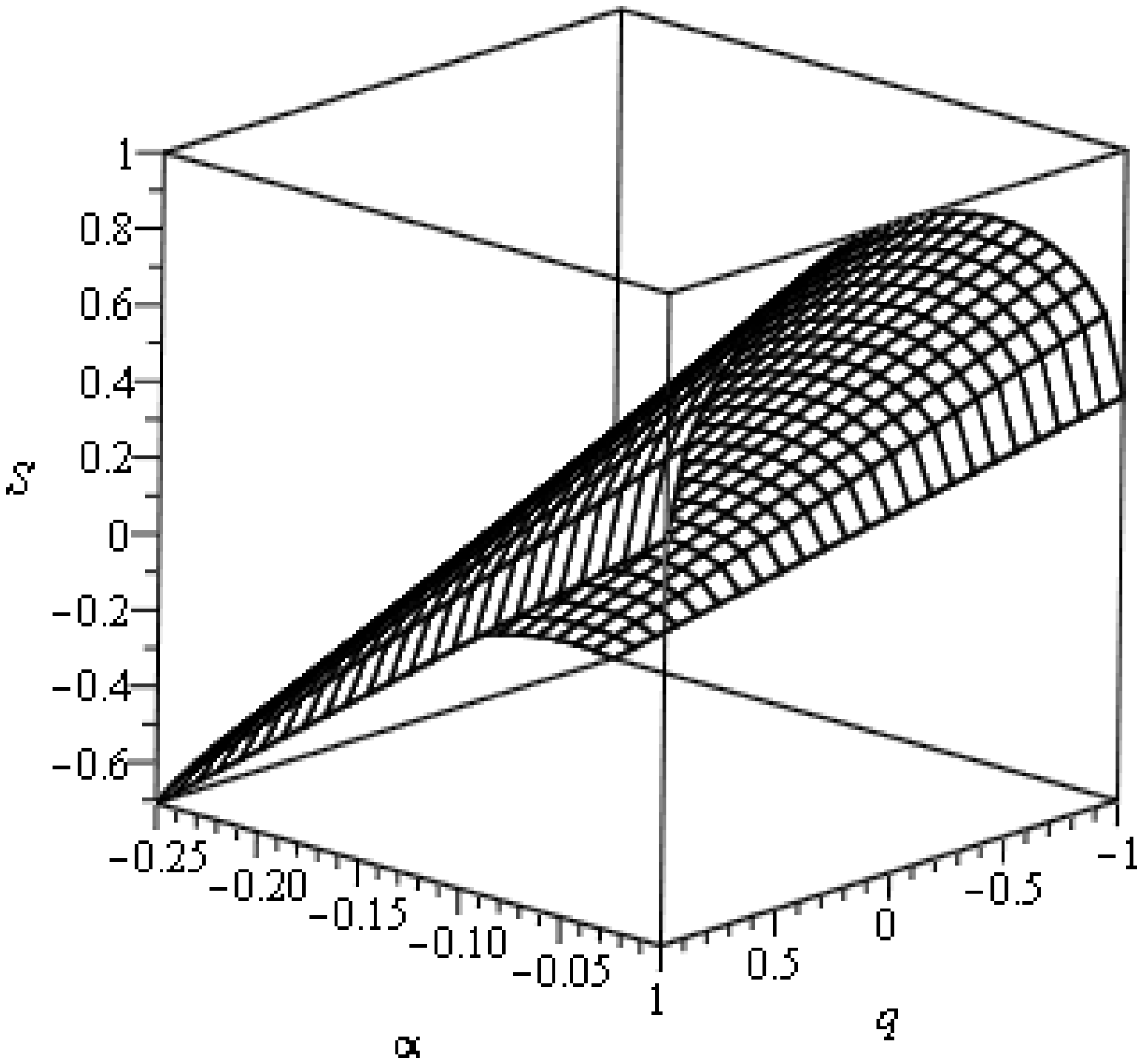}
  \caption{Modified temperature, heat capacity $C_Q$, and entropy as functions of the
  specific charge $q$ and the coupling parameter $\tilde\alpha$
 for a fixed value of the mass $M=1$.}
 \label{fig2}
\end{figure}

\begin{equation}
\frac{1} {T} =\frac{\partial S}{\partial M}\ ,\quad \frac{\phi}{T}
=-\frac{\partial S}{\partial Q}\ .
\end{equation}
From the above expressions one obtains the
temperature and electric potential of the black hole on the event
horizon as
\begin{equation}
T=\frac{8}{3}\frac{\sqrt{M(1-q^2)}}{(4\tilde \alpha + M
+M\sqrt{1-q^2} ) (\sqrt{1+q}+\sqrt{1-q})} \ ,
\end{equation}
\begin{equation}
\phi= \frac{2}{\sqrt{3}}\frac{\sqrt{1+q}-\sqrt{1-q}}{\sqrt{1+q}+\sqrt{1-q}}\ ,
\end{equation}
where $q=\frac{2Q}{\sqrt{3}M}$ represents a rescaled specific charge
that satisfies the condition $q^2\leq 1$. Furthermore, to find out
the phase transitions structure we must find the points where the
heat capacity $(C_Q=T\partial S/\partial T |_Q= - (\partial
S/\partial M)^2/(\partial^ 2 S /\partial M ^2) |_Q )$
\begin{equation}
C_{Q}=-\frac{3}{4} \frac{\sqrt{M}\sqrt{1-q^2}(\sqrt{1+q}+\sqrt{1-q})
(4\tilde\alpha + M + M\sqrt{1-q^2}) ^2}{M(1-3q^2) + (4\tilde \alpha
+M)\sqrt{1-q^2} - 8\tilde\alpha}
\end{equation}
diverges. Since the coupling constant $\tilde \alpha$ is negative,
the temperature function turns out to be positive definite only for
certain ranges of $\tilde \alpha$ and $q$. In Fig.\ref{fig2}, we
choose a particular range of values of $\tilde \alpha \in [-1/4,0]$
in which the modified temperature is positive. We also plot the
modified heat capacity  and entropy in the same range of values.
Notice that the entropy is not positive definite in this interval;
however, it is possible to choose the arbitrary constant $S_0$ in
Eq.(\ref{smodf})  so that the modified entropy is always positive
and the expressions for the modified temperature and heat capacity
remain unchanged.

\begin{figure}
  \includegraphics[width=6cm]{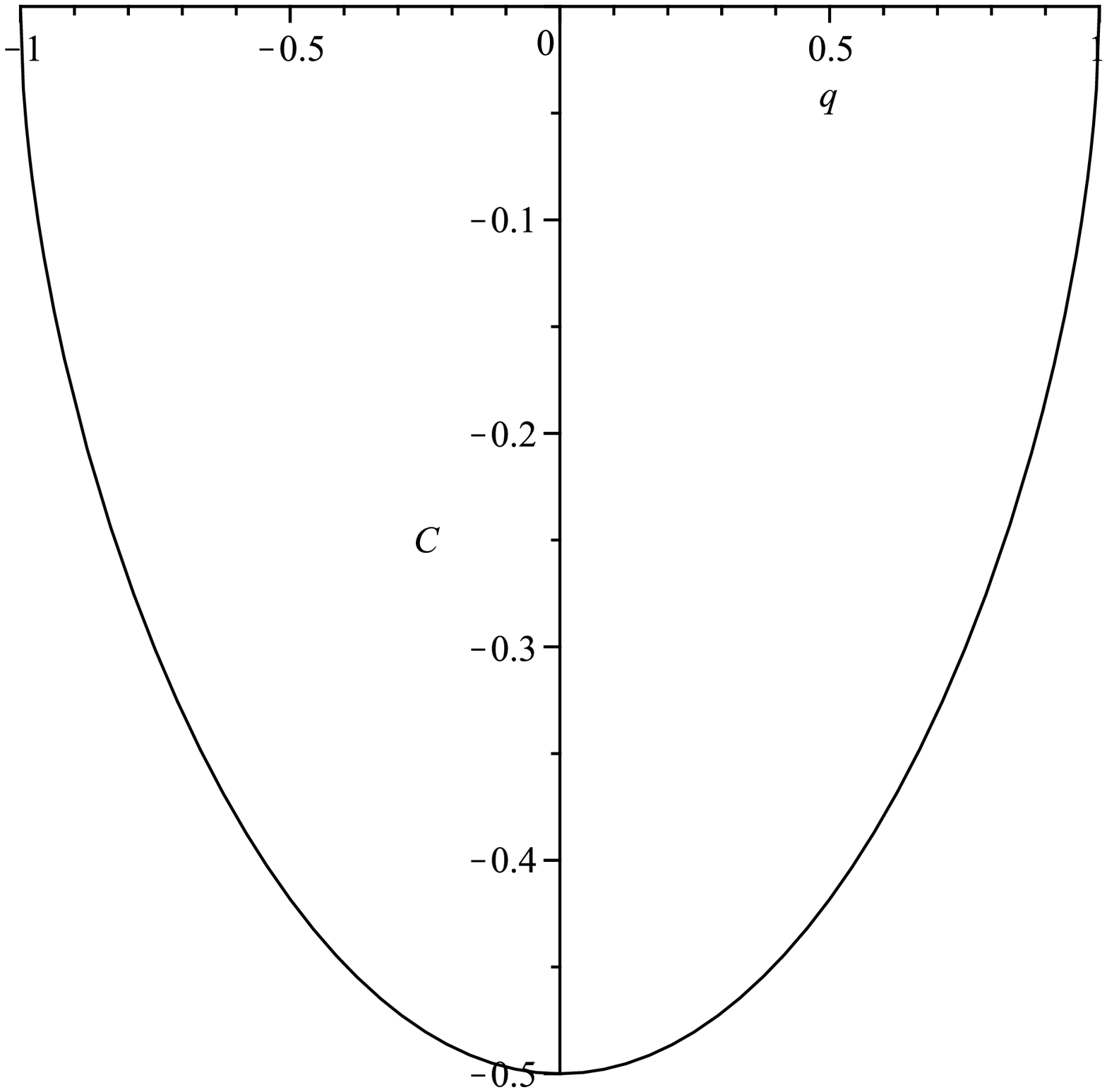}
  \includegraphics[width=6cm]{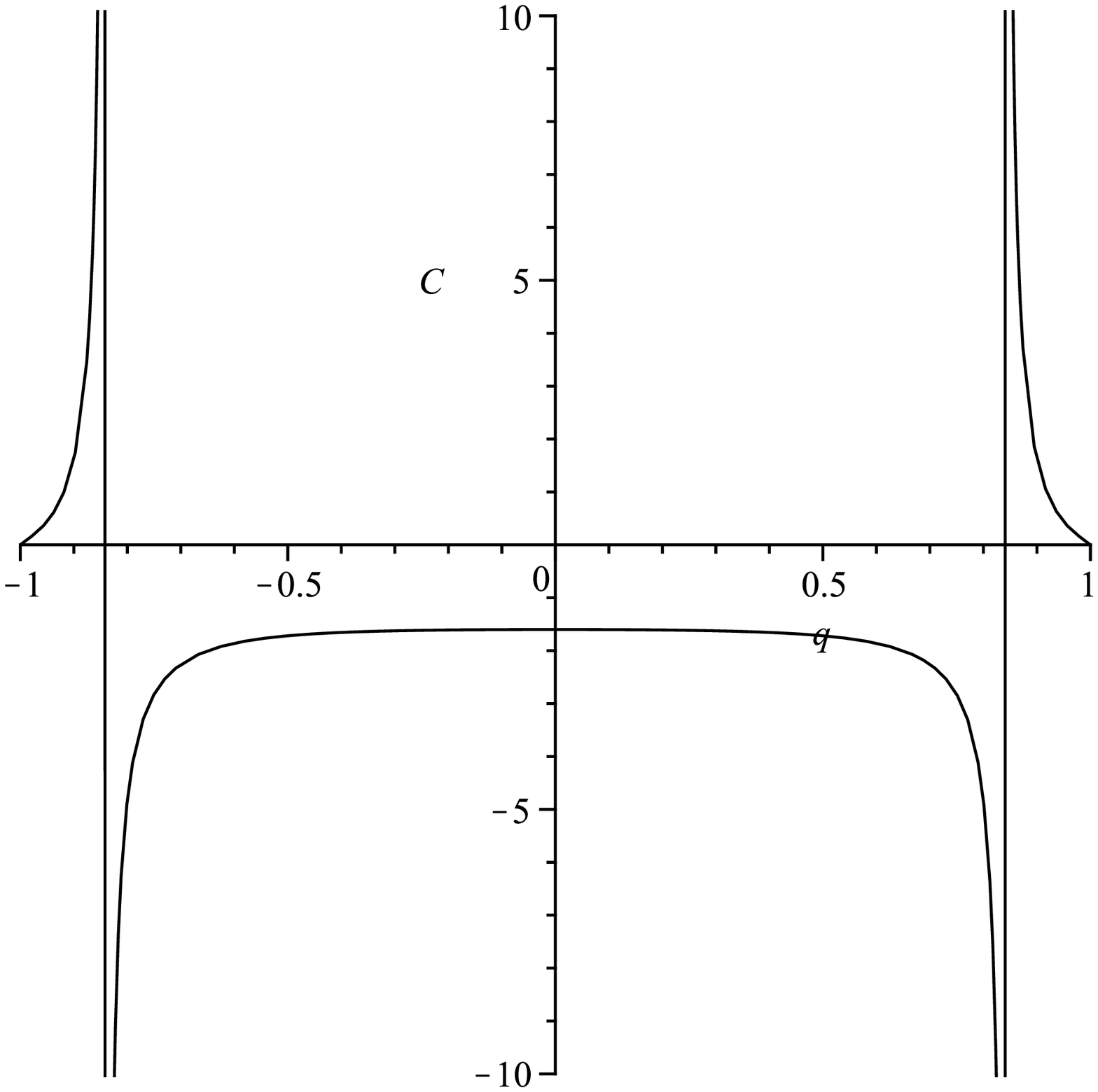}
  \caption{The heat capacity $C_Q$ as a function of the specific charge $q$ for
  $\tilde \alpha=-1/4$ (left plot) and $\tilde\alpha=-1/10$ (right plot).
  In both cases the mass is $M=1$. }
 \label{fig3}
\end{figure}

An interesting result of using the modified entropy is that the
phase transition structure now depends on the value of the coupling
constant $\tilde\alpha$. For instance, if we choose it as
$\tilde\alpha=-1/4$, the heat capacity is as illustrated in Fig.
\ref{fig3} (left plot). In this case, the heat capacity is
represented by a negative smooth function with no singularities in
the interval $-1\leq q\leq 1$, indicating that the black hole  is a
completely unstable thermodynamic system with no phase transition
structure. This behavior changes drastically, if we choose a
different value of the coupling constant. In fact, Fig.\ref{fig3}
(right plot) illustrates the singular behavior of the heat capacity
in the case $\tilde\alpha=-1/10$. We can see that at $q\approx \pm
0.82$, the black hole undergoes a second order phase transition. In
the interval $-0.82\leq q \leq 0.82$, the configuration is unstable
because the heat capacity is negative. Outside this interval,
however, the black hole is stable. We conclude that the coupling
constant $\tilde\alpha$ can induce a second order phase transition
in an unstable black hole in such a way that the resulting
configuration is a stable black hole for certain values of the
specific charge $q$.

\begin{figure}
  \includegraphics[width=6cm]{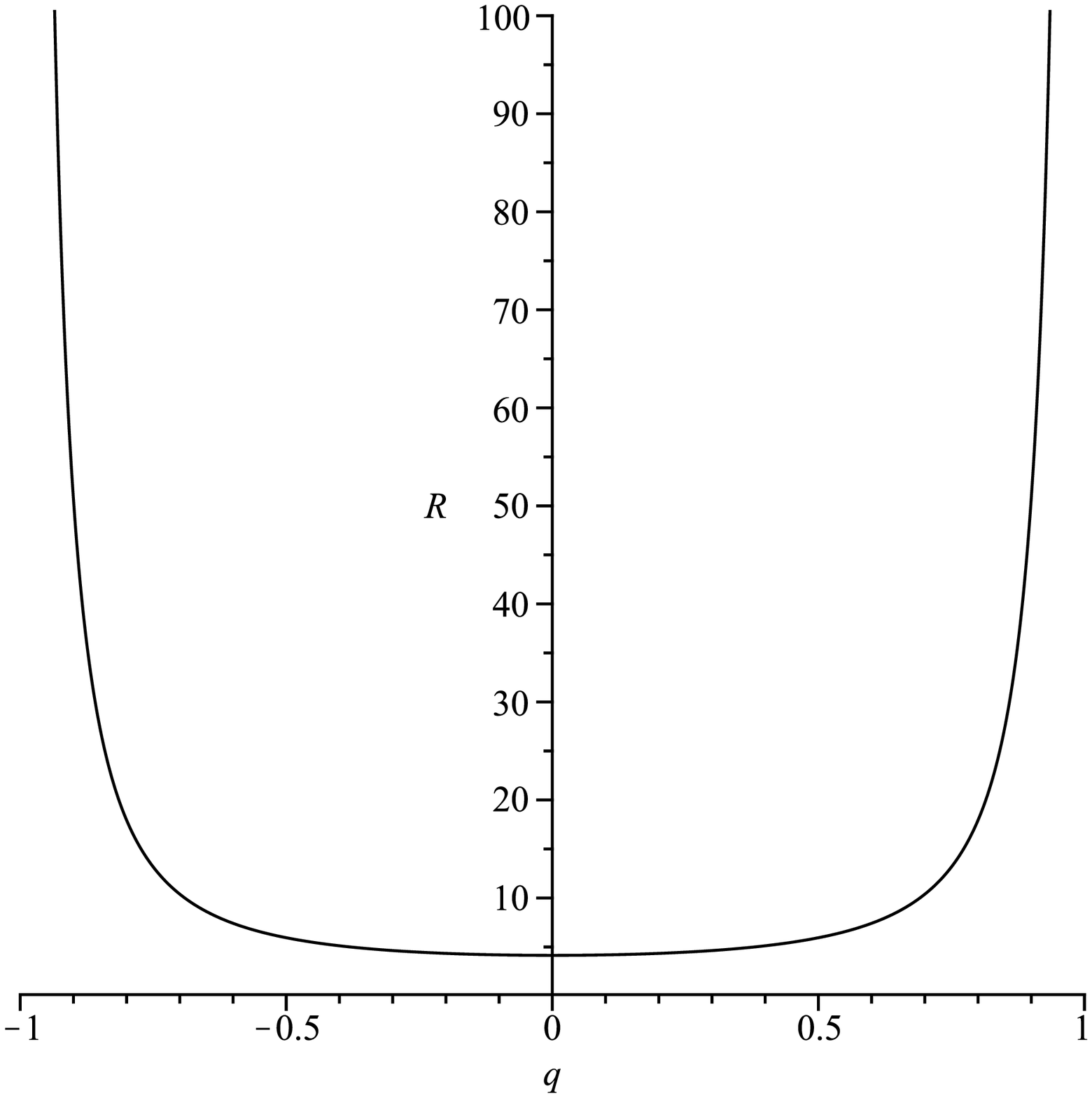}\quad
  \includegraphics[width=6cm]{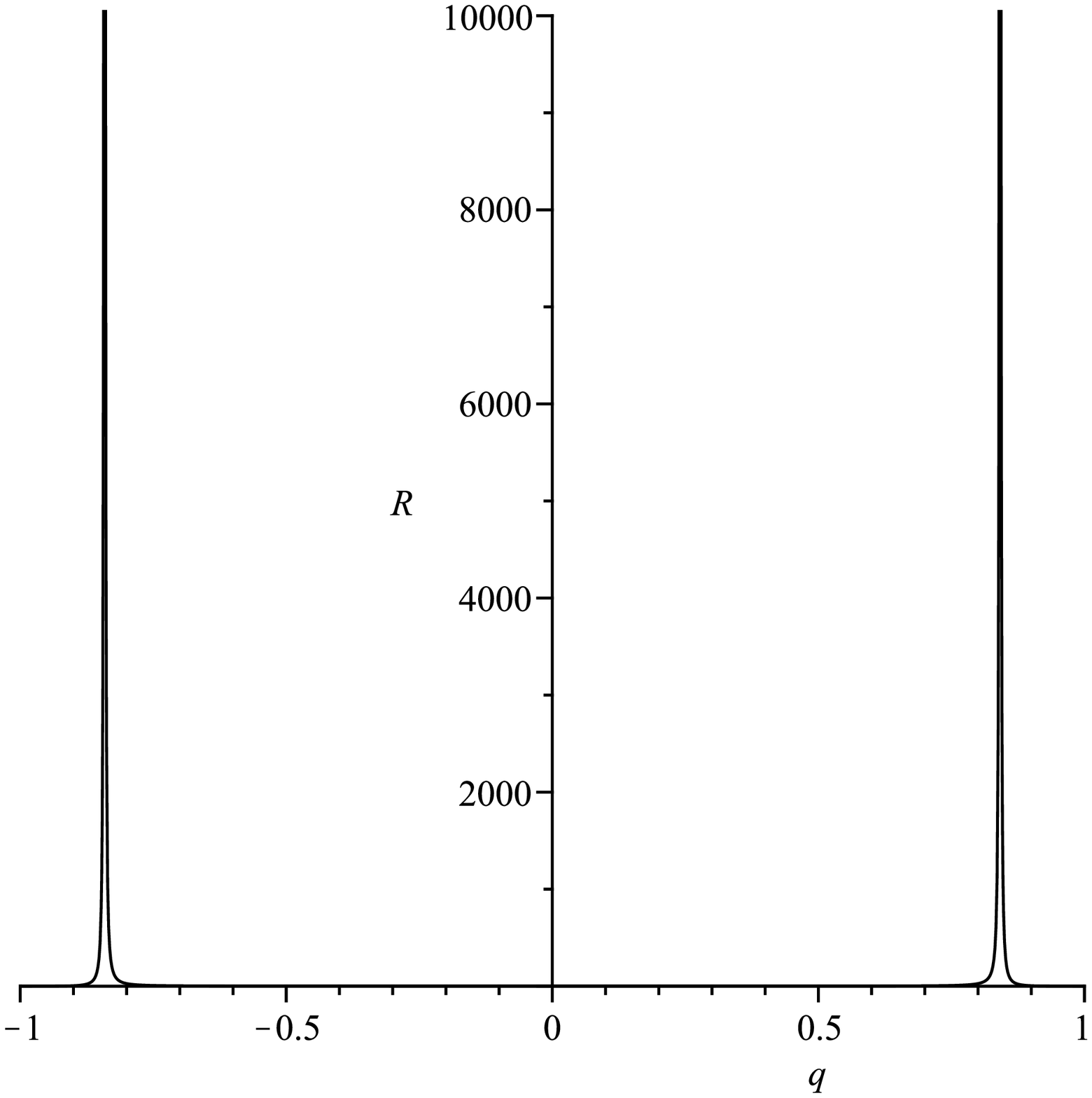}
  \caption{Thermodynamic curvature as a function of the rescaled specific
  charge $q$. For $\tilde\alpha=-1/4$ (left plot) the curvature is completely
  regular, except at $q=\pm 1$. Curvature singularities appear for
  $q\approx 0.82$ and $\tilde\alpha=-1/10$ (right plot). }
\label{fig4}
\end{figure}

Now we investigate this black hole configuration in the context of
GTD. As mentioned above, the coordinates of the equilibrium manifold
are $E^a=(M,Q)$ and the thermodynamic potential $\Phi=S$ is given by
means of the fundamental equation (\ref{feqmod1}). Then, the
thermodynamic metric for the equilibrium manifold ${\cal E}$ is
given by
\begin{equation}
\begin{split}
g= -  \frac{3 r_H (r_H^2+2\tilde\alpha)}{ 8\sqrt{1-q^2} \ M^2}
\bigg\{ & \frac{3r_H [2r_H^2-3q^2 M + 4\tilde\alpha (\sqrt{1-q^2}
-2)]} {2(1-q^2)^{3/2}} \ dM^2\\&+\sqrt{M} \left[\frac{4\tilde\alpha
+ (2+q)M}{(1+q)^{3/2}} + \frac{4\tilde\alpha + (2-q)M}{(1-q)^{3/2}}
\right] dQ^2\bigg\}\ .
\end{split}
\end{equation}
The behavior of the corresponding  scalar curvature is shown in
Fig.\ref{fig4} for two different values of the coupling constant.
The plot on the left shows the case $\tilde\alpha=-1/4$ and
corresponds to the case of an unstable configuration as shown in
Fig. \ref{fig3} (left plot). We can see that the curvature is
represented by a smooth function that is free of singularities in
the entire domain of the specific charge, except at $q=\pm 1$ where
the temperature vanishes (see Fig.\ref{fig2}). The  plot on the
right illustrates the behavior for $\tilde\alpha=-1/10$ and shows
two curvature singularities at $q\approx \pm 0.82$ which are the
points where the phase transition occurs (see right plot in
Fig.\ref{fig3}). In this case, it is also possible to show that an
additional curvature singularity (not plotted) exists in the
limiting case $q\rightarrow \pm 1$, indicating the blow up of the
approach as $T\rightarrow 0$.

\section{Spherically symmetric  black hole in EMGB gravity with cosmological constant} 
\label{sec:qm}

In the case of the Einstein--Maxwell-Gauss-Bonnet theory with
cosmological constant,the matter component of the action
(\ref{egbaction}) is given by \be
 L_{matter}= F_{\alpha\beta}F^{\alpha\beta} - 2 \Lambda
 \ , \quad F_{\alpha\beta}=A_{\beta,\alpha}-A_ {\alpha,\beta}\ ,
\ee
where $\Lambda$ is the cosmological constant, and $F_{\alpha\beta}$ represents
the electromagnetic Faraday tensor.

A five dimensional spherically symmetric solution in EMGB gravity
with $\Lambda$ was obtained by Wiltshire \cite{wilt}, using the
metric ansatz (\ref{lelemgb}) with $k=+1$ and the metric function
\begin{equation}
f(r)=1+\frac{r^{2}}{4\alpha }-\frac{r^{2}}{4\alpha
}\sqrt{1+\frac{8\alpha M }{ r^{4}}-\frac{8\alpha Q^{2}}{3r^{6}}+\frac{4\alpha
\Lambda }{3}}\ .
\label{fungl}
\end{equation}%
The two parameters $M(>0)$ and $Q$ are identified as the mass and
electric charge of the system. The limit of vanishing cosmological
constant generates a solution contained in Eq.(\ref{fung}) with the
minus sign in front of the square root and a redefined mass
parameter. In this limit, however, the resulting solution does not
describe a black hole, but a naked singularity.

In order for the solution (\ref{fungl}) to describe a black hole
spacetime, it is necessary that the expression inside the square
root be positive and the function $f(r)$ vanish on  the horizon
radius, i. e., \be 1+\frac{8\alpha M }{ r_H^{4}}-\frac{8\alpha
Q^{2}}{3r_H^{6}}+\frac{4\alpha \Lambda }{3}
> 0\ ,\quad
1+\frac{r_H^{2}}{4\alpha }-\frac{r_H^{2}}{4\alpha
}\sqrt{1+\frac{8\alpha M }{ r_H^{4}}-\frac{8\alpha Q^{2}}{3r_H^{6}}+
\frac{4\alpha \Lambda }{3}} = 0\ . \ee Moreover, to guarantee that
the mass of the black hole is always positive (see below) we must
demand that the coupling constant $\alpha$ be positive and the
cosmological constant $\Lambda$ be positive definite. In this
section we will limit ourselves to this range  of parameters, so
that the black hole determined by the function (\ref{fungl}) turns
out to be asymptotically anti de Sitter.

\subsection{Analysis with the Bekenstein-Hawking entropy
relation}
\label{sec:bh3}

The condition $f(r_H)=0$ implies that
\begin{equation}
\frac{\Lambda}{3}r_{H}^{6}-2r^{4}_{H}+2\left(M-2\alpha\right)r_{H}^{2}-
\frac{2}{3} Q^{2}=0\ .
\end{equation}%
Moreover, as we mentioned in Sec.(\ref{sec:bh1}), with the
appropriate choice of units the Bekenstein-Hawking entropy of the
black hole is given by $S=r_{H}^{3}$. Then, the corresponding
thermodynamic fundamental equation in the mass representation
becomes
\begin{equation}
M=2 \alpha+ S^{2/3} + \frac{Q^2}{3 S^{2/3}} -\frac{\Lambda}{6} S^{4/3} \ .
\label{feqlam}
\end{equation}%
Notice that to guarantee the positiveness of the mass,  we must
assume that $\alpha >0$ and $\Lambda < 0$.

\subsubsection{Thermodynamics}

Using the energy conservation law of the black hole (i. e. $dM = TdS
+\phi dQ$),  one obtains the temperature and electric potential of the
black hole on the event horizon as
\begin{equation}
T=\frac{2}{9}\,{\frac {3\,{S}^{4/3}-\Lambda\,{S}^{2}-{Q}^{2}}{{S}^{5/3}}}\ ,
\label{temgbl}
\end{equation}%
and
\begin{equation}
\phi=\frac{2 Q}{3S^{\frac{2}{3}}}.
\end{equation}%
Now, for the grand canonical ensemble the heat capacity has the form
\begin{equation}
C_{Q}= 3S\left(\frac{3S^{\frac{4}{3}}-\Lambda
S^{2}-Q^{2}}{5Q^{2}-3S^{\frac{4}{3}}-\Lambda S^{2}}\right) \ .
\label{CQeymgb}
\end{equation}%
The expression for the temperature (\ref{temgbl}) shows that it is
positive only in the range $3\,{S}^{4/3}-\Lambda\,{S}^{2}> {Q}^{2}$.
Consequently, the heat capacity can take either positive or negative
values, indicating the possibility of stable and unstable states for
the black hole. In fact, the expression for the heat capacity
exhibits a very rich structure in the range where the temperature is
positive. In Fig.{\ref{fig8}, a particular range was chosen to show
the behavior of the temperature and the heat capacity. The plot on
the right shows for the particular value $Q=1/2$ two different phase
transitions at $S\approx 0.6$ and $S\approx 4.9$. The first one
corresponds to a transition from a stable state $(C_Q>0)$ to an
unstable state $(C_Q<0)$. The second one represents a second order
phase transition in which the black hole becomes a stable system
again.

\begin{figure}
  \includegraphics[width=5cm]{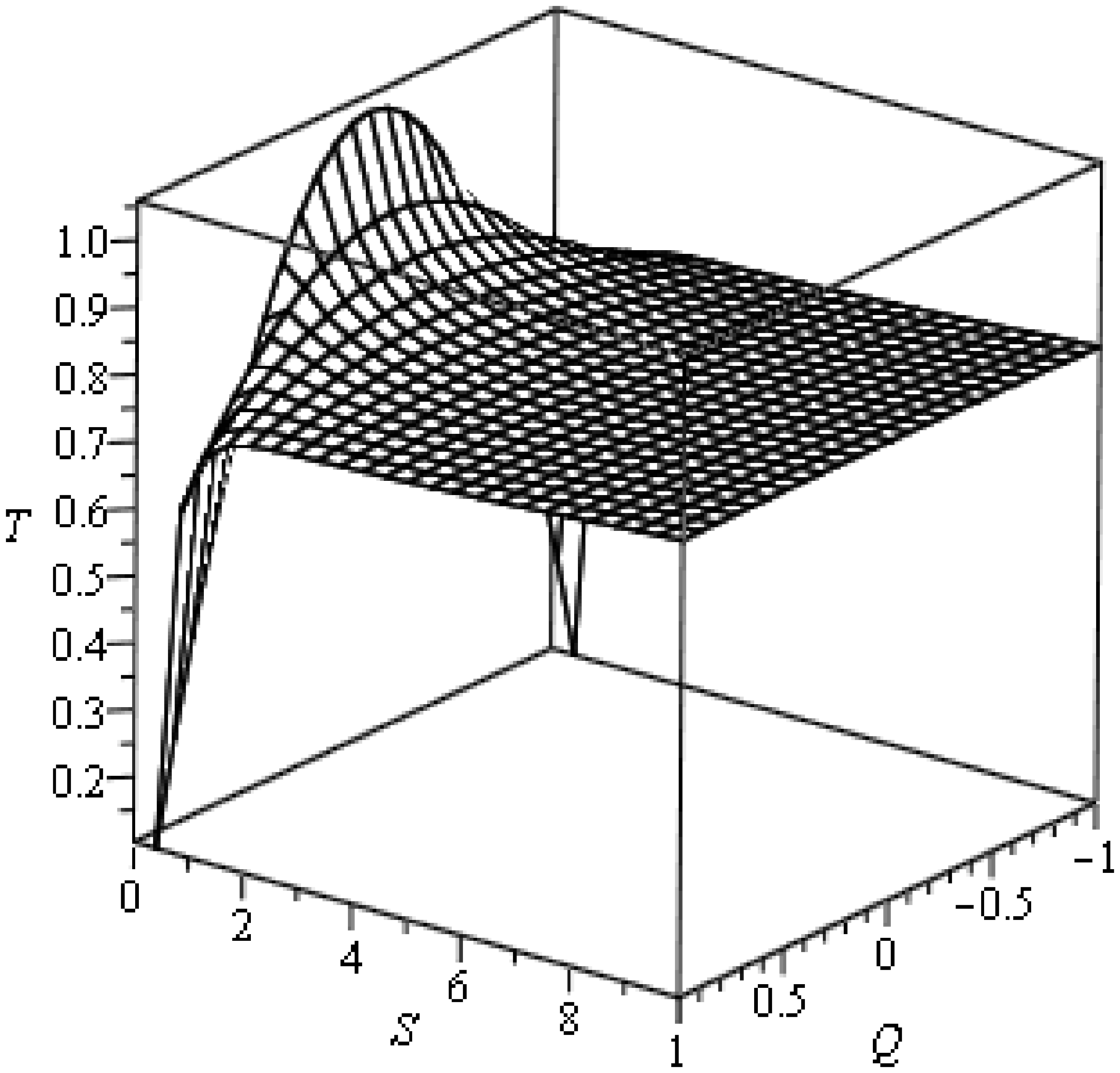}
  \includegraphics[width=5cm]{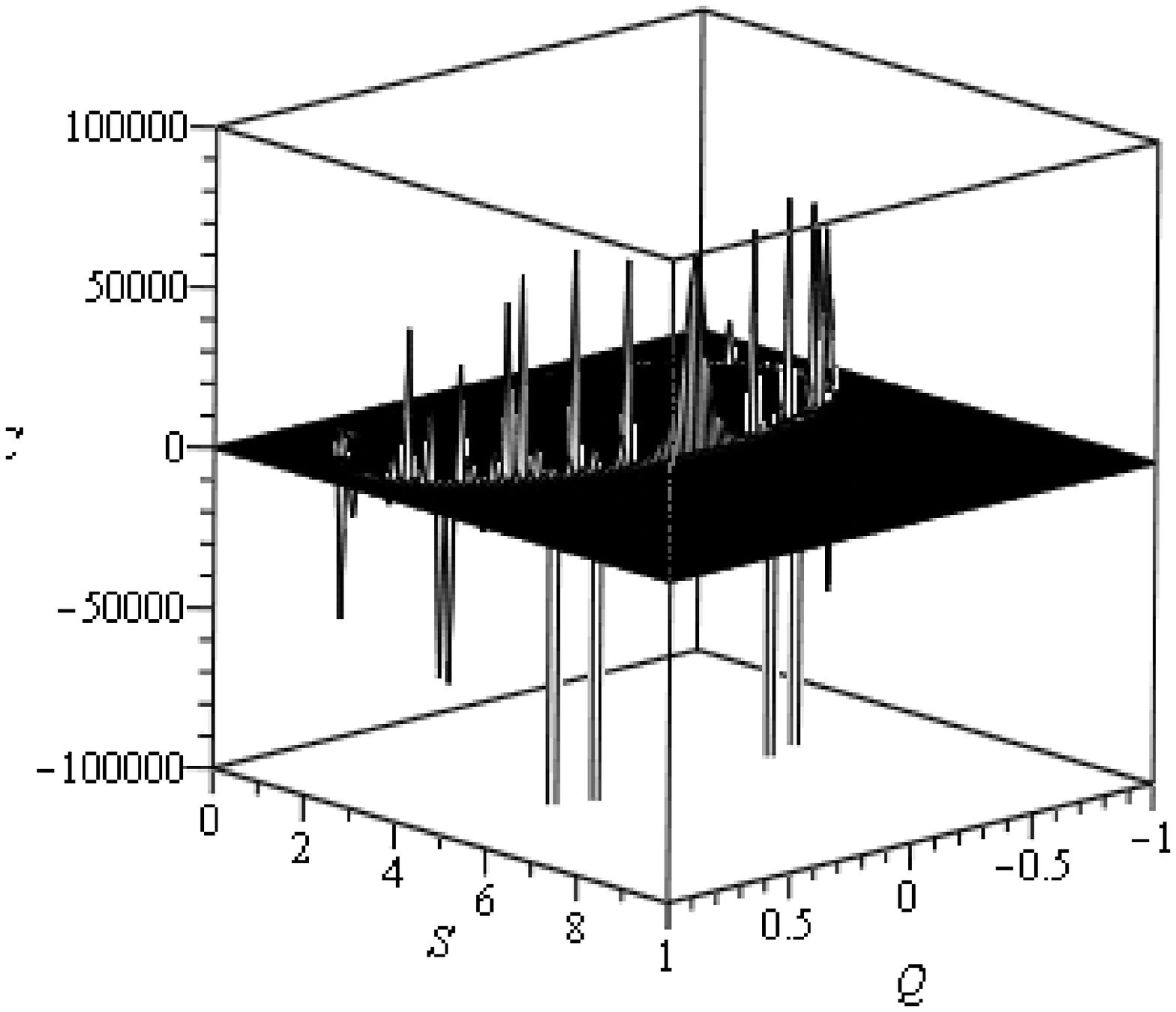}
  \includegraphics[width=5cm]{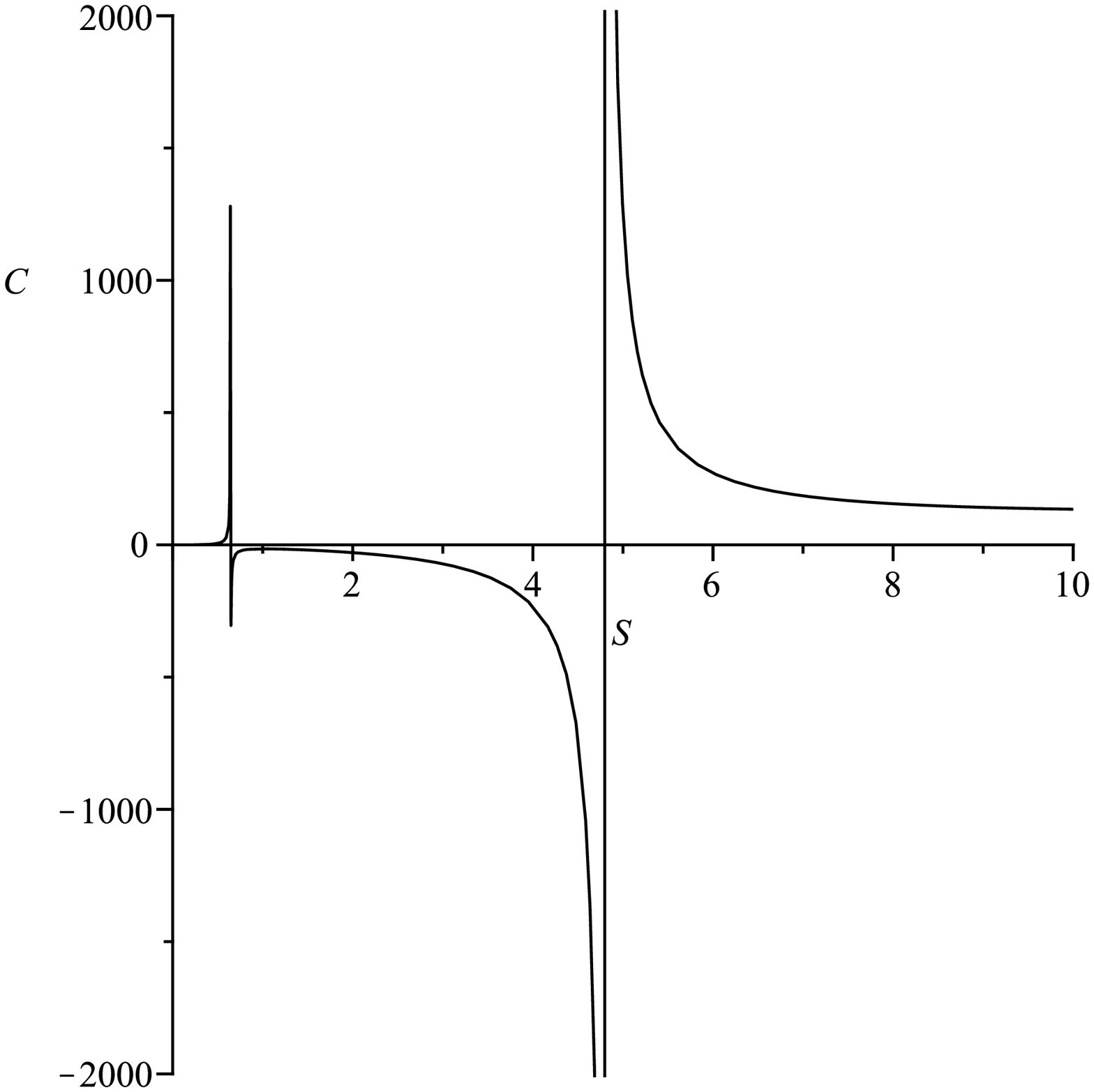}
  \caption{Behavior of the temperature and the heat capacity $C_Q$ in the
  range $Q\in [-1,1]$ and $S\in [0,10]$ for a fixed value of the cosmological
constant $\Lambda=-1$. The right plot shows the details of the phase
transition structure for the particular charge $Q=1/2$. }
  \label{fig8}
\end{figure}

Let us now consider the thermodynamic potential
\be
H = M-\phi Q=2 \alpha+ S^{2/3} - \frac{3}{4} \phi^2 S^{2/3} -\frac{\Lambda}{6} S^{4/3} \ .
\label{feqlm1}
\ee
for the canonical ensemble in which the first law of thermodynamics reads $dH=dQ_{heat} - Q d\phi$ with $dQ_{heat}=TdS$, as before.
In this ensemble, we can define the heat capacity at fixed electric potential, i. e.
\be
C_\phi \equiv \left(\frac{\partial H}{\partial T}\right)_\phi = - 3S \frac{ 12S^{1/3} - 9 \phi^2 S^{1/3} - 4\Lambda S}{12S^{1/3} - 9 \phi^2 S^{1/3} + 4\Lambda S} \ ,
\label{cphilambda}
\ee
whose behavior strongly depends on the value of the cosmological constant. Figure \ref{fig9b} shows that for a given value of $\Lambda$, the heat capacity $C_\phi$ can be either positive or negative with a quite complex singularity structure. 
\begin{figure}
  \includegraphics[width=7cm]{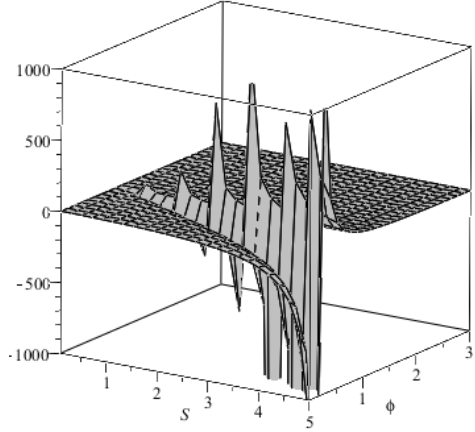}
 \caption{The behavior of the heat capacity $C_\phi$ as a function of the entropy $S$ for different values of the electric potential $\phi$ and $\Lambda=-1$.}
  \label{fig9b}
\end{figure}
If we take into account the condition $12S^{1/3} - 9 \phi^2 S^{1/3} - 4\Lambda S >0$, which follows from the condition $T>0$, we find that stable states  $(C_\phi>0)$ are allowed for entropies in the range
\be
S^{2/3} > \frac{3}{4|\Lambda|} (4 - 3 \phi^2) \ .
\ee
If we choose $\phi^2>4/3$, this condition is always satisfied for $S>0$, indicating that stable states always exist in this case. Moreover, from the heat capacity (\ref{cphilambda}) we see that the roots of the equation $12S^{1/3} - 9 \phi^2 S^{1/3} + 4\Lambda S=0$ determine the locations where second order phase transitions occur. This behavior is illustrated in Fig. \ref{fig9c}.      
\begin{figure}
  \includegraphics[width=7cm]{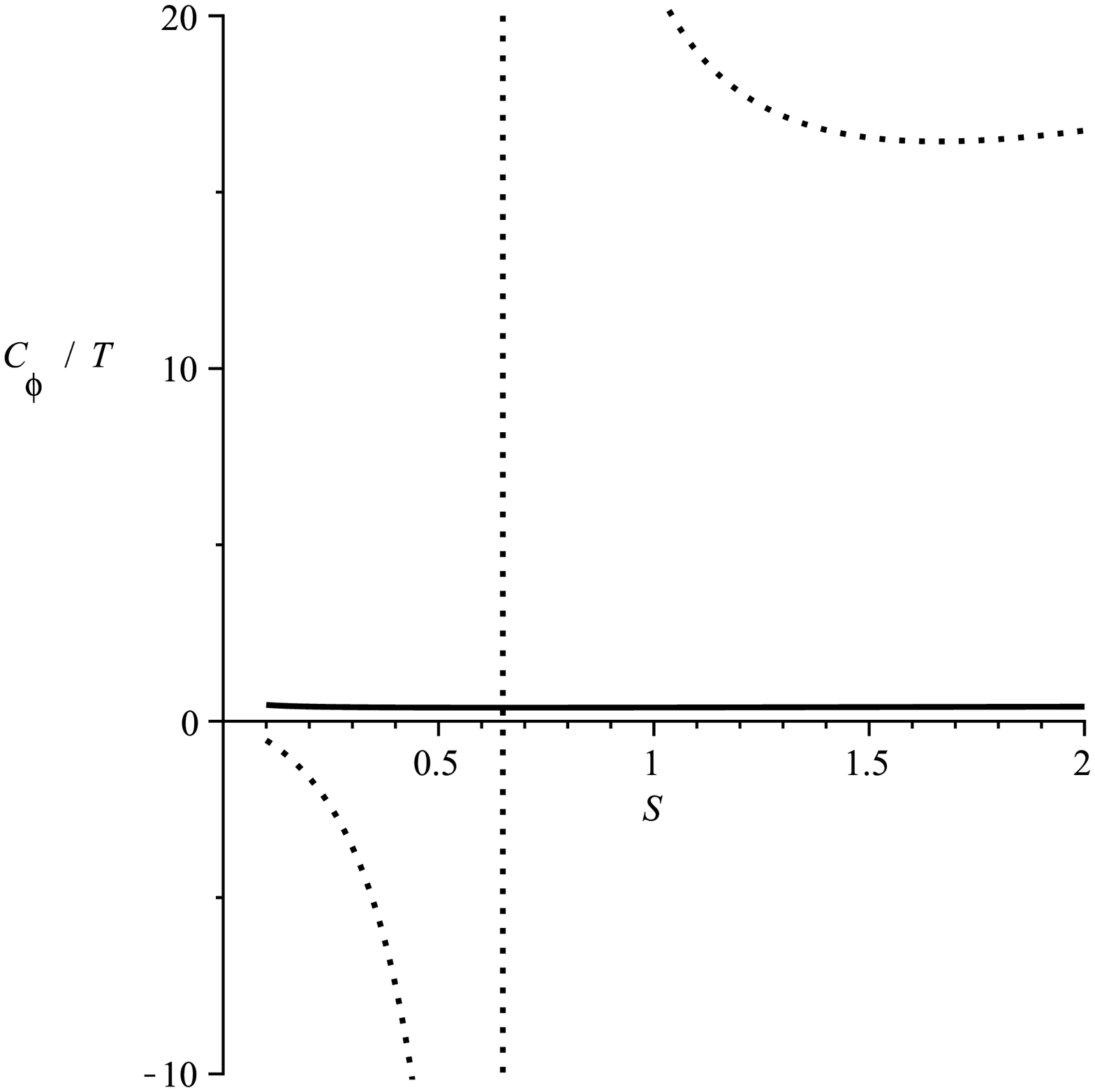}
  \includegraphics[width=7cm]{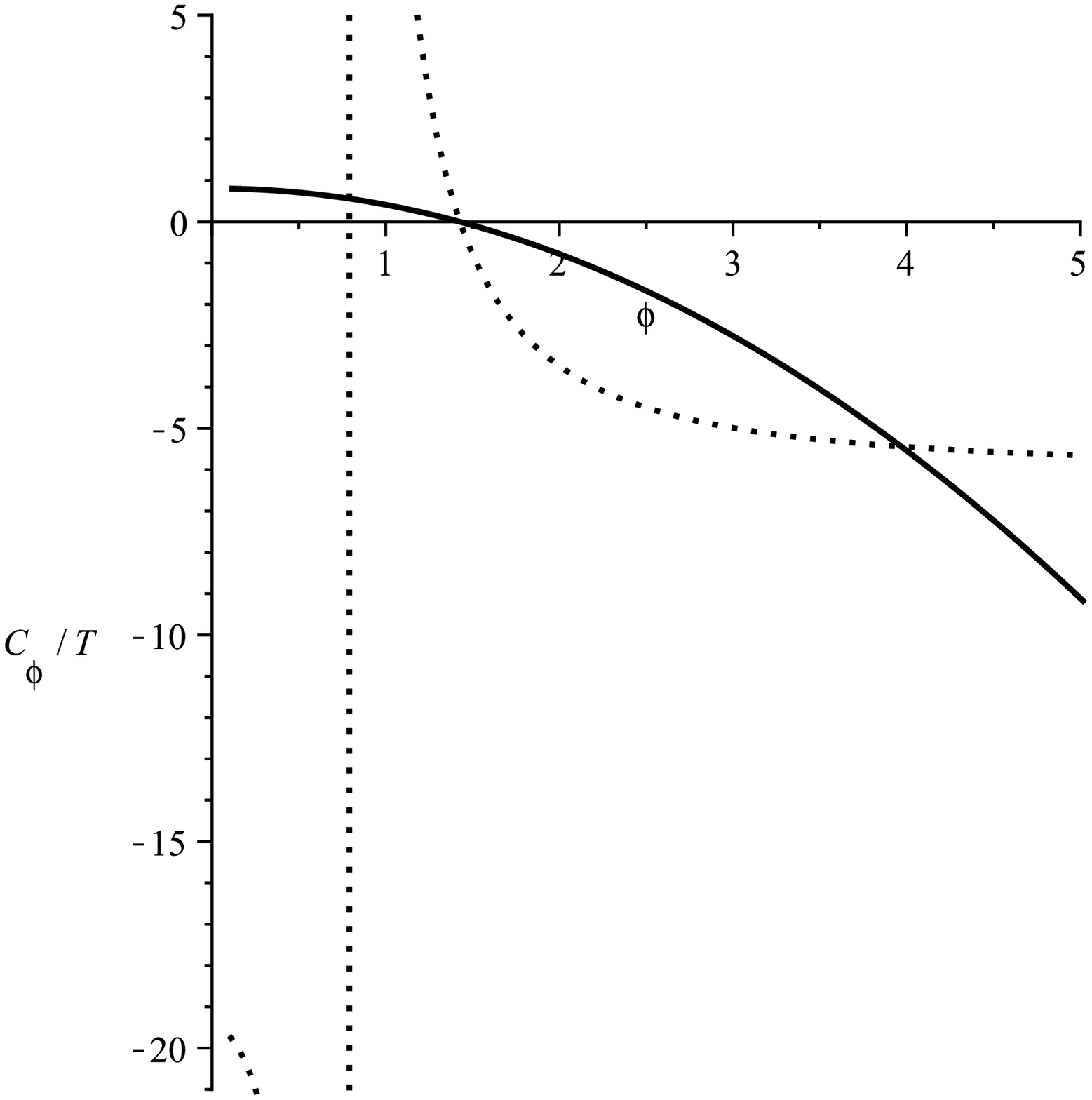}
  \caption{The behavior of the heat capacity (dotted curve) and the temperature (solid curve) for a fixed value of the electric potential, $\phi=1$, (left plot) and for a fixed value of the entropy, $S=2$ (right plot). In both plots we set $\Lambda =-1$. }
  \label{fig9c}
\end{figure}
The left plot, where the interval has been chosen such that $T>0$ is always satisfied, shows a phase transition at $S_c \approx 0.61$. The black hole is stable for $S>S_c$, and unstable in the interval $0<S<S_c$. The right plot shows that for a given value of the entropy, it is possible to find a range of values for $\phi$ in which the temperature is positive definite and the heat capacity is positive, singular or negative, indicating the existence of stable and unstable black holes.

Notice that the singularities of $C_\phi$ are different from those of $C_Q$; consequently, the corresponding phase transition structures do not coincide.
In fact, in Fig. \ref{fig9e} we show the behavior of the heat capacities and all thermodynamic potentials for this case. 
\begin{figure}
  \includegraphics[width=7cm]{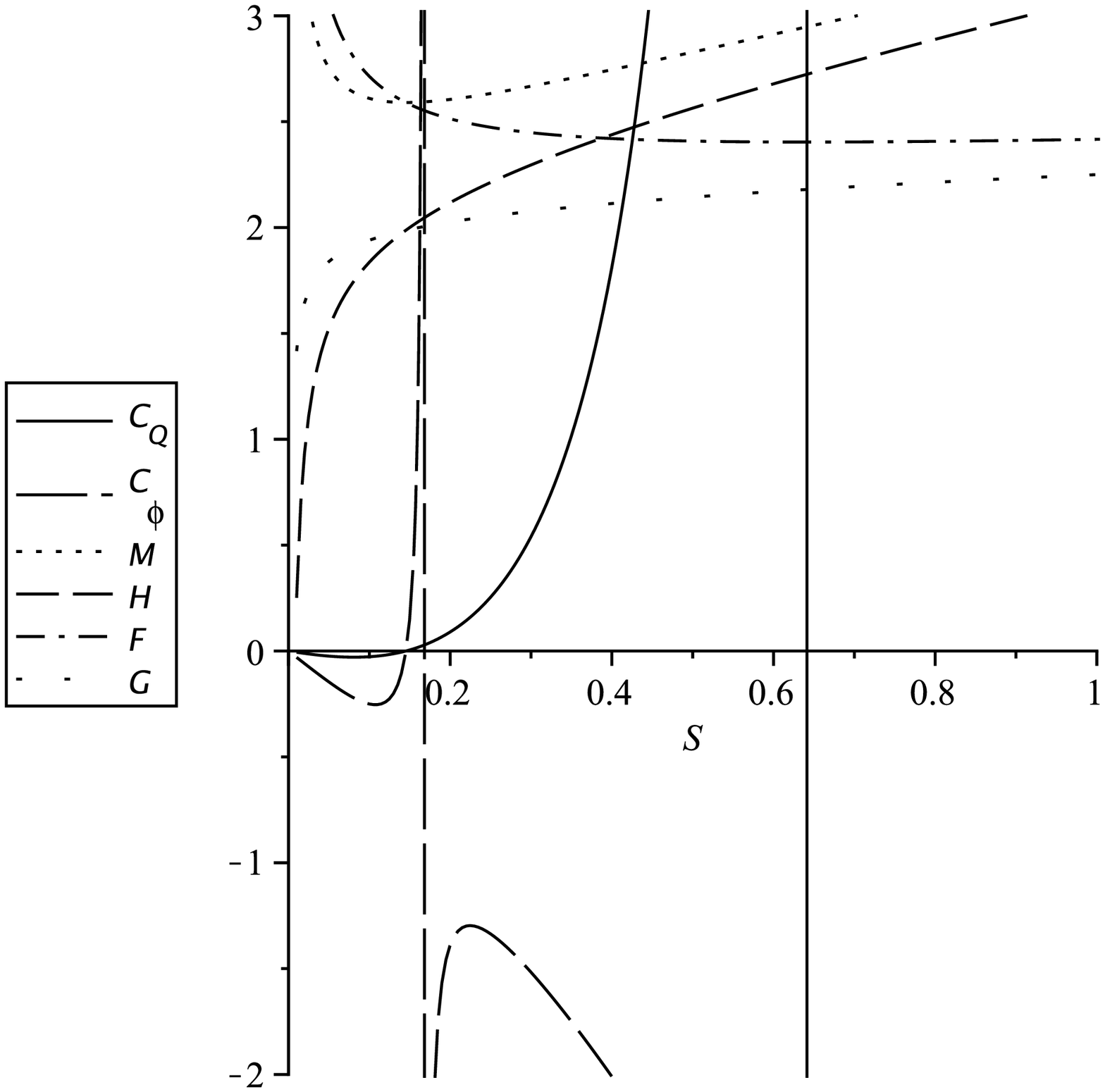}
  \includegraphics[width=7cm]{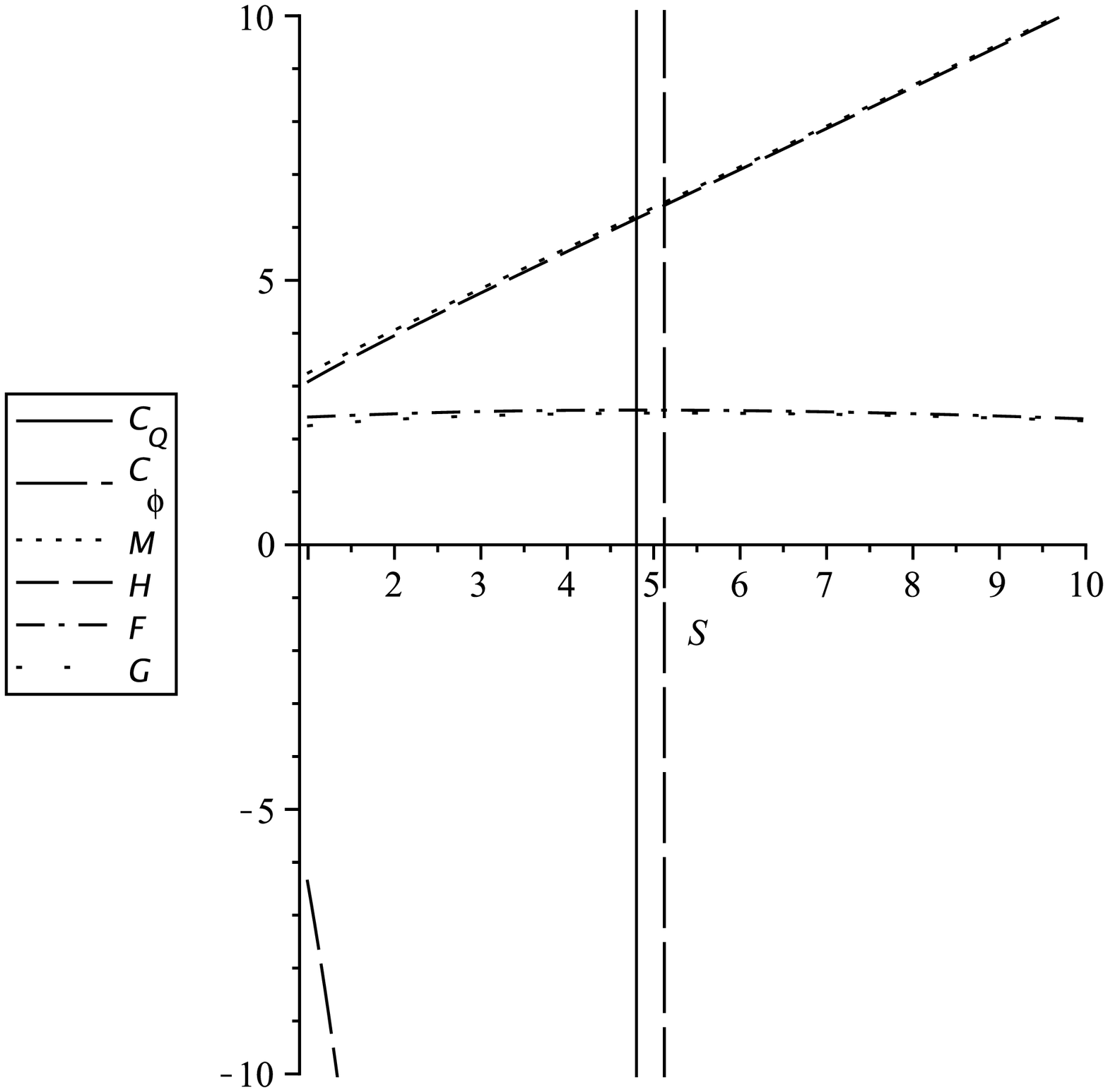}
  \caption{Heat capacities and thermodynamic potentials in terms of the entropy for $\alpha=1$, $\Lambda=-1$, and $Q=1/2$. For clarity, two different intervals of $S$ are depicted with different scales.}
  \label{fig9e}
\end{figure}
First, we see that at the points of phase transitions in $C_\phi$ all the potentials are well behaved and no critical points are observed. Moreover, the first divergence of $C_Q$ (at $S\approx 0.62$)  is situated on a local minimum of $F$ whereas the second singularity (at $S\approx 4.8$) is located on a local maximum. This situation is also illustrated in Fig. \ref{fig9g} where the Helmholtz free energy is plotted in terms of the entropy for different values of the electric charge.    
\begin{figure}
  \includegraphics[width=7cm]{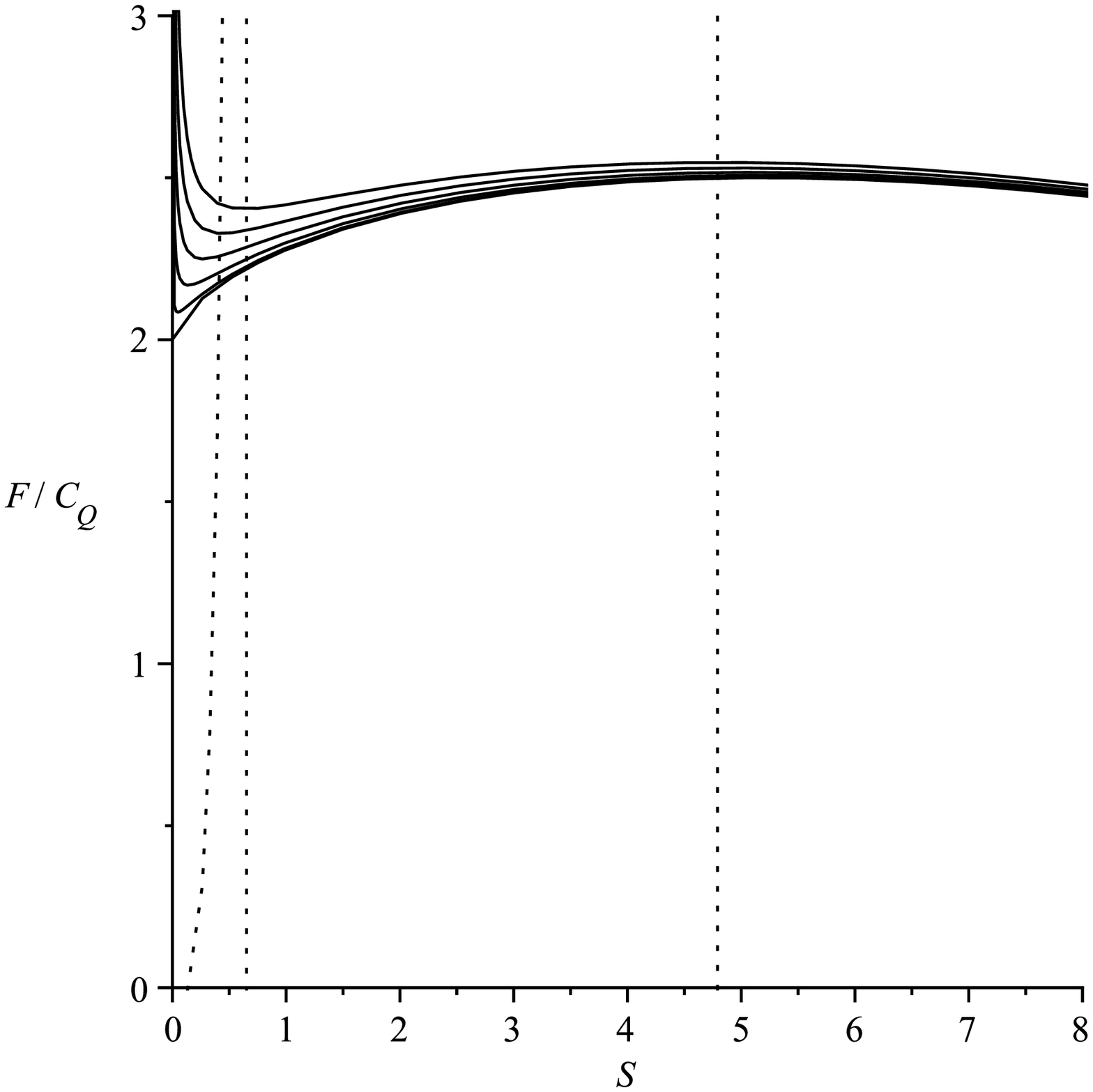}\quad
  \includegraphics[width=7cm]{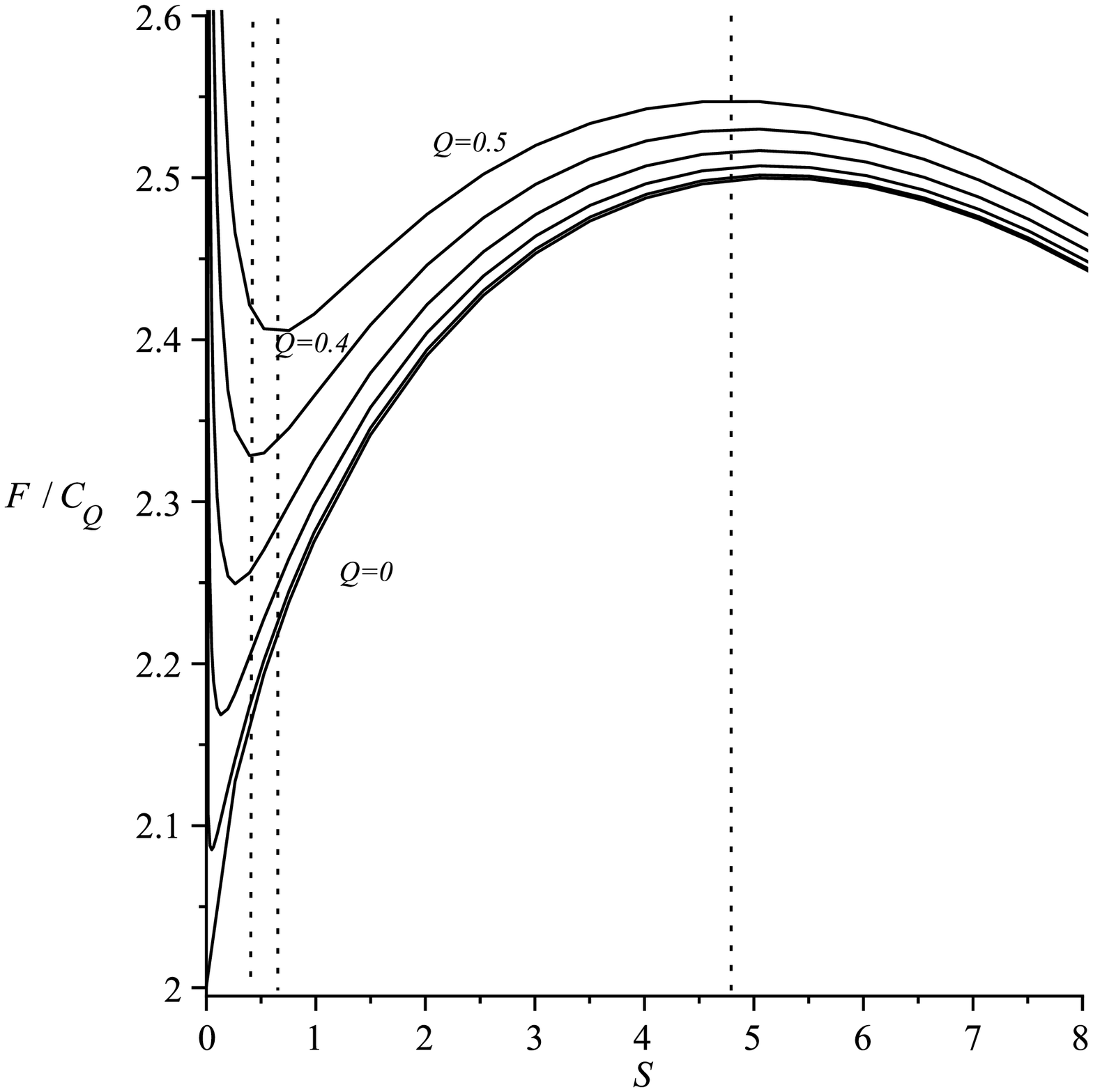}
  \caption{The Helmholtz free energy as a function of the entropy for different values of the electric charge. The heat capacity $C_Q$ (dotted curve) is also depicted for $Q=0.5$ to illustrate that phase transitions occur on the minimum and maximum of $F$ for the same value of $Q$. The left plot covers a larger interval of the vertical axis to illustrate the behavior of $C_Q$. The right plot illustrates in more detail the behavior of $F$. }
  \label{fig9g}
\end{figure}
The first phase transition occurs on a metastable point of $F$ and describes a configuration in which a stable black hole transforms into an unstable one. On the contrary, the second phase transition occurs on an unstable value $F$ and corresponds to a black hole that passes from an unstable state to a stable state.

\subsubsection{Geometrothermodynamics}

In this subsection, we derive the thermodynamic metrics for the equilibrium manifold. In the case of the grand canonical ensemble, the 
fundamental equation is given as $M=M(S,Q)$ in
Eq.(\ref{feqlam}). Then, we associate the coordinates $E^a=(S,Q)$ to
the equilibrium manifold ${\cal E}$ and $\Phi=M$ is the
thermodynamic potential. The thermodynamic metric (\ref{gdown}) can then be
written as 
\be g= \frac{4}{27 S^{4/3}}(3S^{4/3} - \Lambda S^2 -
Q^2)\left[ \frac{1}{9S^2}( 3S^{4/3} + \Lambda S^2 - 5 Q^2) dS^2 +
dQ^2\right] \ . 
\ee 
A straightforward computation results in the
following scalar curvature: 
\be R= \frac{27}{2}\frac{S^{7/3}
N(S,Q,\Lambda)}{ \left( 3\,{S}^{4/3}-{Q}^{2}-\Lambda\,{S}^{2}
\right) ^{3} \left( 3\,{S}^{4/3}-5\,{Q}^{2}+\Lambda\,{S}^{2} \right)
^{2} } , 
\ee with 
\bea N(S,Q,\Lambda)= &
&42\,{Q}^{2}{S}^{7/3}\Lambda-34\,S{Q}^{4}\Lambda-5\,{S}^{3}{Q}^{2}{
\Lambda}^{2}-18\,{Q}^{4}\sqrt [3]{S}\nonumber\\
&-&7\,{S}^{5}{\Lambda}^{3}+36\,{S}^{
11/3}\Lambda+15\,{S}^{13/3}{\Lambda}^{2}-162\,{S}^{3}+108\,{Q}^{2}{S}^
{5/3} \ .
\eea

From the expression for the scalar curvature it is obvious that the
singularities are located at the points satisfying the equation
$3\,{S}^{4/3}-{Q}^{2}-\Lambda\,{S}^{2} =0$, which coincide with the
points where $T\rightarrow 0$, and at the points satisfying the equation
$3\,{S}^{4/3}-5\,{Q}^{2}+\Lambda\,{S}^{2}=0$, which are the points
where $C_Q\rightarrow \infty$. For instance, for the particular case
$\Lambda=-1$ and $Q=1/2$ the singularities are shown in
Fig.\ref{fig9}; their locations clearly coincide with the points
where second order phase transitions occur (see right plot in
Fig.\ref{fig8}).

\begin{figure}
  \includegraphics[width=7cm]{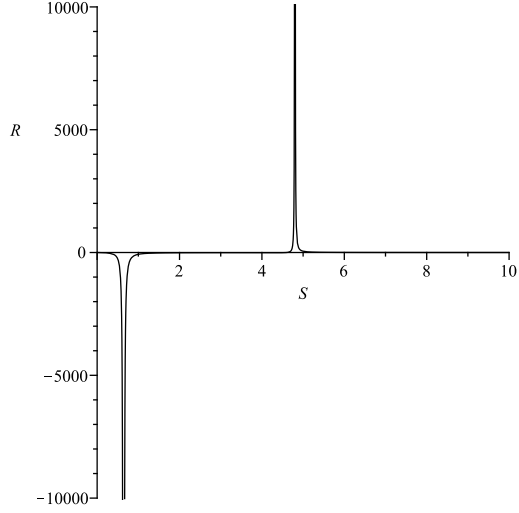}
 \caption{Curvature singularities in the equilibrium manifold of the
 black hole (\ref{fungl}) with $\Lambda=-1$ and $Q=1/2$. The
 singularities are located at
 $S\approx 0.6$ and $S\approx 4.9$. }
  \label{fig9}
\end{figure}

To see if the phase transitions predicted by $C_\phi$ can also be described in the context of GTD, let us consider the thermodynamic metric (\ref{gdown}) with the the  canonical. In this case, the thermodynamic potential is $\Phi = H$ and $E^a=(S,\phi)$ are the coordinates of the equilibrium manifold. Then, using the fundamental equation (\ref{feqlm1}), we obtain from (\ref{gdown}) the metric 
\be
g=\left(1-\frac{3}{4}\phi^2-\frac{\Lambda}{3}S^{2/3}\right)\left[\frac{4}{27}S^{-2/3}\left(1-\frac{3}{4}\phi^2+\frac{\Lambda}{3}S^{2/3}\right)dS^2 - S^{4/3}d\phi^2\right] \ ,
\ee
and the corresponding curvature scalar
\be
R = \frac{N(S,\phi,\Lambda)}{(12S^{1/3} - 9\phi^2S^{1/3} - 4\Lambda S)^3(12S^{1/3} - 9\phi^2S^{1/3} + 4\Lambda S)^2}\ ,
\ee
with
\be
N(S,\phi,\Lambda) = (12S^{1/3} - 9\phi^2S^{1/3} - 4\Lambda S)^2 +\Lambda S^{4/3}[27\phi^2(4-3\phi^2)+2\Lambda S^{2/3}(9\phi^2-4\Lambda S^{2/3}-4)]\ .
\ee
The curvature singularities situated at the roots of $12S^{1/3} - 9\phi^2S^{1/3} + 4\Lambda S=0$ determine the phase transition structure of the black hole, because they coincide with the singularities of $C_\phi$. The second set of singularities for which  $12S^{1/3} - 9\phi^2S^{1/3} - 4\Lambda S=0$ corresponds to the limit $T\rightarrow 0$ and indicates the break down of the thermodynamic description of the black hole.

\subsection{Geometrothermodynamics with a modified entropy relation}
\label{sec:mod3}

The modified entropy relation (\ref{smod5}), with $k=+1$, cannot be
solved in this case to obtain an explicit fundamental equation
$M=M(Q,S)$. We must therefore consider the implicit fundamental
equation determined by the relationships \be S=r_H^3 + 6\tilde\alpha
r_H\ , \quad M = \frac{\tilde\alpha}{3} + \frac{Q^2}{3r_H^2} + r_H^2
- \frac{\Lambda}{6} r_H^4 \ . \ee Then, the main thermodynamic
variables can then be expressed as
\begin{equation}
T=\frac{ 2 (3r^{4}_{H}-Q^{2}-\Lambda r^{6}_{H})}
{9r^{3}_{H}(r^{2}_{H}+2\tilde{\alpha})},
\end{equation}
\begin{equation}
\phi=\frac{2 Q}{3r^{2}_{H}}\ ,
\end{equation}
\begin{equation}
C_{Q}=\frac{3r_{H}(r^{2}_{H}+2\tilde{\alpha})^{2}(3r^{4}_{H}-Q^{2}-
\Lambda r^{6}_{H})}{6\tilde{\alpha}Q^{2}+5Q^{2}r^{2}_{H}+6\tilde
{\alpha}r^{4}_{H}- 3r^{6}_{H}(1+ 2\tilde{\alpha}\Lambda)-\Lambda
r^{8}_{H}}\ . 
\label{hcmod3}
\end{equation}

For a physically reasonable configuration we demand the positiveness
of the temperature; this implies that $3r_H^4  -\Lambda  r_H^6 >
Q^2$. For a given value of $\Lambda$ and $Q$, this condition
determines a minimum horizon radius $r_H^{min}$ for which the
temperature is positive. Moreover, from the expression for the heat
capacity (\ref{hcmod3}) and from the condition of positive
temperature, it follows that if the condition \be \tilde \alpha
|\Lambda| \geq \frac{1}{2} \ \label{condlambda} \ee is satisfied,
the heat capacity is positive and, consequently, all possible  black
hole configurations are stable. This is an interesting condition
that relates two fundamental constants, namely, the tension of the
string, proportional to  $\tilde{\alpha}^{-1}$, and the cosmological
constant $\Lambda$.

For the range $\tilde \alpha |\Lambda|< 1/2$ where unstable states
in principle can exist, let us consider the parameters $\Lambda=-1$
and $Q=1$. This choice together with the positiveness condition of
the temperature fix the value of $r_{H}^{min}\approx 0.73$ (see left
plot in Fig.\ref{fig10}). Notice that the value of $r_H^{min}$ does
not depend on the value of the coupling constant $\tilde\alpha$. We
explore the behavior of the heat capacity in Fig.\ref{fig10} for the
entire range $\tilde\alpha \in (0,1/2)$, according to the condition
$\tilde \alpha |\Lambda|< 1/2$. One can see that the heat capacity
is represented by a smooth positive function in the entire domain.
We conclude that also in this case all the black hole configurations
are stable.

\begin{figure}
  \includegraphics[width=5cm]{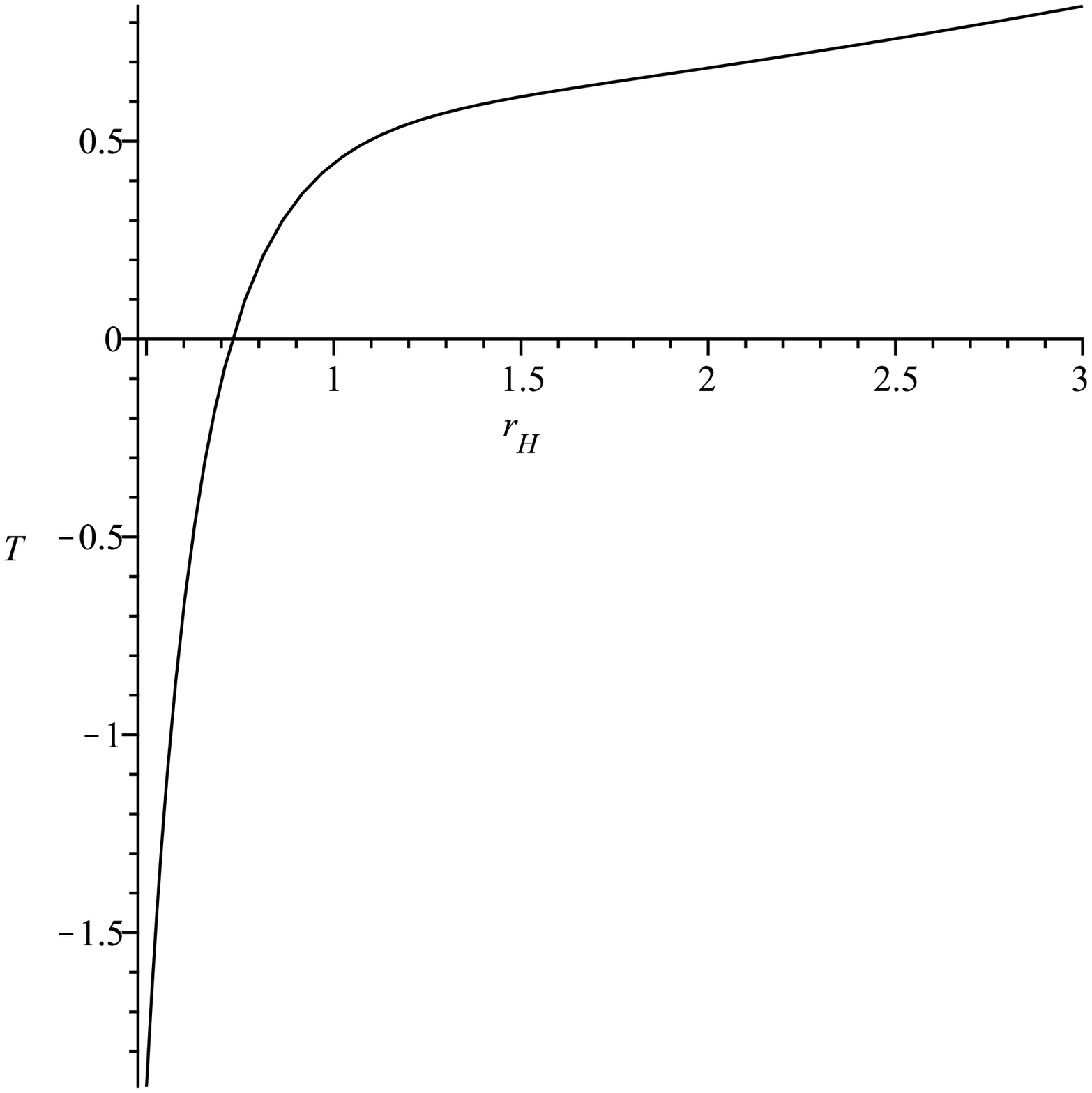}
  \includegraphics[width=5cm]{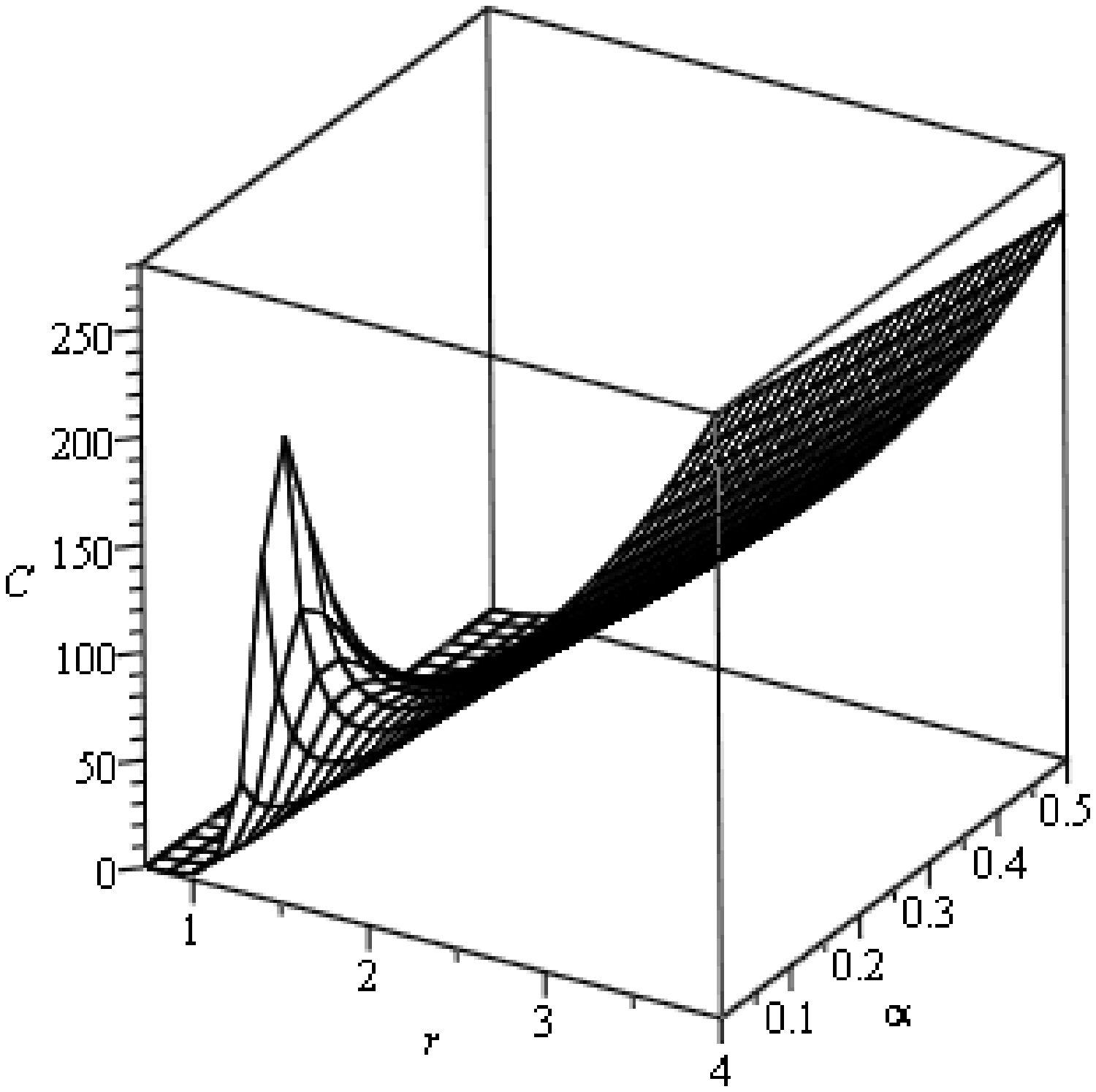}
  \includegraphics[width=5cm]{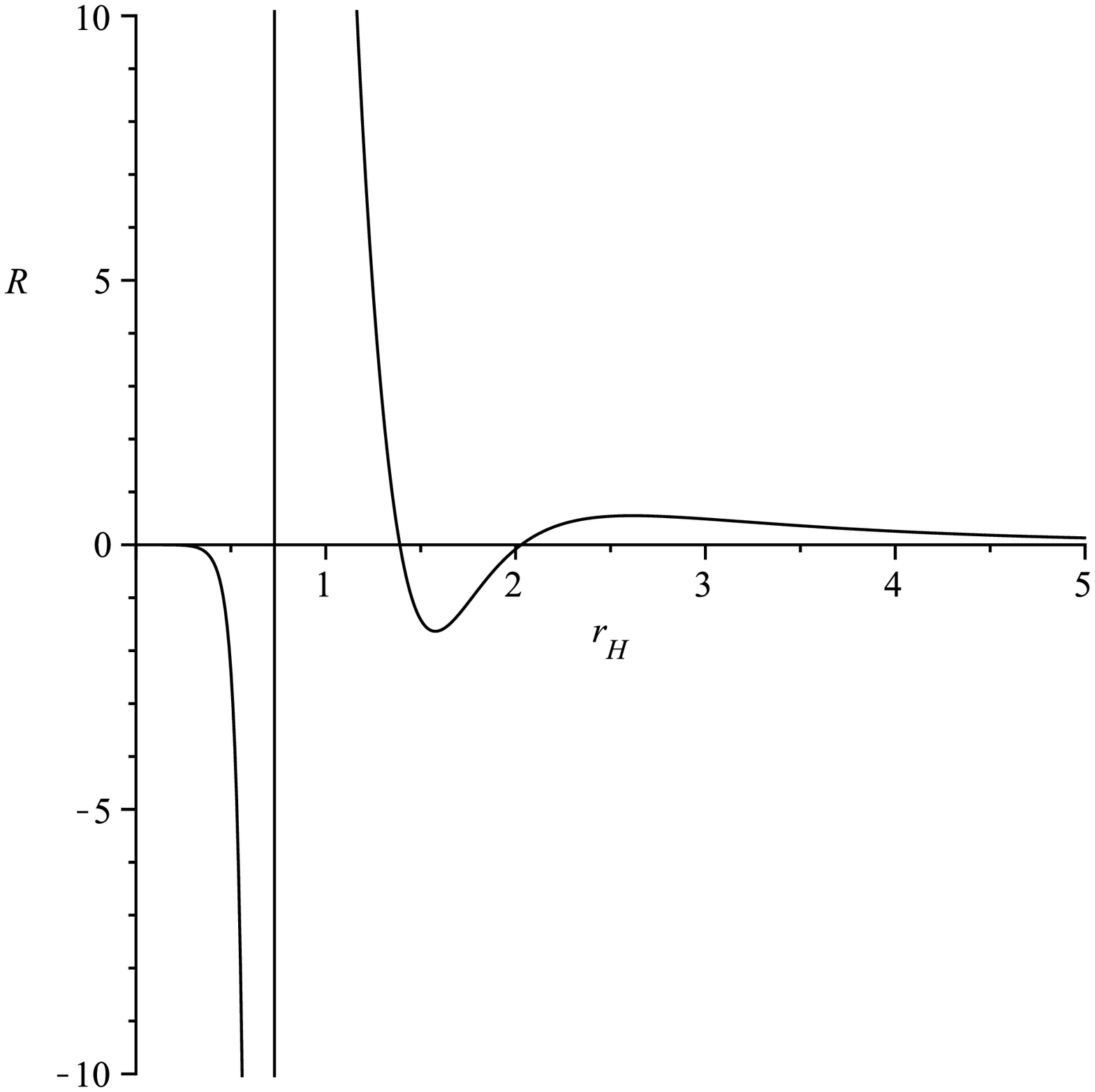}
 \caption{Temperature, heat capacity $C_Q$, and thermodynamic curvature
  for a black hole of the EMGB theory with cosmological constant
  $\Lambda=-1$, charge $Q=1$, and coupling constant $\tilde\alpha=1/4$
  (for the temperature) and $\tilde\alpha \in (0,0.5)$ (for the
 heat capacity).  The temperature is positive for $r_H> r_H^{min}\approx 0.73$. }
  \label{fig10}
\end{figure}

We now investigate the geometric properties of the equilibrium
manifold. According to the implicit fundamental equation, the
thermodynamic metric (\ref{gdown}) can be written as
\begin{equation}
g=-f_{1}(r_H,Q,\tilde{\alpha},\Lambda)\bigg\{\left(5Q^{2}r^{2}_{H}+6\tilde{\alpha}
r^{4}_{H}-3r^{6}_{H}(1+2\tilde{\alpha}\Lambda)-\Lambda
r^{8}_{H}\right)dS^{2}+9r_H\left(r_H^2+2\tilde{\alpha}\right)^2dQ^2\bigg\},
\end{equation}
where
\begin{equation}
f_{1}(r_H,Q,\tilde{\alpha},\Lambda)=\frac{4}{243}\frac{(r_H^2+6\tilde\alpha)
(3r^{4}_{H}-
Q^{2}-\Lambda r^{6}_{H})}{r_H^6(r_H^2+2\tilde\alpha)^4}.
\end{equation}
The expression for the scalar is quite cumbersome but it can
schematically be represented as \be R= \frac{N(r_{H}, \Lambda, Q,
\tilde \alpha )} { (3r^{4}_{H}-Q^{2}-\Lambda r^{6}_{H}) ^3
\left(6\tilde{\alpha}Q^{2}+5Q^{2}r^{2}_{H}+6\tilde{\alpha}r^{4}_{H}-3r^{6}_{H}(1+
2\tilde{\alpha}\Lambda)-\Lambda r^{8}_{H}\right)^2
(r_H^2+6\tilde{\alpha})^3} \ , \ee where $N(r_{H}, Q, \Lambda,
\tilde \alpha )$ is a finite function in the entire domain of
definition. From the expression for the scalar curvature, the
temperature and heat capacity given above, it follows that
singularities can take place only at those points where
$T\rightarrow 0$ or $C\rightarrow \infty$. In Fig.\ref{fig10}, the
behavior of the scalar curvature is shown for a particular choice of
the parameters. We see that a singularity occurs at the point where
the temperature vanishes. The singularity situated at $
(r_H^2+6\tilde{\alpha})=0$ corresponds to the limit $S\rightarrow 0$
which indicates the breakdown of the thermodynamic picture of the
black hole and, hence of GTD. No other singularities exist because
the heat capacity is finite in this domain.

\section{Spherically symmetric  black hole in EYMGB gravity}
\label{sec:fock}

In this section we first describe the black hole solution \cite{A24}
and its properties in Einstein-Yang-Mills-Gauss-Bonnet (EYMGB)
gravity, and then study the geometry of the black hole
thermodynamics in the subsequent sections.

The 5D spherically symmetric solution obtained recently by
Mazharimousavi and Halisoy \cite{A24} has the metric
\begin{equation}
ds^{2}=-f(r)dt^{2}+\frac{dr^{2}}{f(r)}+r^{2}d\Omega _{3}^{2}\ ,
\end{equation}%
where
\begin{equation}
f(r)=1+\frac{r^{2}}{4\alpha } - \sqrt{1 +\frac{M}{2\alpha} +
\frac{r^{4}} {16\alpha^2 } +\frac{Q^{2}}{\alpha} \ln r }\ .
\end{equation}%
Here $M$ is an integration constant to be identified as the mass,  and
$Q$ is the only non-zero gauge charge. The Yang-Mills field 2-form
$F_{\alpha \beta }^{i}F^{i\alpha \beta }=6Q^{2}/r^{4}$ represents in
this case the matter Lagrangian in the general action
(\ref{egbaction}). The metric on the unit three sphere $d\Omega
_{3}^{2}$ is given by
\begin{equation}
d\Omega _{3}^{2}=\frac{1}{4}(d\theta ^{2}+d\phi ^{2}+d\psi
^{2}-2\cos \theta d\phi d\psi ),
\end{equation}%
with $\theta \in \lbrack 0,\pi ]$, and $(\phi ,\psi )\in \lbrack 0,2\pi]$.
The event horizon radius  $r_{H}$ satisfies the equation $f(r_H)=0$ and
is given by
\begin{equation}
M=r_{H}^{2}-2Q^{2}\ln r_{H} \ .
\label{feqymgb0}
\end{equation}%

This black hole solution of the EYMGB theory is
well defined for all $r$ if the Gauss--Bonnet coupling parameter
$\alpha $ is positive definite. For $\alpha <0$ , the spacetime
has a curvature singularity at the hypersurface $r=r_{s},$ where
$r_{s}$ is the largest root of $f(r)=0$.

\subsection{Analysis with the Bekenstein-Hawking entropy relation}
\label{sec:bh2}

In suitable units,  the entropy $S$ of the black hole is given by
$S=r_{H}^{3}$, and $A=2\pi ^{2}r_{H}^{3}$ is the surface area of
the event horizon. Then, according to Eq.(\ref{feqymgb0}),
the thermodynamic fundamental equation in the $M$-representation becomes
\begin{equation}
M=S^{\frac{2}{3}}-\frac{2}{3}Q^{2}\ln S\ .
\label{feqeymgb}
\end{equation}%
This is the main relationship from which all the thermodynamic properties of this black hole can be derived.

\subsubsection{Thermodynamics}

The expressions for the main thermodynamic quantities,  namely, the
temperature and the electric potential are given by
\begin{equation}
T=\frac{2}{3} \frac{S^{2/3}-Q^2} {S } \ ,\quad
\phi =- \frac{4}{3}Q\ln S\ .
\end{equation}%
It follows that for the temperature of the black hole to be positive
the charge must satisfy the condition $Q<S^{1/3}$. Moreover, for a
fixed value of the entropy, the maximum temperature is reached at
the value $Q=0$, indicating that the  Yang-Mills charge reduces the
temperature of the black hole. This  behavior is illustrated in
Fig.\ref{fig5}.

Now, in the grand canonical ensemble  
the heat capacity is given by  the expression
\begin{equation}
C_Q=- 3S\frac{S^{2/3}-Q^{2}}{ S^{2/3} -3 Q^{2} } \ .
\label{hceymgb}
\end{equation}%
In the region $Q<S^{1/3}$, where the temperature is positive, the
heat capacity diverges at those points where $Q=S^{1/3}/\sqrt{3}$,
indicating the existence of a second order phase transition. In the
interval $S^{1/3}/\sqrt{3} < |Q| < S^{1/3}$, the heat capacity is
positive (and $T>0$), i. e., the black hole configuration is stable
in this interval. Furthermore,  the heat capacity is negative within
the interval $0 < |Q| < S^{1/3}/\sqrt{3}$ which corresponds to an
unstable thermodynamic configuration. Since the heat capacity at
$Q=0$ is negative, we conclude that the addition of a Yang-Mills
charge $Q$ to an unstable neutral black hole not only reduces its
temperature, but also changes its heat capacity until it becomes
positive and the system becomes stable, if the charge is
sufficiently large. The transition from an unstable state to a
stable state is accompanied by a second order phase transition. This
thermodynamic behavior is illustrated in Fig.\ref{fig5}.

\begin{figure}
  \includegraphics[width=6cm]{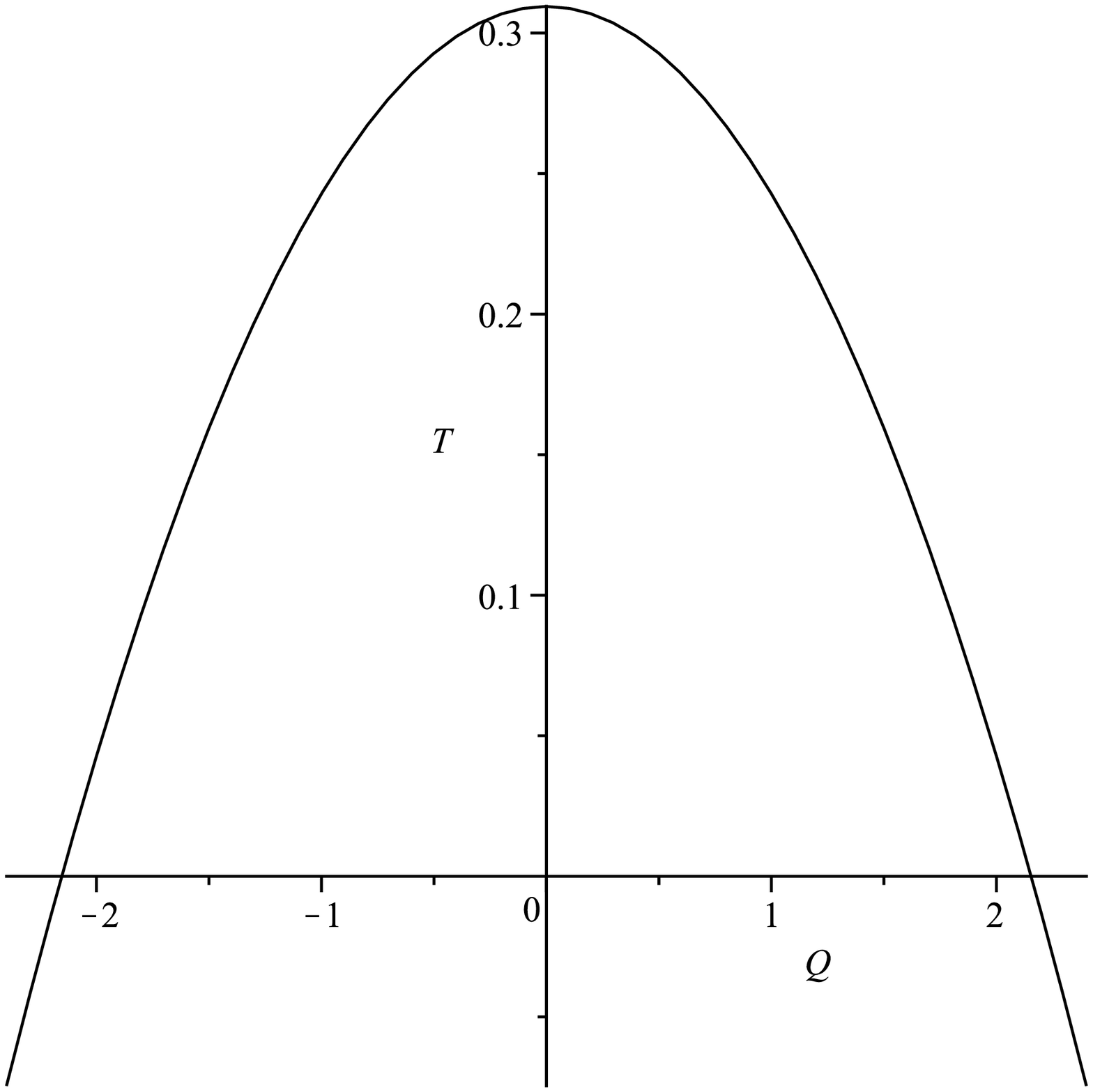}
  \includegraphics[width=6cm]{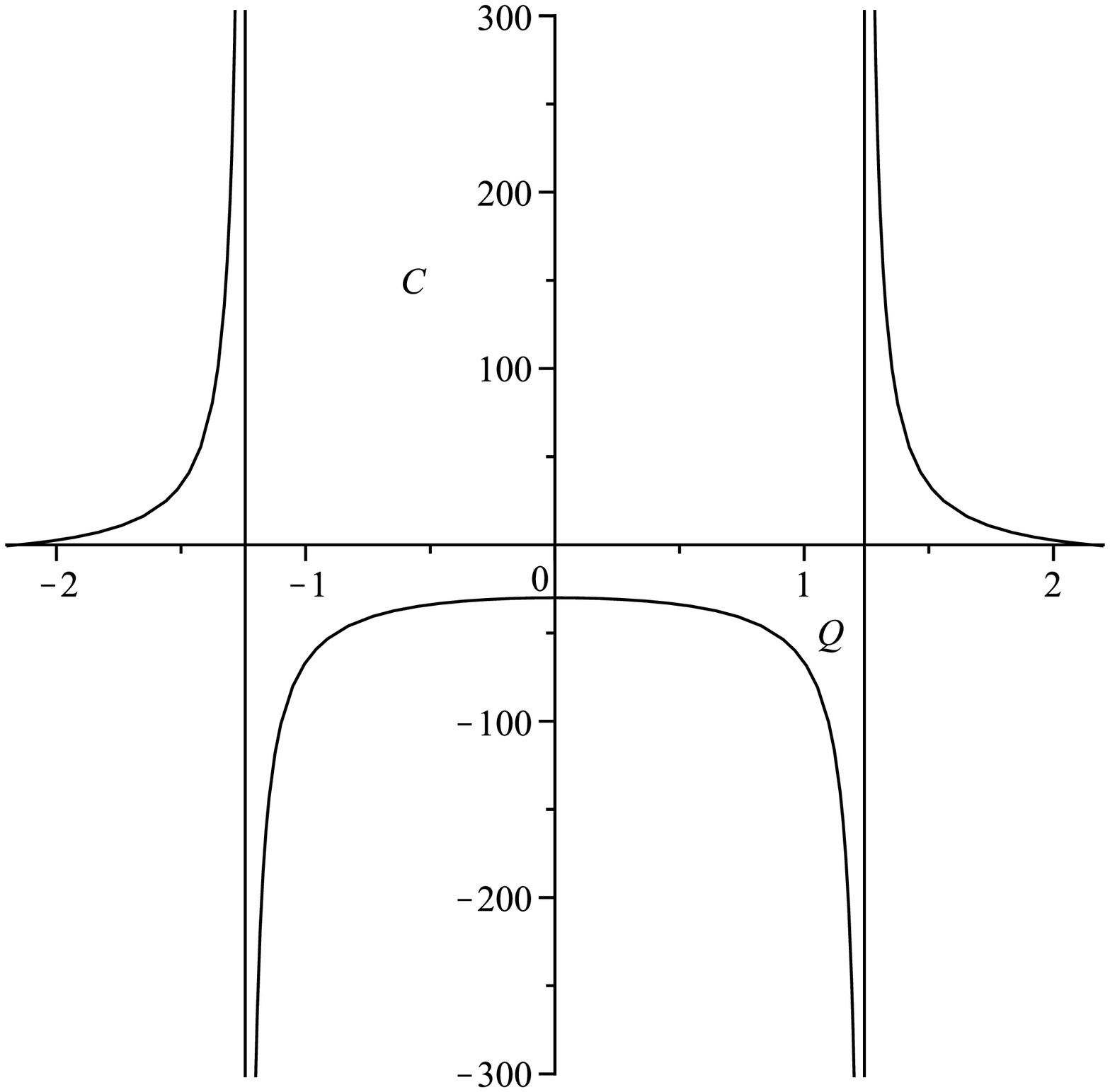}
  \caption{Temperature and the heat capacity $C_Q$ in terms of the Yang-Mills
   charge $Q$. To illustrate the thermodynamic behavior, the Bekenstein-Hawking
entropy was chosen
  as $S=10$.}
  \label{fig5}
\end{figure}

Consider now the canonical ensemble  that is determined by the thermodynamic potential
\be
H \equiv M - \phi Q = S^{2/3} + \frac{3}{8}\frac{\phi^2}{\ln S}\ ,
\label{MHeymgb}
\ee 
from which the dual thermodynamic variables are obtained as
\be
T=\frac{16S^{2/3}\ln^2 S - 9 \phi^2 }{24 S\ln^2 S}\ ,\quad Q = -\frac{3\phi}{4 \ln S}  \ .
\ee
Furthermore, the heat capacity at fixed potential is  given by
\be
C_\phi=\left(\frac{\partial H}{\partial T}\right)_\phi= 
\frac{3S\ln S \, (16S^{2/3} \ln^2 S - 9\phi^2 )}
{-16S^{2/3}\ln^3 S + 27\phi^2(2+ \ln S)}\ ,
\ee
and predicts a phase structure different from that of $C_Q$ as given in Eq.(\ref{hceymgb}). Indeed, in Fig. \ref{fig6a} we illustrate the behavior of $C_\phi$ and the temperature as functions of the entropy. 
\begin{figure}
  \includegraphics[width=7cm]{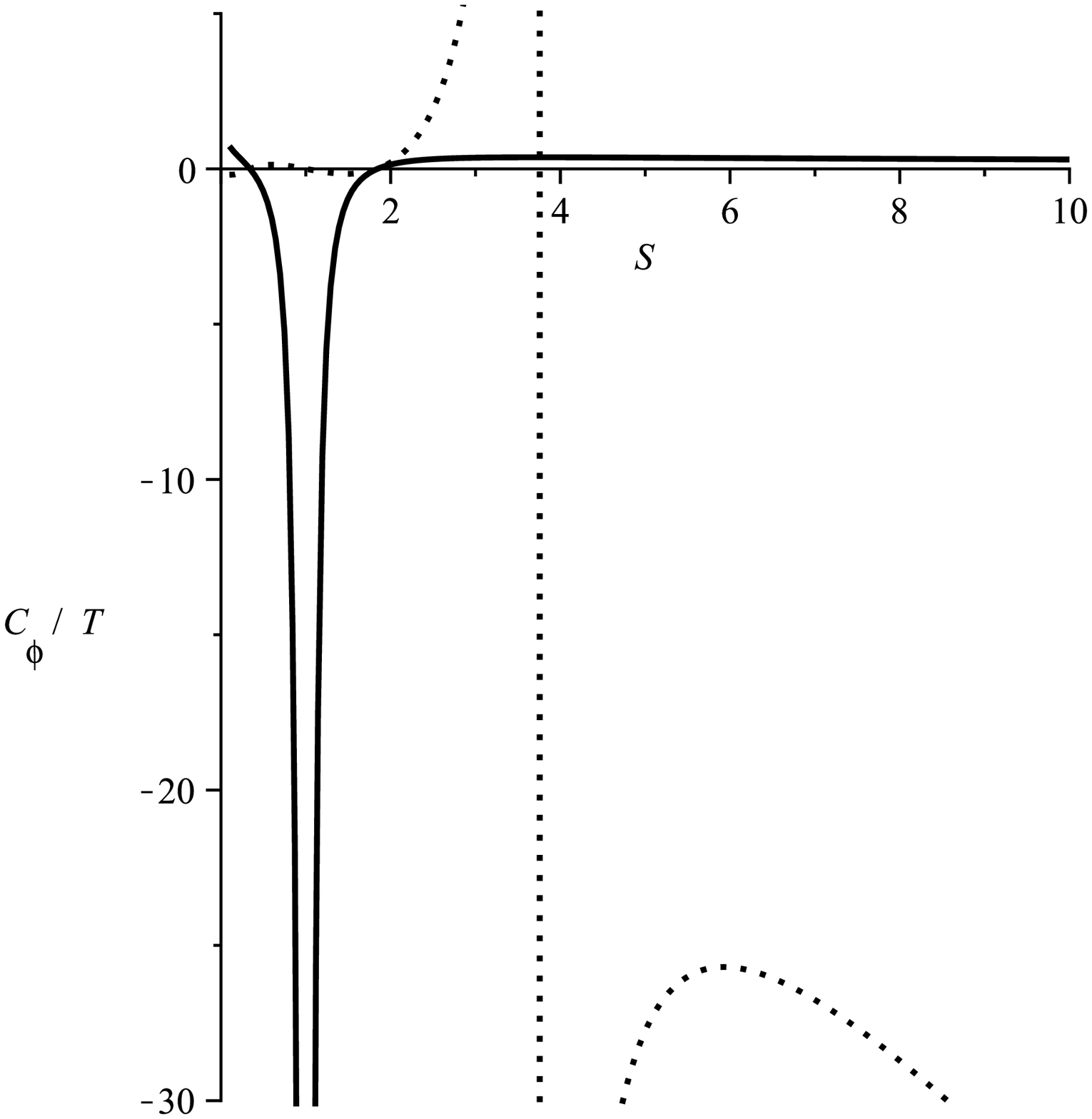}\quad
  \includegraphics[width=7cm]{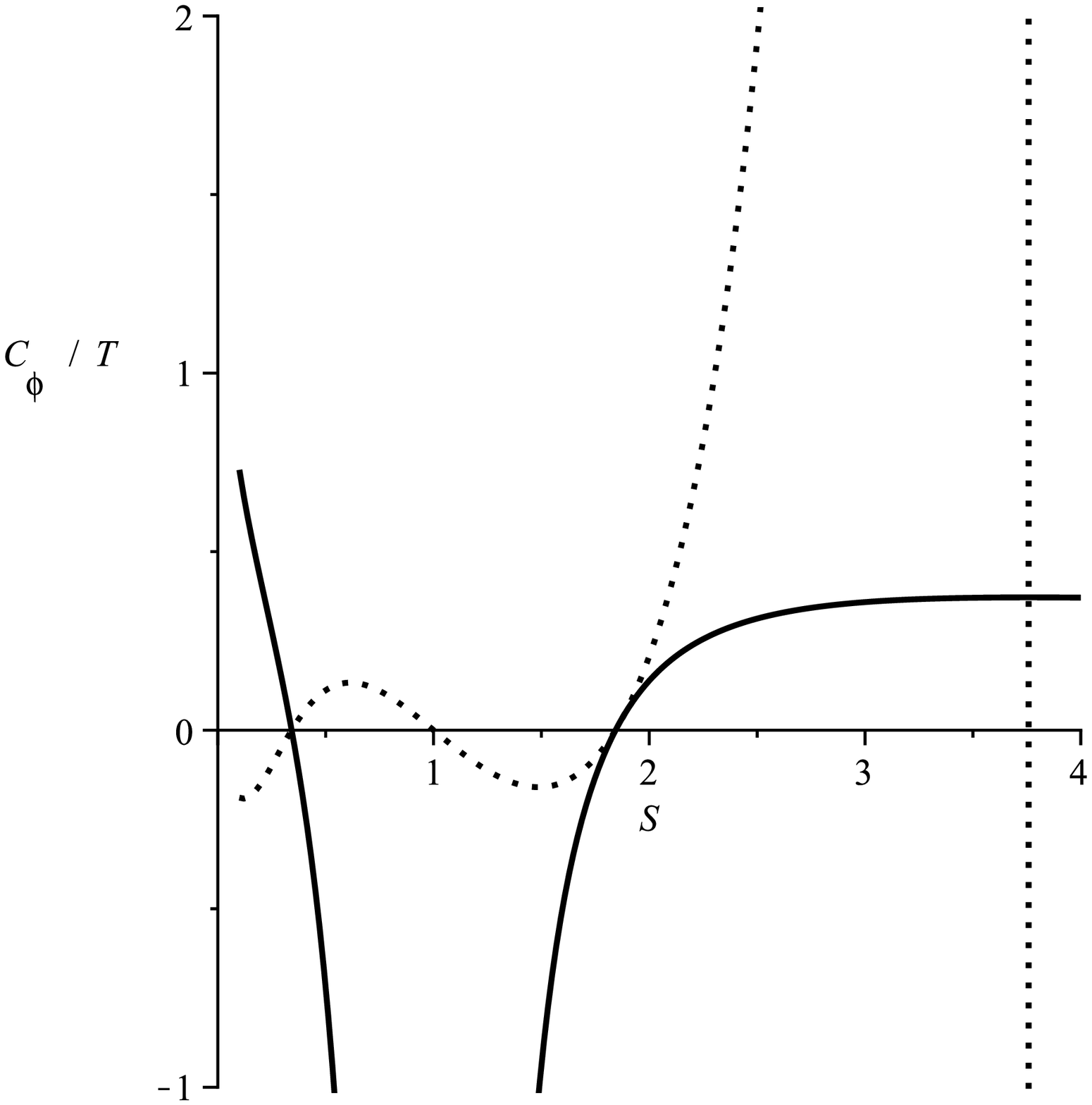}
  \caption{The heat capacity $C_\phi$ (dotted curve) and the temperature (solid curve) in terms of the entropy. Here $\phi=1$ and $\Lambda=-1$. }
  \label{fig6a}
\end{figure}
In the interval $S\in (0,0.4)$ with positive temperature, the black hole is unstable. Then, in the interval $S\in (0.4, 1.8)$ no black holes can exist with positive temperature. In the following interval $S\in ( 1.8, 3.7)$ all black holes are stable and have positive temperature. The vertical dotted line situated at $S\approx 3.7$ denotes the singularity of $C_\phi$ that corresponds to a second order phase transition during which the stable black hole changes into an unstable state. 

To investigate the position of the phase transitions in the parameter space we consider all the thermodynamic potentials $M$, $H$, $F$, and $G$. Their behavior is depicted in Fig. \ref{fig6c}.
\begin{figure}
  \includegraphics[width=7cm]{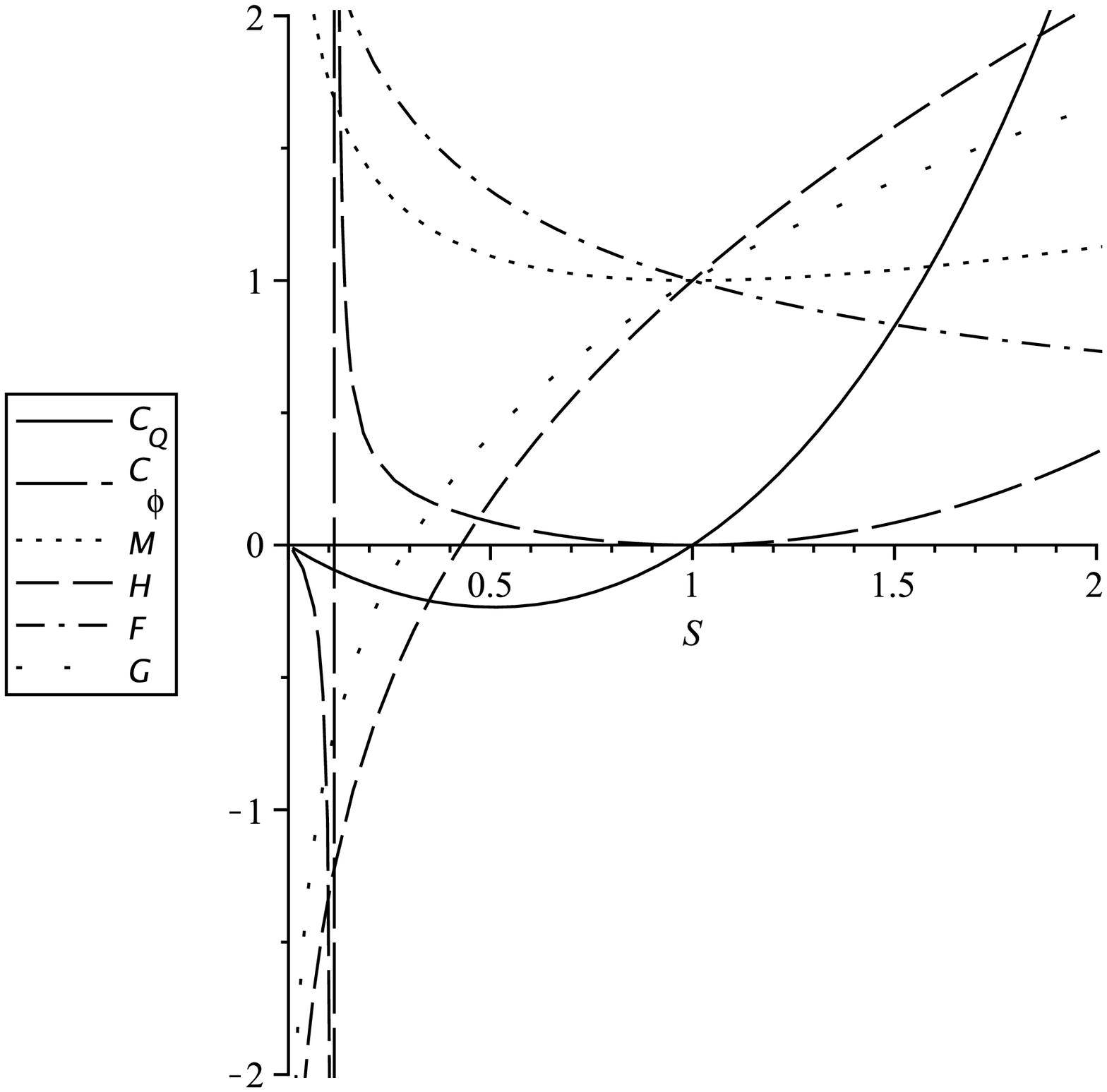}\quad
  \includegraphics[width=7cm]{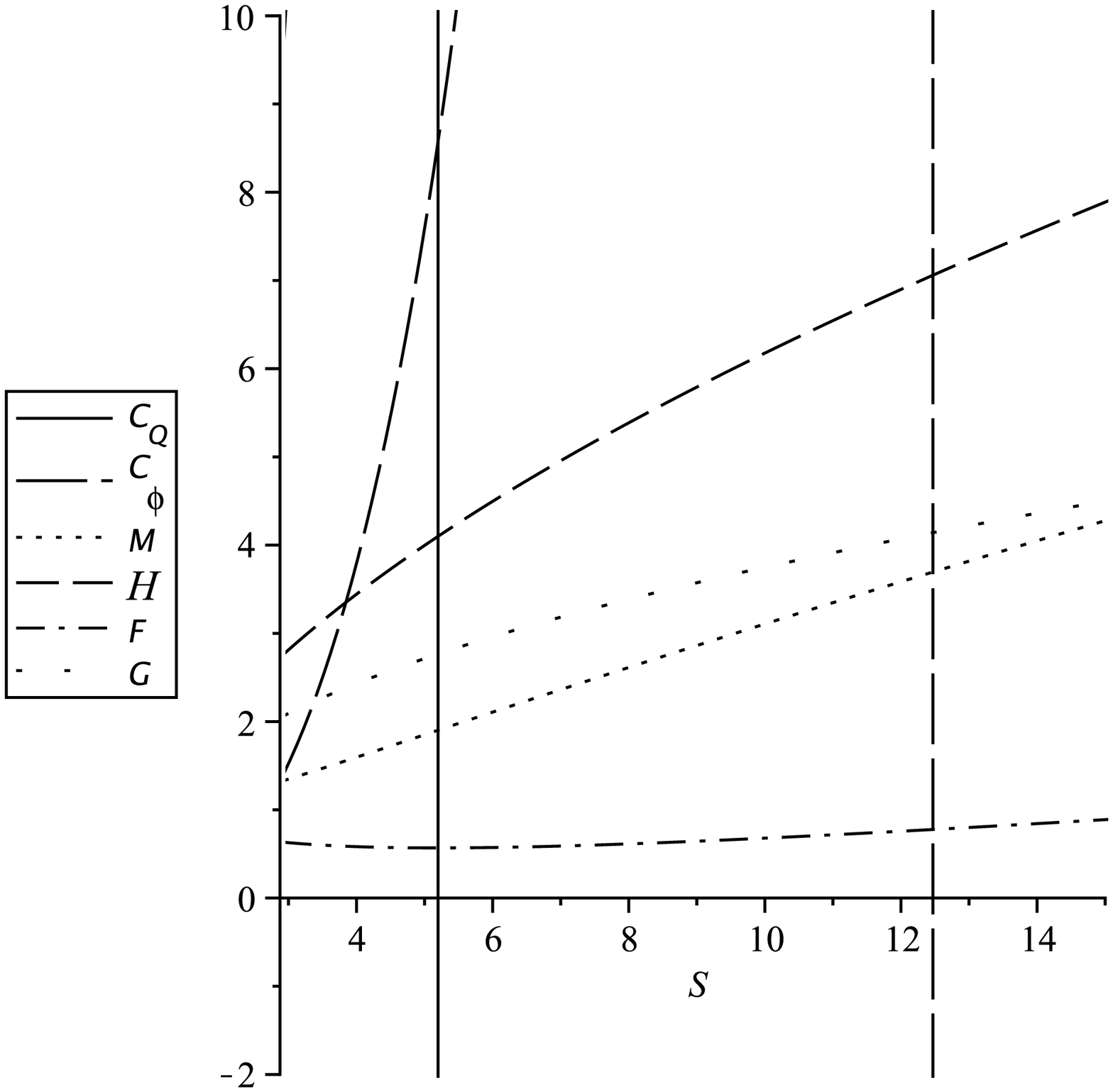}
  \caption{Heat capacities and thermodynamic potentials of the EYMGB black hole for the particular value $Q=1$. }
  \label{fig6c}
\end{figure}
It can be seen that at the point where $C_\phi$ diverges, all the potentials are well-behaved with no extrema. As for the heat capacity $C_Q$, the location of the singularity coincides with the only minimum of $F$. In Fig. \ref{fig6e}, the behavior of $F$ and $C_Q$ are depicted for different values of the charge. 
\begin{figure}
  \includegraphics[width=7cm]{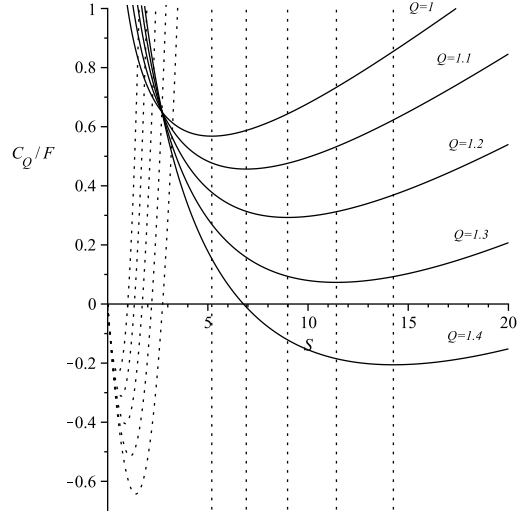}
  \caption{The Helmholtz free energy $F$  (solid curves) and the heat capacity $C_Q$ (dotted durves) in terms of the entropy for different values of the entropy. 
  In each case, the minimum of $F$ coincides with the divergence of $C_Q$.}
    \label{fig6e}
\end{figure}
It follows that the phase transitions take place in regions where the Helmholtz potential is stable.

\subsubsection{Geometrothermodynamics}

To investigate the geometry of the corresponding equilibrium
manifold in the grand canonical  ensemble, we use the general metric (\ref{gdown}) with $\Phi=M$ and $E^a=(S,Q)$, 
and the fundamental equation $M=M(S,Q)$ as given in Eq.(\ref{feqeymgb}).
Then, we obtain 
\be 
g= \frac{4}{27}\left(S^{2/3}-Q^2\right)\left(
\frac{S^{2/3}-3Q^2}{S^2} dS ^2 - 6 \ln S \, dQ^2\right)\ . 
\ee 
The corresponding thermodynamic curvature scalar can be represented as
\be R= \frac{N(S,Q)}{ (S^{2/3}-Q^2)^3(S^{2/3}-3Q^2)^2 \ln^2 S } \ ,
\ee
where $N(S,Q)$ is a well-behaved function of its arguments. We see
that there are several places where true curvature singularities can
exist. First, if $Q=S^{1/3}$ the  curvature scalar diverges and, as
described above, the temperature vanishes. Then, at
$Q=S^{1/3}/\sqrt{3}$ there exists a singularity whose location
coincides with the values at which the heat capacity $C_Q$ diverges and
second order phase transitions occur. Finally, if $\ln S \rightarrow
0$ the curvature scalar diverges. We interpret this additional
singularity as related to a second order phase transition which is
not contained in $C_Q$. In fact, in analogy to the heat capacity
$C_Q$ defined in Eq.(\ref{hceymgb}), we can introduce the
capacitance $C_S \equiv (\partial Q/\partial \Phi)_S = (\partial
\Phi/\partial Q)_S^{-1} =(\partial^2M/\partial Q^2)_S^{-1}$. Then,
from the fundamental equation (\ref{feqeymgb}), we obtain
$C_S=-3/(4\ln S)$ so that in the limit $S\rightarrow 1$ a second
order phase transition occurs. This proves the physical origin of
the additional singularity of the thermodynamic curvature. The
behavior of this thermodynamic curvature is depicted in
Fig.\ref{fig6}.

\begin{figure}
  \includegraphics[width=6cm]{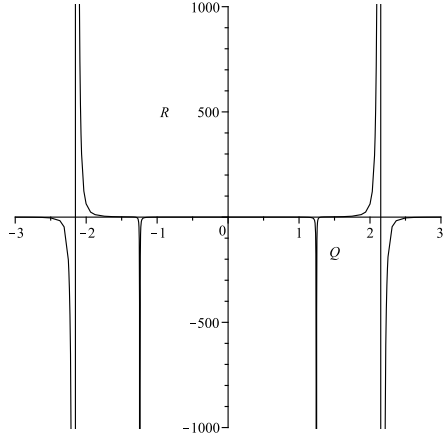}
  \caption{The thermodynamic curvature with $S=10$ as a function of the
  Yang-Mills charge $Q$. Singularities are due to the existence of a second
  order phase transition at $Q\approx \pm 1.24$ or to the vanishing of the
temperature at $Q\approx \pm 2.15$.}
  \label{fig6}
\end{figure}

We now investigate the geometric properties of the equilibrium manifold of this black hole in the canonical ensemble. 
The fundamental equation $H=H(S,\phi)$ is given by  Eq.(\ref{MHeymgb}) so that we can identify the thermodynamic potential as $\Phi=H$ and $E^a=(S,\phi)$. Then, from the general thermodynamic metric (\ref{gdown}) we obtain
\be
g=\frac{16S^{2/3}\ln^2 S - 9\phi^2}{\ln^3 S}\left[\frac{-16S^{2/3}\ln^3 S + 27\phi^2(2+\ln S)}{54 S^2 \ln^2 S}dS^2 -d\phi^2\right]\ ,
\ee
whose curvature scalar can be represented as 
\be
R=\frac{N(S,\phi)}{(-16S\ln^3 S + 54\phi^2 S^{1/3}+27\phi^2 S^{1/3}\ln S)^2(16S \ln^2 S - 9\phi^2S^{1/3})^3}\ .
\ee
The first term in the denominator of $R$ coincides with the denominator of $C_\phi$ and, consequently, reproduces exactly the phase transition structure of the black hole. The second term in the denominator of $R$ vanishes as $T\rightarrow 0$, indicating the break down of the thermodynamic and geometrothermodynamic descriptions.

\subsection{Geometrothermodynamics with a modified entropy relation}
\label{sec:mod1}

In this case, the thermodynamic fundamental equation cannot be
written explicitly. Therefore, we use the implicit equation
$S=S(M,Q)$ determined by the expressions
\be 
S=r_H^3 + 6 \tilde
\alpha r_H\ , \quad \mbox{and} \quad M=r_H^2 - 2 Q^2 \ln r_H \ .
\label{feqmodym}
\ee
Then, we find the following thermodynamic
variables
\begin{equation}
T=\frac{2}{3}\frac{r_{H}^{2}-Q^{2}}{r_{H}(r_{H}^{2}+2\tilde{\alpha})}\ ,
\end{equation}
\begin{equation}
\phi=4Q\ln{r_{H}}\ ,
\end{equation}
\begin{equation}
C_{Q}=\frac{3r_{H}(r_{H}^{2}-Q^{2})(r_{H}^{2}+2\tilde{\alpha})^{2}}
{-r_{H}^{4}+(2\tilde{\alpha}+3Q^{2})r_{H}^{2}+2Q^{2}\tilde{\alpha}}\ .
\label{hcymmod}
\end{equation}

Notice that in this case the condition for a positive definite
temperature reads $r_H^2 > Q^2$. Moreover, the explicit presence of
the coupling constant $\tilde \alpha$ in the heat capacity leads to
the possibility of modifying the phase transition structure of the
black hole by changing the value of the GB coupling constant.
Indeed, the expression for the heat capacity (\ref{hcymmod})
diverges for 
\be 
r_H^2=\frac{3}{2}\,{Q}^{2}+\tilde{\alpha}\pm
\frac{1}{2}\,\sqrt {9\,{Q}^{4}+20\,\tilde{\alpha}\,{Q}^{2}+4\,{
\tilde{\alpha}}^{2}}\ , 
\label{roots} 
\ee indicating that for a
given value of the Yang-Mills charge it is possible to find a range
of values of $\tilde \alpha$ for which second order phase
transitions take place. This behavior is illustrated in
Fig.\ref{fig7}

\begin{figure}
 \includegraphics[width=5cm]{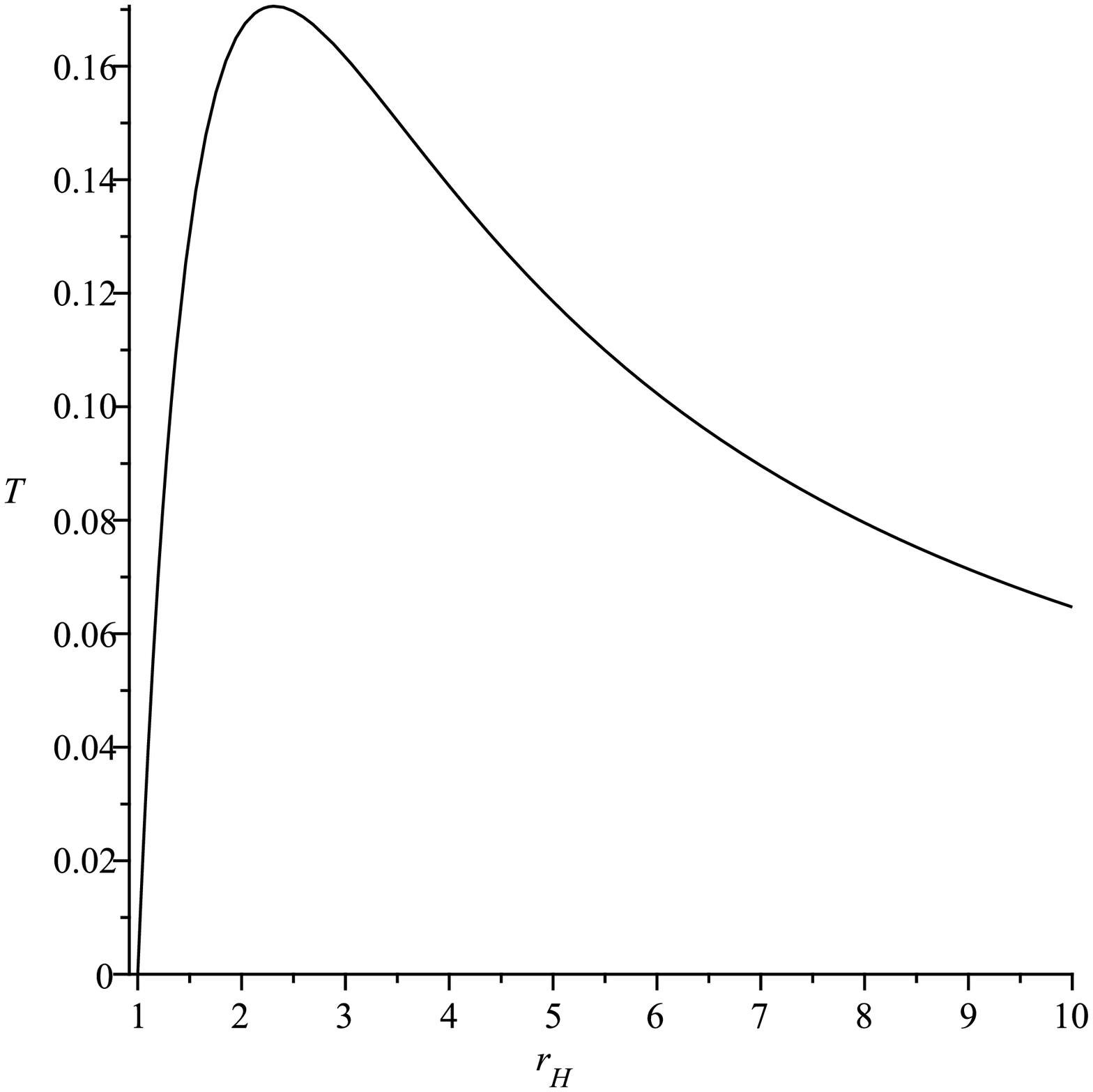}
 \includegraphics[width=5cm]{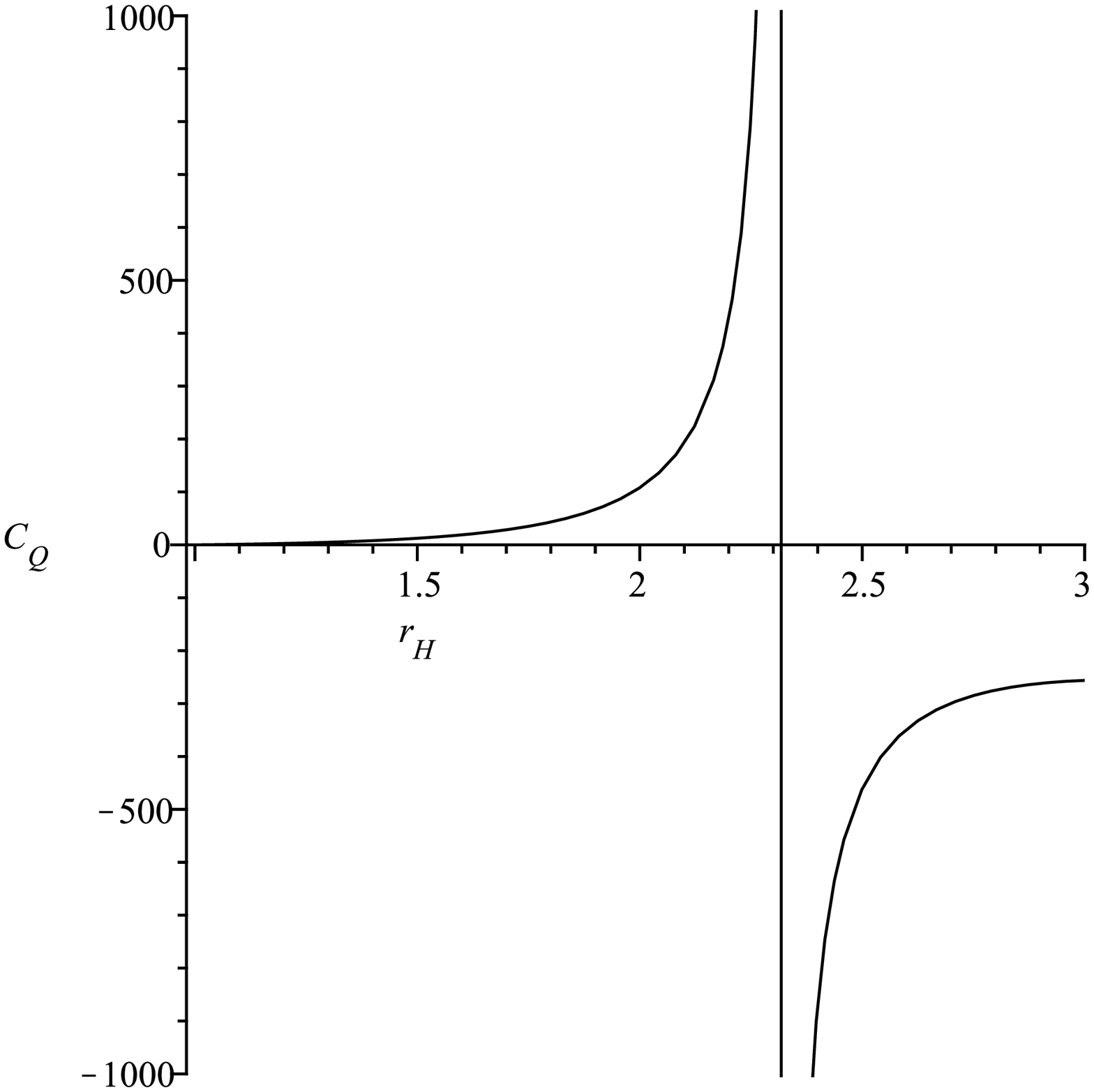}
 \includegraphics[width=5cm]{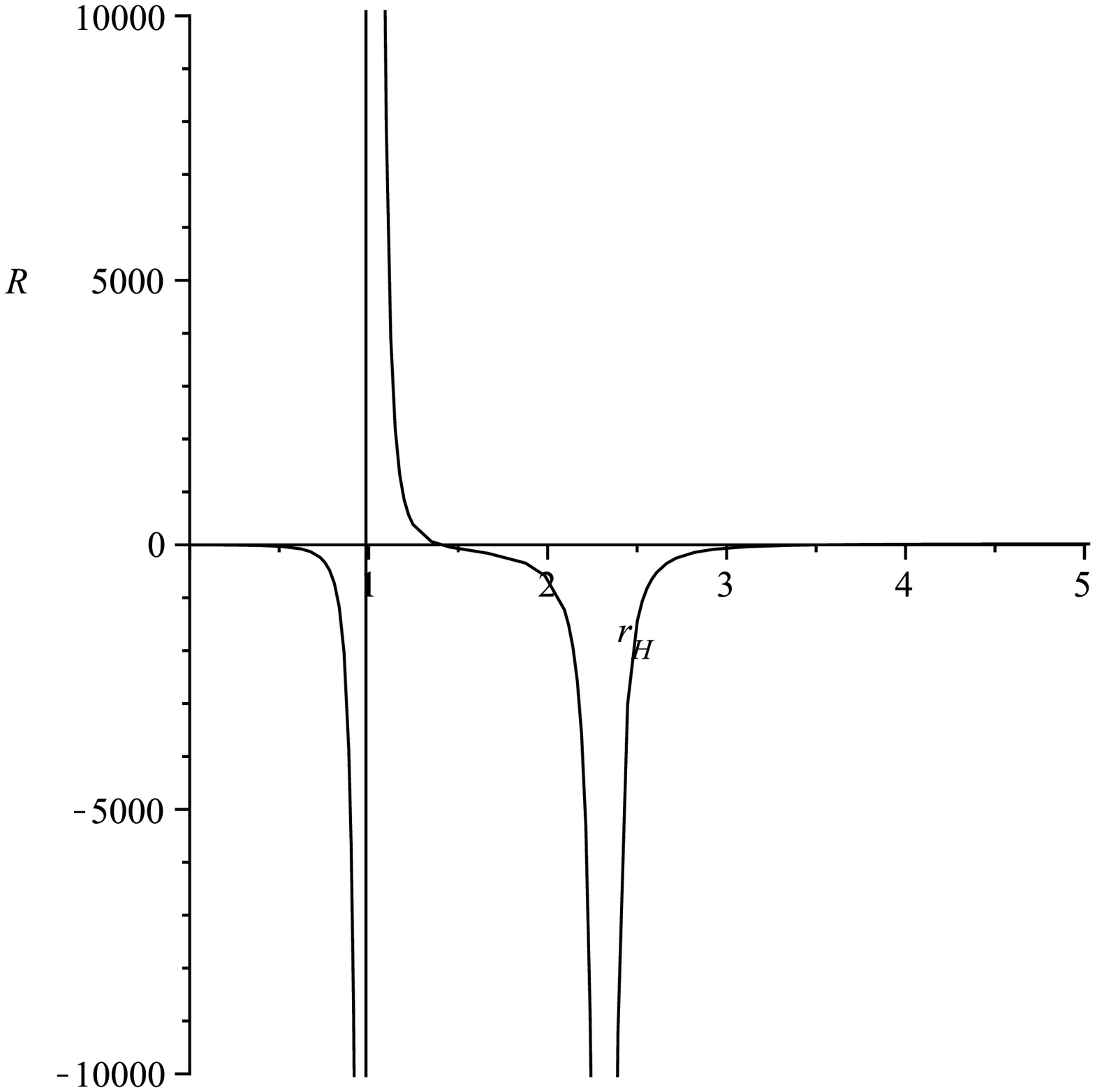}
  \caption{Temperature, heat capacity, and scalar curvature in terms of the
  horizon radius $r_{H}$ of a black in the EYMGB theory. Here the modified
entropy relation is used with $Q=1$ and $\tilde \alpha =1$. }
\label{fig7}
\end{figure}

For the thermodynamic system determined by  the fundamental equation
(\ref{feqmodym}), the Legendre invariant metric (\ref{gdown}) is
given by
\begin{equation}
g=-\frac{4}{27}\frac{(r^2_{H}+6\tilde{\alpha})(r^2_{H}-Q^2)}
{r^2_{H}(r^2_{H}+2\tilde{\alpha})^4}\bigg
\{\left[-r_{H}^{4}+(2\tilde{\alpha}+3Q^{2})r_{H}^{2}+2Q^{2}\tilde{\alpha}\right]dS^{2}
+18r^2_{H}(r^2_{H}+2\tilde{\alpha})^3 \ln {r_{H}}dQ^{2}\bigg\},
\end{equation}
The expression for the scalar curvature can be schematically written
as \be R= \frac{N(r_{H}, Q, \tilde \alpha
)}{\left[-r_{H}^{4}+(2\tilde{\alpha}+3Q^{2})r_{H}^{2}+2Q^{2}\tilde{\alpha}\right]^2
(r_H^2-Q^2)^3(r_H^2+6\tilde{\alpha})^3(\ln{r_H})^2 } \ , 
\ee where
$N(r_{H}, Q, \tilde \alpha )$ is a function that is finite at those
points where the denominator vanishes. We see that curvature
singularities occur at $r_H^2=Q^2$, which is the point where the
temperature vanishes, and at the roots of Eq.(\ref{roots}), which determine
the points where $C_Q\rightarrow\infty$ and second order phase transitions occur.  The
singularity situated at $\ln r_H \rightarrow 0$ corresponds to a
second order phase transition determined by the capacitance
$C_S\equiv (\partial Q/\partial \Phi)_S = -1/(4\ln r_H)$, according
to Eq.(\ref{feqmodym}). Finally, the singularity situated at $
(r_H^2+6\tilde{\alpha})=0$ corresponds to the limit $S\rightarrow 0$
which indicates the breakdown of the thermodynamic picture of the
black hole and, consequently, of GTD. A particular example of
the location of these curvature singularities is depicted in
Fig.\ref{fig7}.

\section{Conclusions}
\label{sec:con}

In this work, we analyzed the thermodynamics of static spherically
symmetric black holes in the five dimensional Einstein-Gauss-Bonnet
theory and its generalizations including an electromagnetic Maxwell
field (EMGB), a cosmological constant (EMGB$\Lambda$), and a
Yang-Mills field (EYMGB). This kind of black holes was also recently investigated in 
\cite{cbmc11} with results which are compatible with the ones obtained in the present analysis.
To investigate the thermodynamics of these
black holes we use two different approaches. The first one is based
upon the Bekenstein-Hawking entropy relation, according to which the
entropy of a black hole is proportional to the area of its event
horizon. The second approach uses as  starting point a modified
entropy relation that follows from the assumption that black holes
satisfy the first law of thermodynamics in higher dimensions. The
two approaches are not equivalent since the corresponding
thermodynamic variables exhibit completely different behaviors. In
particular, we noticed that the thermodynamics of black holes based
upon the modified entropy formula depends on the value of the
coupling constant of the Gauss--Bonnet term that appears in the
action of the theory. Phase transitions appear that depend on the
explicit value of the coupling constant, and change the stability
properties of the black holes. This kind of phase transitions is
absent when the Bekenstein-Hawking entropy relation is used.

To study the phase transition structure that follows from the Bekenstein-Hawking entropy we used the original 
Davies' definition, according to which divergences of the heat capacity at fixed charge $C_Q$ represent second order phase transitions, and an
alternative approach based on the analysis of the divergences of $C_\phi$, where $\phi$ is the electric potential dual to $Q$. For all three black holes studied in this work, we showed that the divergences of $C_Q$ do not coincide with the divergences of $C_\phi$ and, consequently, the corresponding phase transition structures are different. Moreover, we analyzed the behavior of all possible thermodynamic potentials in the parameter space $S-Q$, and found that the divergences of $C_\phi$ do not correspond to any particular point in the parameter space. On the contrary, the divergences of $C_Q$ are always situated on points where the Helmholtz free energy $F$ possesses an extremum. The remaining thermodynamic potentials do not show any special behavior at the singular points. To be more specific, in the case of the EMGB black holes we found that $C_Q$ predicts the existence of stable and unstable states with a phase transition that occurs in a region of stability of $F$, and during which the black holes undergoes a transition from a stable to an unstable state. For this black hole, the heat capacity $C_\phi$ implies the existence of only unstable states with no phase transitions. In the case of the EMGB black hole with cosmological constant, $C_Q$ predicts two phase transitions. During the first transition the black hole goes from a stable state to an unstable one whereas at the second divergence of $C_Q$ the black hole undergoes a phase transition from an unstable state to a stable state. In the parameter space, the first phase transition turns out to be located in a metastable region of $F$, and the second transition is situated in an unstable region of $F$. The alternative heat capacity $C_\phi$ predicts stable and unstable states  with two different divergences. At the first divergence, the black hole undergoes a transition from a stable state to an unstable state, and at the second divergence it becomes stable again. None of the thermodynamic potentials present extrema on the divergences of $C_\phi$.  As for the EYMGB black hole, the corresponding heat capacity $C_Q$ contains a singularity that describes the transition from a stable state to an unstable one. The singularity is located on a stable point of $F$. Finally, the capacity $C_\phi$ predicts two divergences the first of which corresponds to a transition from an unstable to a stable state whereas the second divergence corresponds to a transition to an unstable region.

For all the black holes analyzed in this work, we use the formalism
of geometrothermodynamics (GTD) to find the geometric properties of
the corresponding manifolds of equilibrium states. Once the
thermodynamic fundamental equation of the black is given, a standard
procedure of GTD allows us to compute the explicit form of the
thermodynamic metric that describes the geometric properties of the
equilibrium manifold. It turns out that the thermodynamic metrics
depend on the entropy relation used to construct the thermodynamics
of the black holes under consideration. 
The thermodynamic metrics obtained from the Bekenstein-Hawking relation 
were derived explicitly for the grand canonical  ensemble, with the thermodynamic potential $M=M(S,Q)$,
and for the canonical ensemble, with potential $H=H(S,\phi)$. In the first case, the curvature of the equilibrium manifold 
reproduces the thermodynamic behavior of the heat capacity at fixed charge $C_Q$ and, in the second case, the phase structure of $C_\phi$ is 
reproduced correctly. We also investigated an alternative thermodynamics based upon the modified entropy relation. In this case, we calculated for all 
the black holes the heat capacity $C_Q$, which predicts a non-trivial phase transition structure, and showed that GTD represents correctly the phase transitions as curvature singularities.
 Moreover, curvature singularities
also appear at those points where the temperature vanishes,
indicating the limit of applicability of black hole thermodynamics
and of GTD.

In the case of the thermodynamics based on the modified entropy relation, in principle, it could be possible to perform a similar analysis with different heat capacities and thermodynamic potentials. The computations, however, cannot be carried out in a similar manner, because it is not possible to write down explicitly the intensive variables in terms of the extensive ones. This means, for instance, that we cannot write explicitly the enthalpy $H$ in terms of $S$ and $\phi$ and so it is not possible to directly compute the thermodynamic metric, using the fundamental equation $H(S,\phi)$.  Of course, one could express all the derivatives with respect to $\phi$ in terms of derivatives with respect to the radius of the horizon $r_H$ that, in turn, is a function of $S$ and of the implicit function $Q(\phi)$. This is not an impossible task, but the computations become rather cumbersome; we do not believe that such an analysis would shed more light on the properties of GTD that has shown already to correctly describe the thermodynamics based on the Bekenstein-Hawking entropy relation.

We conclude that the formalism of GTD can be used in the EGB theory
in five dimensions to describe in an invariant manner the
thermodynamic properties of black holes in terms of geometric
concepts, regardless of the entropy relation used to formulate the
thermodynamics of black holes. In particular, we found that all the
black holes we analyzed in the EGB theory are characterized by
non-flat equilibrium manifolds. This means that all those black
holes possess an intrinsic non-trivial thermodynamic interaction.
Moreover, since we represent the thermodynamic interaction by means
of the curvature of the equilibrium manifold, the points where the
heat capacity diverges and, consequently, second order phase
transitions occur, are represented in GTD by curvature
singularities, indicating the limit of applicability of the
thermodynamic approach to black holes and of GTD.

We have seen that the phase transition structure of EMG black holes depend upon the thermodynamic ensemble, 
because the corresponding heat capacities have different singular behaviors. It is not clear which structure is 
the right one. Moreover, the investigation of thermodynamic potentials does not seem to clarify this question, because  
no obvious relationship was found between the critical points of the potentials and the divergences of the heat capacities. 
Although we found that the divergences of the heat capacity at fixed charge always occur at points where the Helmholtz free energy
has extrema, it is not clear why we should not take into account divergences of heat capacities that are not related to extrema of the potentials.
We believe that a classification of phase transitions that would take into account the Legendre invariance of ordinary thermodynamics
could shed some light into this issue. We expect to consider this problem in the near future.

\section*{Acknowledgements}
The authors are grateful to ICRANet for warm hospitality and
support. This work was supported in part by DGAPA-UNAM, grant No. IN106110, and Conacyt-Mexico, grant No. 166391.


\end{document}